\documentclass[prd,aps]{revtex4}
\usepackage{color}
\usepackage{mathtools}
\usepackage{array}
\usepackage{tabularx}
\usepackage{diagbox}
\usepackage{tikz}
\usetikzlibrary{decorations.pathmorphing}
\usepackage{bm}
\usepackage{amsmath,bm,amssymb,amsfonts,dcolumn,color,graphicx,latexsym,epsfig}
\usepackage{rsfso}
\usepackage[pagebackref=false, colorlinks=true]{hyperref}
\definecolor{reddish}{rgb}{0.9,0.3,0.0}  
\definecolor{blueish}{rgb}{0.1,0.1,1}
\hypersetup{
colorlinks=true,
linkcolor=reddish,
filecolor=reddish,      
urlcolor=magenta,
citecolor=magenta,
}
\usepackage{subfigure}
\usepackage{multirow}
\usepackage{natbib}
\usepackage{float}
\usepackage{booktabs}
\usepackage{pifont}
\usepackage{mathrsfs}
\usepackage{caption}
\usepackage{gensymb}
\usepackage[normalem]{ulem}
\usepackage{caption, threeparttable}
\usepackage{mathrsfs}

\DeclareMathOperator{\sn}{sn}
\DeclareMathOperator{\cn}{cn}
\DeclareMathOperator{\dn}{dn}

\begin{document}

\title{ Gravitational wave radiation from periodic orbits in regular black holes}
\author{Rishav Agrawal}
\email[Email address: ]{rishavrste@gmail.com}
\affiliation{Department of Physics, National University of Singapore, 2 Science Drive 3, Singapore 117551}
\author{Anjan Kar}
\email[Email address: ]{anjankar.phys@gmail.com}
\affiliation{Department of Physics, Indian Institute of Technology, Kharagpur 721 302, India}
\author{Soumya Jana}
\email[Email address: ]{soumyajana.physics@gmail.com}
\affiliation{Department of Physics, Sitananda College, Nandigram, 721 631, India}
\author{Sayan Kar}
\email[Email address: ]{sayan@phy.iitkgp.ac.in}
\affiliation{Department of Physics, Indian Institute of Technology, Kharagpur 721 302, India}
\begin{abstract}
\noindent Gravitational wave radiation from periodic orbits in
some standard regular black hole spacetimes is studied, primarily
using known methods (numerical and analytic). We demonstrate specific differences with the
singular Schwarzschild geometry by analysing orbit characteristics, gravitational wave strain profiles, and the corresponding power spectrum
density, for different values of the regularising parameter `$g$'. Further, 
we assess our results  {\em vis-a-vis}  the LISA sensitivity curves and show how our
results may be useful while developing templates for detecting regular black holes
as viable alternatives to the singular ones. The appendices to our article contain details on errors
in our estimates and provide for the first time, some exact analytical expressions
on gravitational wave radiation from different types of periodic orbits in Schwarzschild spacetime.

\noindent 
\end{abstract}


\maketitle


\section{Introduction}
\noindent It is well-known from early work that the energy and spectrum of emitted gravitational radiation due to a point test particle of mass $m$ falling radially towards a Schwarzschild black hole of mass $M>>m$ can be estimated using standard tools of gravitational wave (GW) physics~\cite{Davis: 1971}. In a similar fashion, any object in an orbit about a black hole will also emit gravitational radiation, which is easily calculable, at least numerically. In this article, following the current upsurge in research on gravitational radiation from periodic orbits in various types of spacetimes (mainly black holes), we investigate and present the details on similar calculations, results and comparisons, primarily for regular black holes. Our central goal is to obtain signatures specific to regular black holes and show how they may be different from similar ones for their singular counterparts. Through this approach and the ensuing results, we hope to provide templates which may be used in future observational ventures such as LISA~\cite{Danzmann:1997, Schutz:1999, Gair:2004, LISA:2017, Maselli:2022}, Tianqin~\cite{Luo:2016, Gong:2021}, Taiji~\cite{Hu:2017}, where we expect to successfully detect GWs from typical systems known as extreme-mass-ratio inspirals (EMRIs in short). 

\noindent The two distinct basic inputs in our work here from
past literature are (a) regular black holes and (b) periodic 
orbits in such spacetimes. Let us now briefly recall qualitatively
the bare essentials on them, which are needed to proceed further.

\noindent Regular black holes were first proposed as nonsingular alternatives to their singular counterparts way back in 1969, first through the seminal work of Bardeen~\cite{Bardeen:1968}. Since then, there have been several proposed spacetimes--the Hayward spacetime~\cite{Hayward:2006}, Fan-Wang geometry~\cite{Fan:2016}, Dymnikova spacetime~\cite{Dymnikova:1992}, and many others~\cite{Roman:1983, Ayon:1998, Ayon:1999, Ayon:1999grg, Ayon:2000, Bronnikov:2001, Ayon:2005, Balart:2014, Bronnikov:2018, Poshteh:2021, Bronnikov:2024, Kar:2024, Kar:2024kk}, including some recent ones~\cite{Borissova:2025, Bueno:2025, Eichhorn:2025, Muniz:2025, Konoplya:2025, Santos:2025, Aydiner:2025, Santos:2025prd, Kar:2025}. A regular black hole has horizon(s), usually more than one (i.e. inner and outer), no singularity but a de Sitter or Minkowski core replacing it. The thermodynamics of regular black holes is not yet fully understood--a major block being the connection between singular behaviour and the non-compatibility with the usual first law~\cite{Myung:2009}. In the more recent past there has been a large body of work on regular black holes with special emphasis on observational signatures~\cite{Wu:2025, Bhattacharjee:2025, Kar:2025xx, Bolokhov:2025, Wang:2025, Saka:2025, Arbelaez:2026, Acharyya:2026}. Our work is mainly centred around the well-known Bardeen and Hayward geometries.

\noindent Periodic orbits in black hole spacetimes 
(essentially the Schwarzschild) were first
characterised in a quantitative way in the work of
Levin~\cite{Levin:2009sk}. Thereafter, periodic orbits have been studied in various spacetimes, including Kerr geometry~\cite{Levin:2008ci, Rana:2019bsn, Bambhaniya:2020zno}, charged black holes~\cite{Misra:2010pu}, Kerr-Sen black holes~\cite{Liu:2018vea}, and other scenarios~\cite{Healy:2009zm, Babar:2017gsg, Pugliese:2013xfa, Wei:2019zdf, Zhou:2020zys, Gao:2020wjz, Deng:2020hxw, Azreg-Ainou:2020bfl, Gao:2021arw, Lin:2021noq, Lin:2022llz, Wang:2022tfo, Zhang:2022psr, Lin:2022wda, Zhang:2022zox, Yao:2023ziq, Lin:2023rmo, Habibina:2022ztd, Chan:2025ocy, Al-Badawi:2025yum, Wang:2025wob, Sharipov:2025yfw, Wei:2025qlh}.
The gravitational waves emitted from periodic orbits in different types of spacetimes are studied in Refs.~\cite{Tu:2023xab, Li:2024tld, QiQi:2024dwc, Yang:2024lmj, Shabbir:2025kqh, Junior:2024tmi, Zhao:2024exh, Meng:2024cnq, Haroon:2025rzx, Lu:2025cxx, Chen:2025aqh, Choudhury:2025qsh, Alloqulov:2025ucf, Wang:2025hla, Alloqulov:2025bxh, Li:2025sfe, Gong:2025mne, Zare:2025aek, Zahra:2025tdo, Li:2025eln, Deng:2025wzz, Ahmed:2025azu, Zhang:2025wni, Alloqulov:2025dqi, Lu:2025xxx, Ahmed:2025xxx, Chen:2025xxx, Hua:2025xxx, Gaete:2026}.

\noindent  We revisit the characterisation of periodic orbits briefly in subsequent Sections~\ref{II} and~\ref{III}. Section~\ref{IV}  recalls standard topics on
gravitational radiation for periodic orbits, in particular, the waveforms.  The power spectrum density and the strain are 
analysed in Section~\ref{V}. Section~\ref{VI} is a conclusion. In the two
appendices, we focus on checks on our numerical work by presenting the
Schwarzschild case explicitly and analytically. We also verify 
that our waveform construction tallies well with the reconstructed
wave found numerically.

\noindent Unless otherwise stated, we work in $G=c=1$ units.

\section{Geodesics and effective potential of regular black hole}\label{II}
\begin{figure}[h]
\centering
\subfigure[\hspace{0.1cm}Bardeen black hole]{\includegraphics[width=0.45\textwidth]{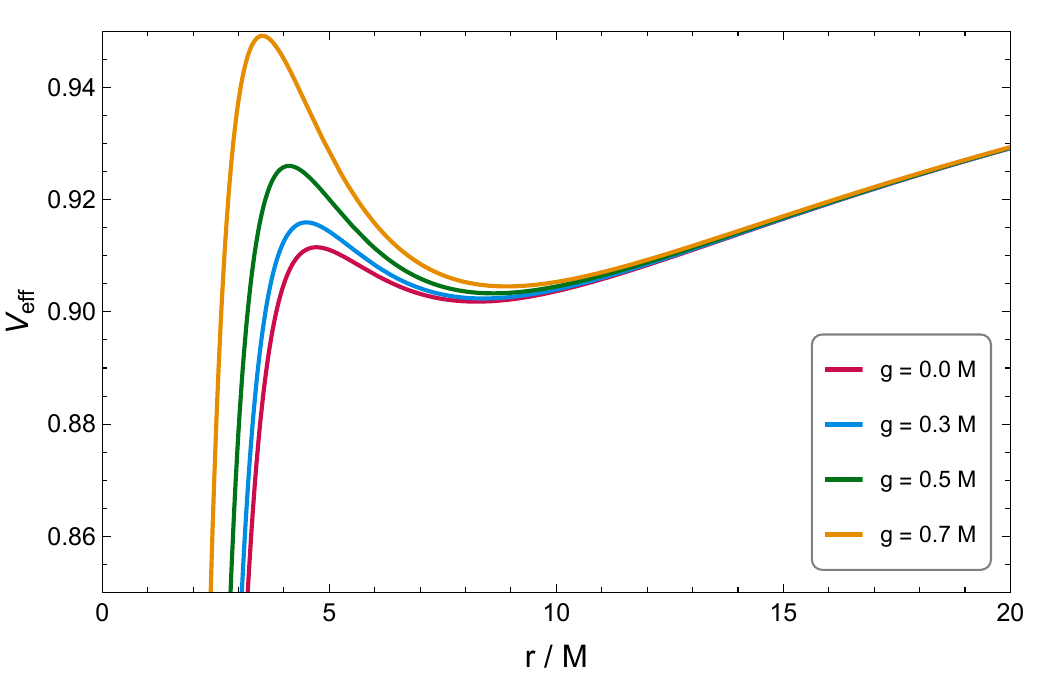}}
\subfigure[\hspace{0.1cm}Hayward black hole]{\includegraphics[width=0.45\textwidth]{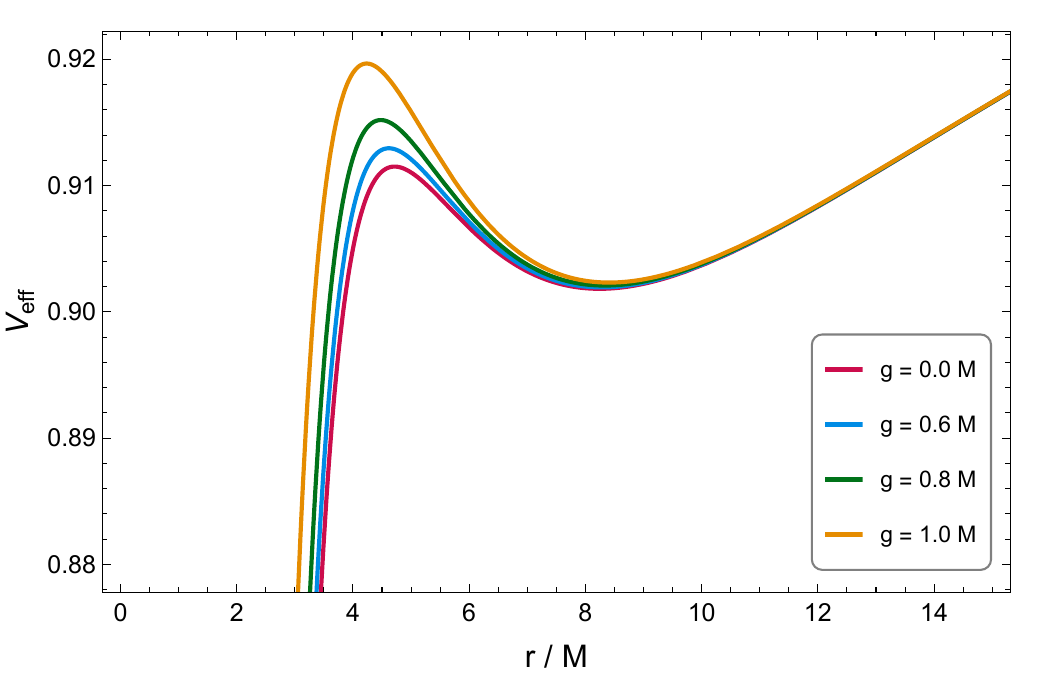}}
\caption{Plot of the effective potentials of the Bardeen and Hayward spacetimes for different values of $g$ with $L=3.6M$.}
\label{fig:effectivepotential}
\end{figure}

\noindent In this section, we provide a brief review of the geodesics around static, spherically symmetric regular black holes. The line element of such a general spacetime is expressed as
\begin{equation}
    ds^2= -f(r) dt^2 + \frac{dr^2}{f(r)} +r^2 \left( d\theta^2 +\sin^2\theta d\phi^2\right),
\end{equation}
where for the Bardeen regular black hole~\cite{Bardeen:1968}
\begin{equation}\label{2.2}
    f(r)=1- \frac{2Mr^2}{(r^2+g^2)^{3/2}}\,.
\end{equation}
and for the Hayward regular black hole we have ~\cite{Hayward:2006}
\begin{equation}\label{2.3}
    f(r)=1- \frac{2Mr^2}{r^3+g^3}\,.
\end{equation}
Here, $M$ is the Arnowitt--Deser--Misner mass and $g$ is the `regularisation parameter'. In the limit of very small $r$, the above geometries behave like de-Sitter spacetime, whereas for $r>>g$ they converge to the Schwarzschild metric and, further away, asymptotically they tend to the flat metrics. The appearance of the parameter $g$ in the metric ensures that all the independent curvature scalars are divergence-free, resulting in a regular spacetime. In the limit $g\to 0$, the above spectimes reduce to the Schwarzschild spacetime. By analysing the causal structure of these two geometries, one finds that the Bardeen metric admits either a double or a single horizon for $0<g^2/M^2\leq16/27$, while for the Hayward black hole, the condition is $0<g^3/M^3\leq32/27$. As we are primarily interested in studying black hole geometries, we strictly follow these conditions throughout this article.
\noindent We consider a massive test particle orbiting around a regular black hole. The Lagrangian of the particle is~\cite{Chandrasekhar:1985kt}
\begin{equation}
    \mathscr{L}=\frac{m}{2}g_{\mu\nu}\,\dot{x}^{\mu}\dot{x}^{\nu}\,,
\end{equation}
where $m$ is the mass of the particle and over-dot represents the derivative with respect to the affine parameter. For simplicity, we consider $m=1$ as this choice does not influence the geodesic motion. The generalised momentum of the particle becomes $p_{\mu}=\partial\mathscr{L}/\partial\dot{x}^{\mu}=g_{\mu\nu}\dot{x}^{\nu}$.
Therefore, the equations of motion of the particle are given as
\begin{equation}
    \begin{aligned}
    p_t=-f(r)\dot{t}=-E\,,\,
    p_{\phi}=r^2\,\sin^2{\theta}\,\dot{\phi}=L\,,\\
    p_{r}=\frac{\dot{r}}{f(r)}\,,\,
    p_{\theta}=r^2\,\dot{\theta}\,,
\end{aligned}
\end{equation}
where $E$ and $L$ are the energy and angular momentum of the particle per unit mass, respectively. Due to the spherical symmetry of the system, we set the orbit in the equatorial plane $(\theta=\pi/2)$, and from the normalisation condition of the four-velocity of a massive particle, we have
\begin{equation}
    \dot{r}^2=E^2-f(r)\left(1+\frac{L^2}{r^2}\right)\,.\label{2.8}
\end{equation}
Thus, the geodesic motion around a regular black hole depends explicitly on the parameter $g$. We rewrite Eq.\eqref{2.8} in the following form
\begin{equation}
    \dot{r}^2=E^2-V_{\text{eff}}\,,
\end{equation}
where $V_{\text{eff}}$ represents the effective potential for radial motion
\begin{equation}\label{2.10}
    V_{\text{eff}}=f(r)\left(1+\frac{L^2}{r^2}\right)\,.
\end{equation}
In Fig.~\ref{fig:effectivepotential}, the radial dependence of the effective potentials for both the Bardeen and Hayward black holes is presented by considering $L=3.6M$. Here, the maximum and minimum of the effective potential correspond to the unstable and stable circular orbits, respectively.
Asymptotically, the effective potential approaches unity. $E=1$ is the critical energy associated with the orbits.  We have a bound orbit when $E\leq1$.
From Fig.~\ref{fig:effectivepotential}, we observe that the values of the effective potential increase as $g$ increases. This increment with $g$ is smaller for the Hayward black hole relative to the Bardeen case.

\noindent The main focus of this article is to study the periodic orbits around a regular black hole. Periodic orbits are bounded in nature. For a particle moving in a bound orbit, its energy and angular momentum satisfy the following conditions,
\begin{equation}
    E_{\text{ISCO}}\leq E\leq E_{\text{MBO}}=1 \quad \text{and}\quad L_{\text{ISCO}}\leq L\,,
\end{equation}
where $E_{\text{ISCO}}$ and $L_{\text{ISCO}}$ are the energy and angular momentum of the particle moving in Innermost Stable Circular Orbit (ISCO), respectively. $E_{\text{MBO}}$ represent the energy of the particle moving on Marginally Bound Orbit (MBO). A particle with energy $E>1$ will escape to infinity, and when $E<E_{\text{ISCO}}$, it will fall into the horizon. Given the effective potential in Eq.\eqref{2.10}, the condition for MBO in a regular black hole is,  
\begin{equation}
    V_{\text{eff}}=E^2=1 , \qquad~\partial_r V_{\text{eff}}=0\,.
\end{equation}
One can numerically find the MBO for the given $M$ and $g$.
Another important bound orbit is the ISCO, which is defined by the following condition,
\begin{equation}
    V_{\text{eff}}=E^2  , \qquad~\partial_r V_{\text{eff}}=0, \qquad~\partial_r^2 V_{\text{eff}}=0\,.
\end{equation}
By solving the above equation, we find the formulae for $r_{\text{ISCO}}$, $L_{\text{ISCO}}$, and the corresponding $E_{\text{ISCO}}$ as follows
\begin{eqnarray}
    r_{\text{ISCO}}&=& \frac{3 f(r_{\text{ISCO}})f'(r_{\text{ISCO}})}{2f'^2(r_{\text{ISCO}})- f(r_{\text{ISCO}})f''(r_{\text{ISCO}})}\,,\label{eq:r_ISCO}\\
    L_{\text{ISCO}}&=& r_{\text{ISCO}}^{3/2}\sqrt{\frac{f'(r_{\text{ISCO}})}{2f(r_{\text{ISCO}})-r_{\text{ISCO}}f'(r_{\text{ISCO}})}}\,, \label{eq:L_ISCO}\\
    E_{\text{ISCO}} &=& \frac{\sqrt{2}f(r_{\text{ISCO}})}{\sqrt{2f(r_{\text{ISCO}})-r_{\text{ISCO}}f'(r_{\text{ISCO}})}}\,.
\end{eqnarray}

\begin{figure}[h]
\centering
\subfigure[\hspace{0.1cm}Bardeen black hole]{\includegraphics[width=0.45\textwidth]{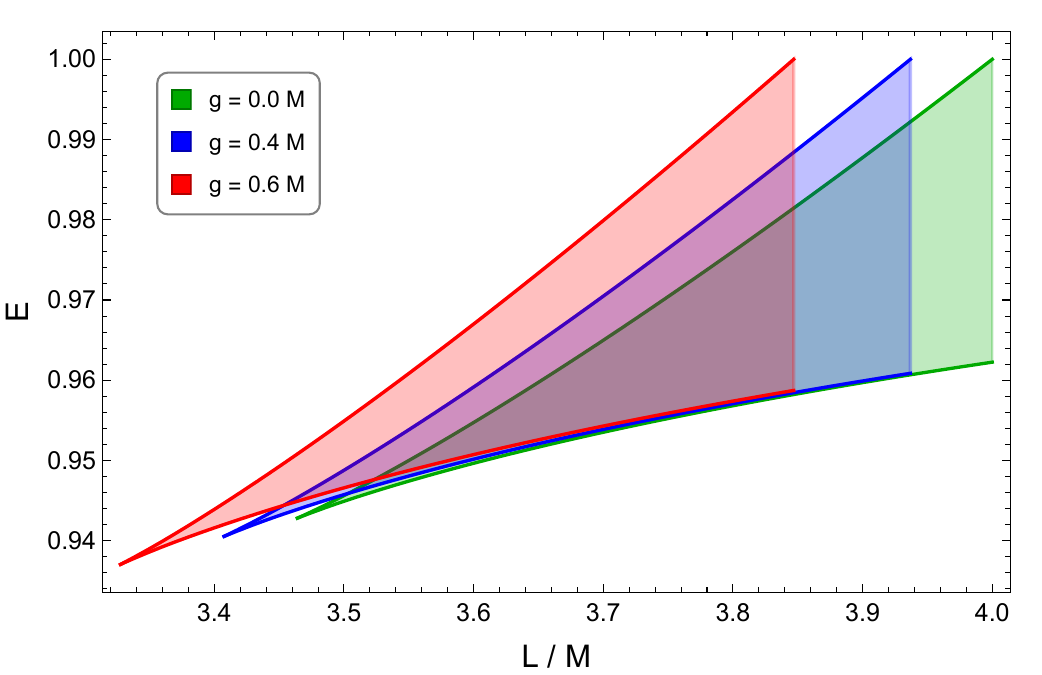}}
\subfigure[\hspace{0.1cm}Hayward black hole]{\includegraphics[width=0.45\textwidth]{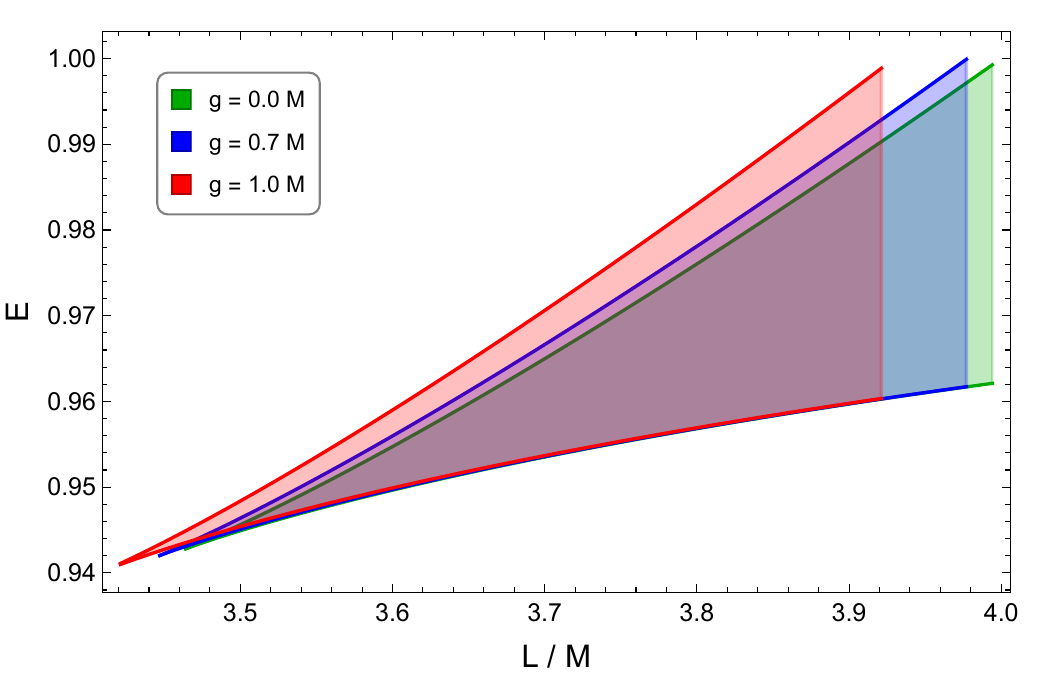}}
\caption{Plot of the allowed values of $E$ and $L/M$ for bound orbits for the Bardeen and Hayward black hole for different values of $g$.}
\label{fig:bound_orbit}
\end{figure}

\noindent To provide further insight, we plot (Fig.~\ref{fig:bound_orbit}) the allowed values of energy and angular momentum for bound orbits in the $E-L/M$ plane for both the Bardeen and Hayward black hole, for different values of $g$. It is found that the allowed region in the $E-L/M$ plane shifts to the left as the parameter $g$ increases. The shift is relatively small in the Hayward case.

\section{Periodic orbits around a regular black hole}\label{III}
\begin{figure}[h]
\centering
\subfigure[\hspace{0.1cm}$\Delta\phi=2\pi$]{\includegraphics[width=0.3\textwidth]{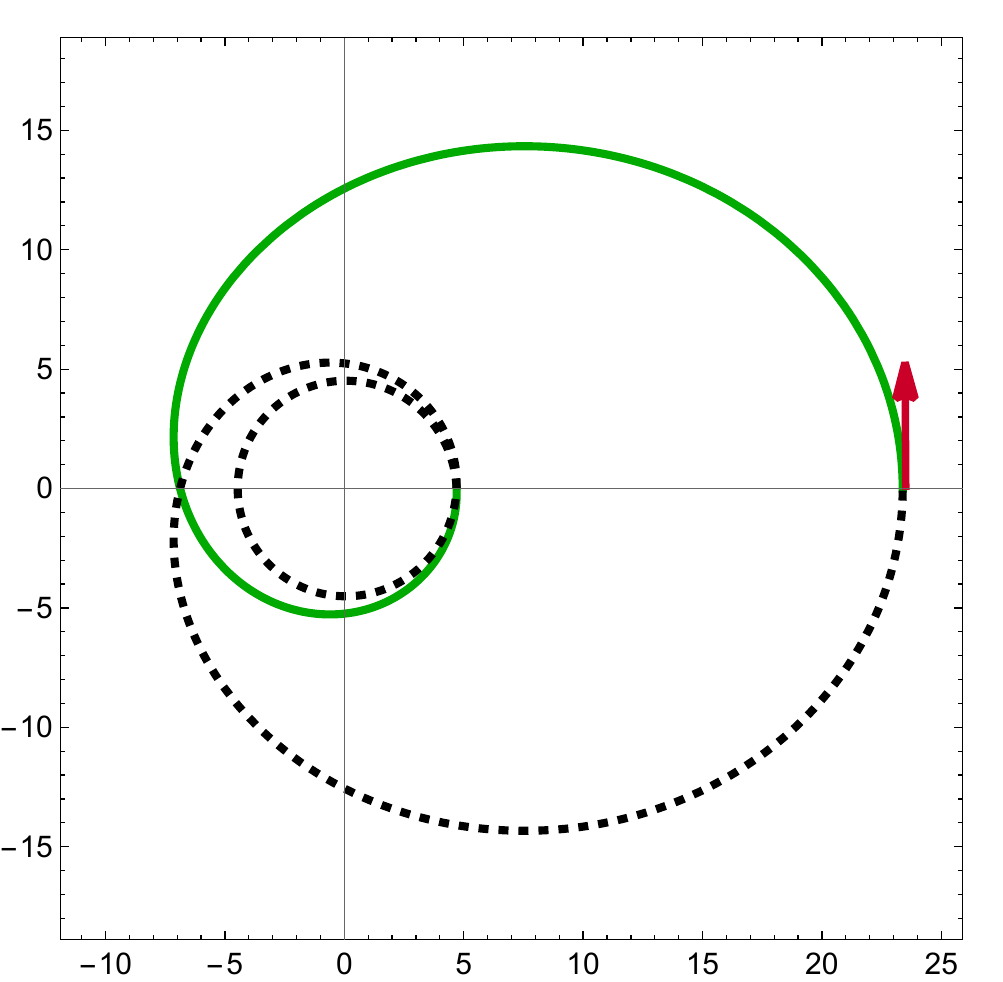}\label{subfig:D1}}
\subfigure[\hspace{0.1cm}$\Delta\phi=4\pi$]{\includegraphics[width=0.3\textwidth]{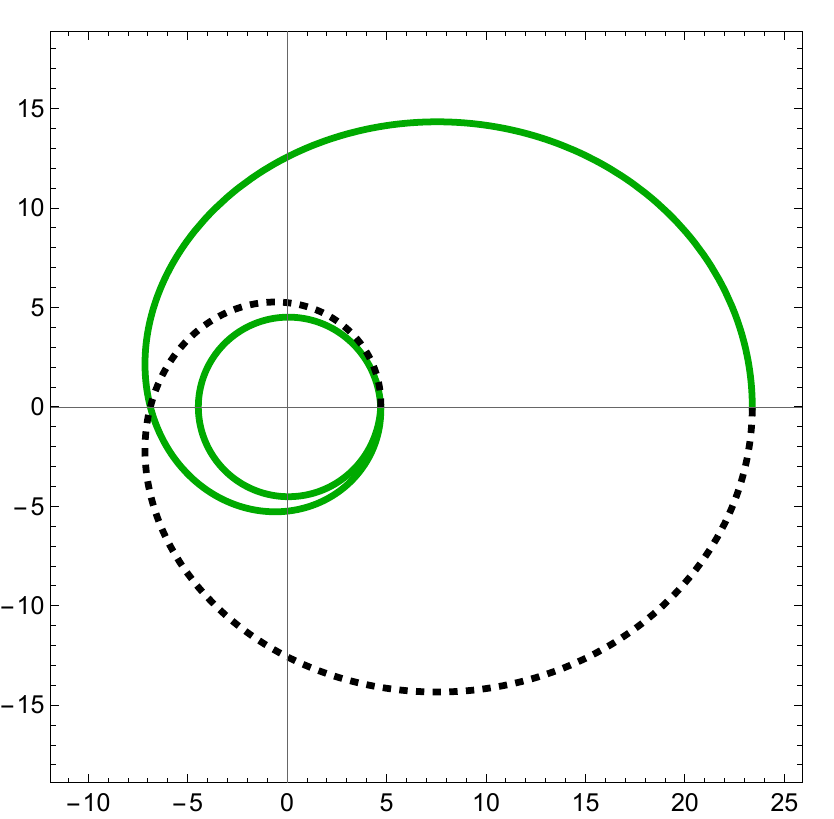}\label{subfig:D2}}
\subfigure[\hspace{0.1cm}$\Delta\phi=6\pi$]{\includegraphics[width=0.3\textwidth]{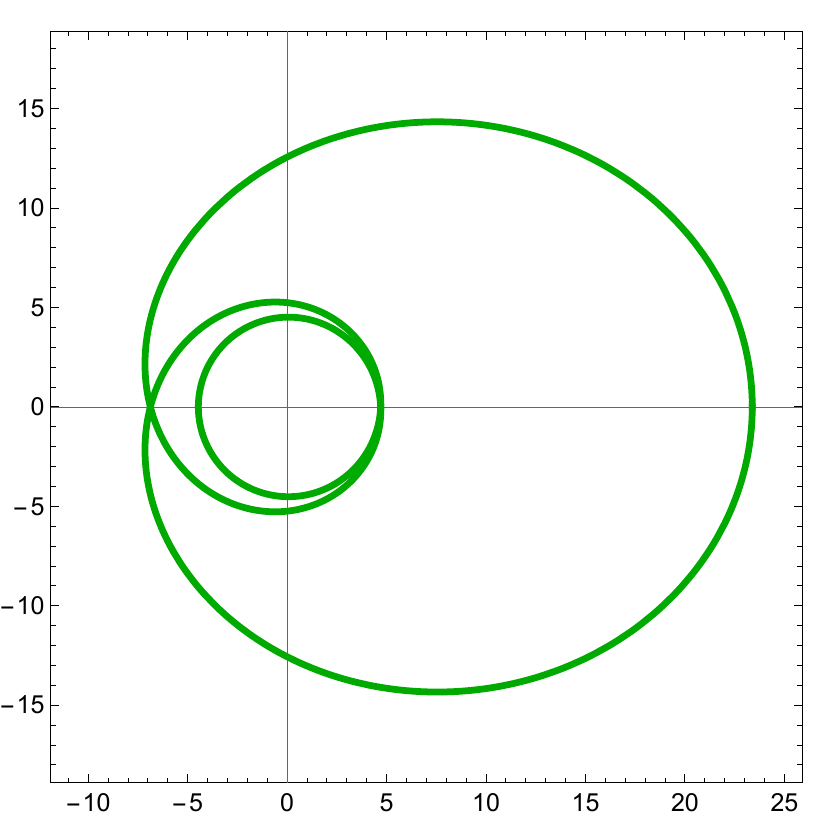}\label{subfig:D3}}
\caption{The trajectory of a typical $(1,2,0)$ orbit after $\phi$ has evolved by $2\pi$, $4\pi$, and $6\pi$, respectively. The starting point is marked with a red arrow.}
\label{fig:Periodic_orbit_demo}
\end{figure}

\noindent In this section, we focus on periodic orbits near a regular black hole.
These orbits are a specific class of bound orbits that return to their initial position after regular intervals of a finite amount of time. In a spherically symmetric black hole spacetime with $\theta=\pi/2$ and $\dot{\theta}=0$, such an orbit is fully determined by the angular velocities in the radial ($r$-motion) and azimuthal ($\phi$-motion) directions. Periodic orbit forms when the ratio of these two frequencies is rational~\cite{Levin:2008}. Following Ref.~\cite{Levin:2008}, we define the ratio as
\begin{equation}\label{3.n1}
    \frac{\Omega_{\phi}}{\Omega_r}= \frac{\Delta \phi_r}{2\pi}= 1+q,
\end{equation}
where $q$ is a rational number. The $\Omega_{\phi}$ and $\Omega_r$ are the angular frequency averaged over one radial period (i.e., the time between two consecutive apastron radial distances) and the radial frequency, respectively.
$\Delta\phi_r$ represents the total azimuthal angle swept by the particle on the equatorial plane during one complete radial cycle. It has to be an integer multiple of $2\pi$ to form a periodic orbit. The azimuthal deviation $\Delta\phi_r$ in the $\phi$-direction over a complete radial period is given as
\begin{equation}\label{3.2}
    \Delta \phi_r = \oint d\phi = 2 \int^{r_a}_{r_p} \frac{\dot{\phi}}{\dot{r}} dr = 2 \int^{r_a}_{r_p} \frac{L dr}{r^2 \sqrt{E^2 - f(r)\left(1+\frac{L^2}{r^2}\right)}},
\end{equation}
where $r_p$ is the periastron radius and $r_a$ is the apastron radius. The factor $2$ appears as one complete radial period consists of motion from apsis to perihelion and then back again to the same initial position. Thus, the total path is twice the separation between two turning points.
Note that $\Omega_r$ is defined as the radial frequency, $\Large \Omega_r= \frac{2\pi}{T_r}$, $T_r$ being the coordinate time elapsed for one radial cycle (not the total time period of an orbit),
\begin{equation}\label{3.3}
    T_r= 2\int^{r_a}_{r_p} \frac{dt}{dr}dr= 2\int^{r_a}_{r_p}\frac{\dot{t}}{\dot{r}}dr= 2\int^{r_a}_{r_p}\frac{E dr}{f(r)\sqrt{E^2- f(r)\left(1+\frac{L^2}{r^2}\right)}}.
\end{equation}
Time averaged value of $\Large \frac{d\phi}{dt}$ over one radial period is defined as the average orbital angular frequency $\Omega_{\phi}$,
\begin{equation}\label{3.4}
    \Omega_{\phi}= \frac{1}{T_r}\int^{T_r}_0 \frac{d\phi}{dt}dt= \frac{\Delta \phi_r}{T_r}.
\end{equation}
From Eq.~\eqref{3.2}, it is evident that the rational number $q$ depends on the particle's energy $E$, its angular momentum $L$, and the regular black hole metric function $f(r)$. To form a bound orbit, $L$ can vary between $L_{ISCO}$ and $L_{MBO}$. Thus, for a given bound orbit, it can be parametrised in the following form
\begin{equation}
    L=L_{ISCO}+\epsilon(L_{MBO}-L_{ISCO})\,,
\end{equation}
where $\epsilon$ is restricted to the range $(0,1)$, with $\epsilon=0$ corresponding to the $L_{ISCO}$ and $\epsilon=1$ corresponding to the $L_{MBO}$.
When $\epsilon>1$, there are no bound orbits.
\noindent Alternatively, the periodic orbits are classified by a set of three parameters ($z,w,v$)~\cite{Levin:2008}. Here $z$ denotes the zoom number--the number of full radial cycles (one apastron to next apastron) in a single orbital period, $w$ denotes the whirl number, i.e. the number of full circular turnings around the central black hole while passing through the periastron, and $v$ is the vertex number denoting the order of approaching the next apastron position.
These parameters are confined in the range $1\leq v \leq z-1$. The rational number $q$ of a periodic orbit can be expressed in terms of $(z,w,v)$ in the following way~\cite{Levin:2008}
\begin{equation}\label{3.1}
    q=  \omega + \frac{v}{z}\,.
\end{equation}
The parameter $q$ characterises the extent of periapsis precession relative to a simple elliptic orbit, which reflects the topological structure of the orbit.
Since a complete orbital cycle contains $z$ full radial cycles, the total azimuthal deviation is $\Delta \phi= z\Delta \phi_r$. In Fig.~\ref{fig:Periodic_orbit_demo}, we present a typical periodic orbit $(1,2,0)$ in the $r-\phi$ plane, which consists of one zoom and two whirl motions with a total $\Delta\phi=6\pi$.
The red arrow marks the initial position of the particle at the apastron.
We divide the trajectory into three segments, each associated with $2\pi$ evolution in $\phi$, colored in green. It demonstrates the orbital progression in a full orbital cycle.
Next, we study the periodic orbit by incorporating both the Bardeen and Hayward mass functions.

\subsection{Bardeen regular black hole}
\begin{figure}[h]
\centering
\includegraphics[width=0.45\textwidth]{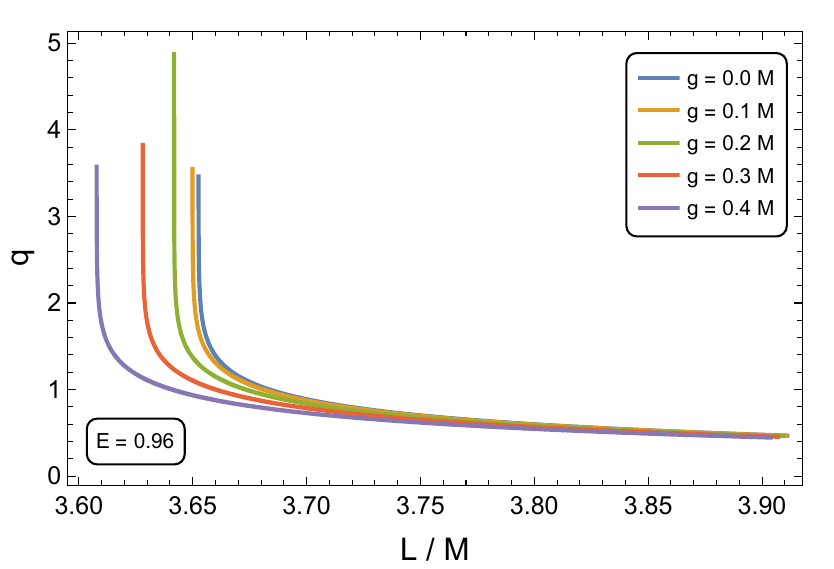}
\includegraphics[width=0.45\textwidth]{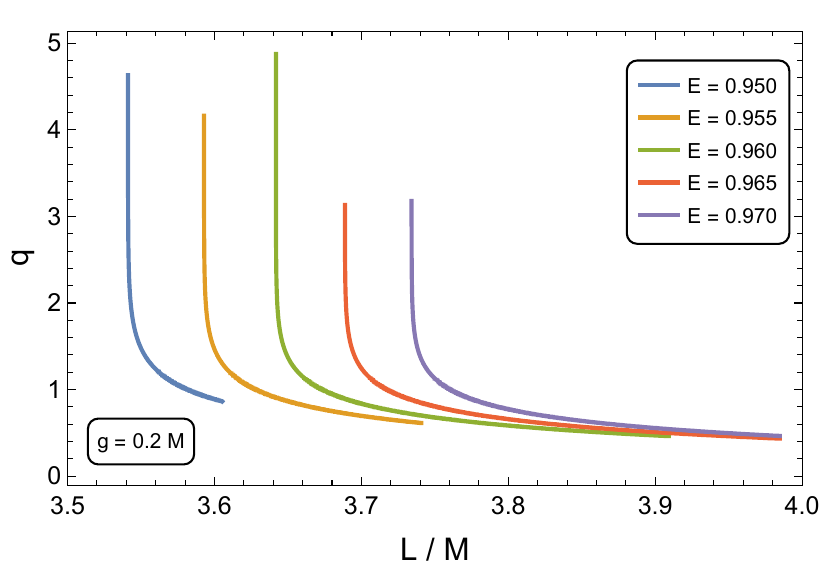}
\caption{Variation of $q$ with $L/M$ for different values of $g$ (left) and $E$ (right) for the Bardeen regular black hole.}
\label{fig:BqvsL}
\end{figure}

\begin{figure}[h]
\centering
\includegraphics[width=0.5\textwidth]{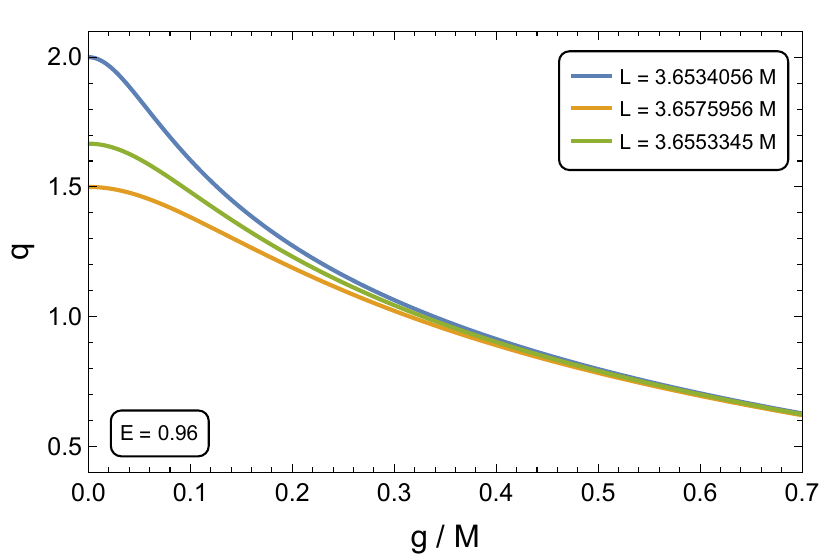}
\caption{Plot of $q$ with $g$ for fixed values of energy $E$ and angular momentum $L$ of the Bardeen regular black hole.}
\label{fig:Bqvsg}
\end{figure}

\begin{figure}[h]
\centering
\subfigure[\hspace{0.1cm}$E=0.9680973$]{\includegraphics[width=0.3\textwidth]{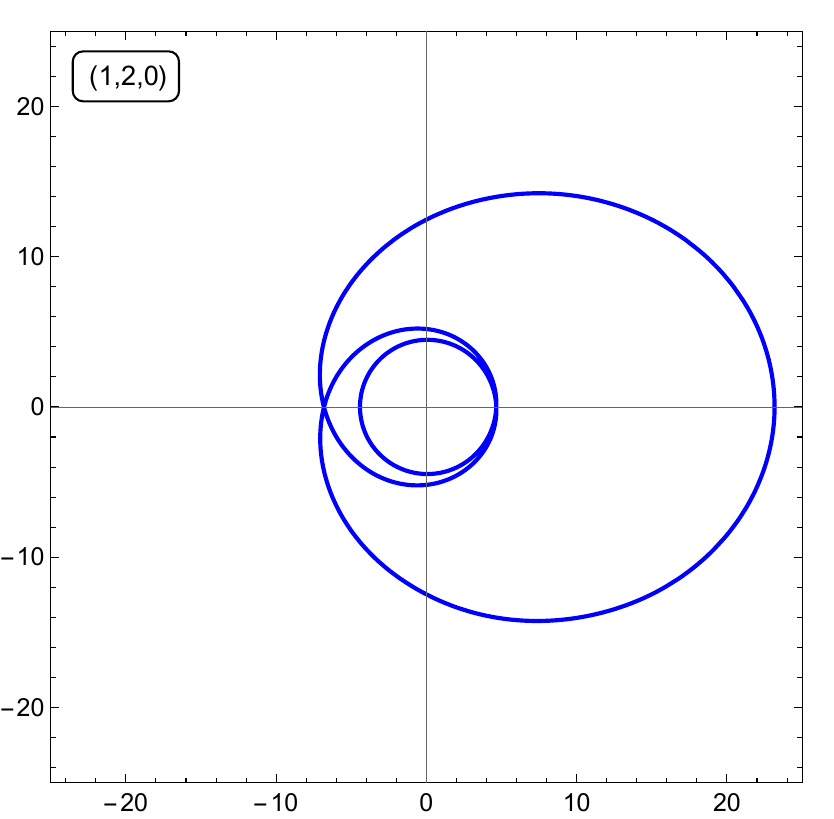}\label{subfig:B21}}
\subfigure[\hspace{0.1cm}$E=0.9677238$]{\includegraphics[width=0.3\textwidth]{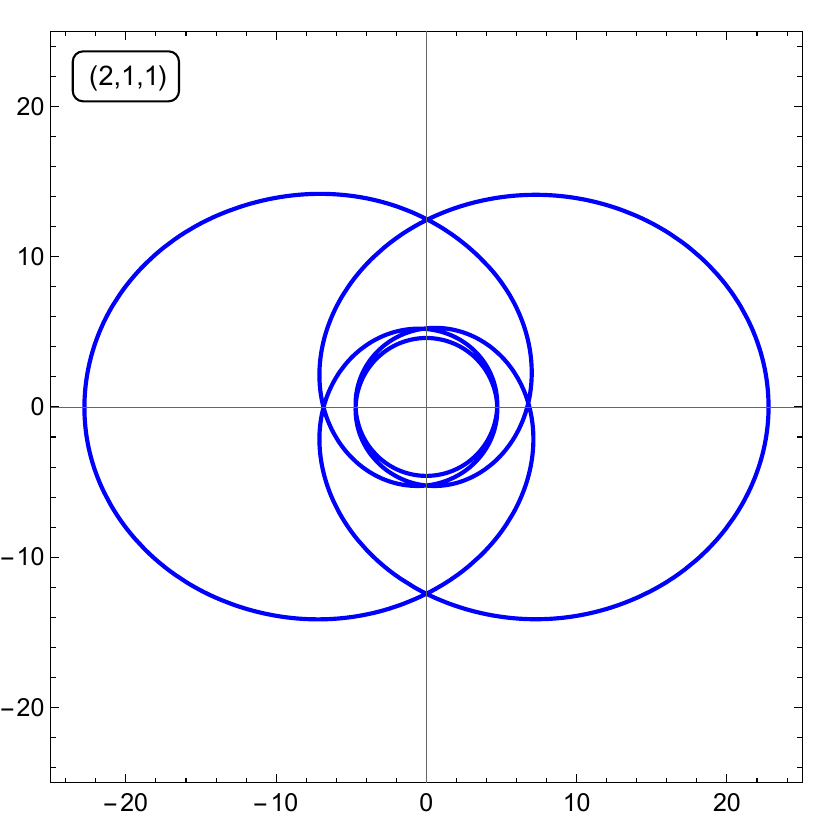}\label{subfig:B22}}
\subfigure[\hspace{0.1cm}$E=0.9681520$]{\includegraphics[width=0.3\textwidth]{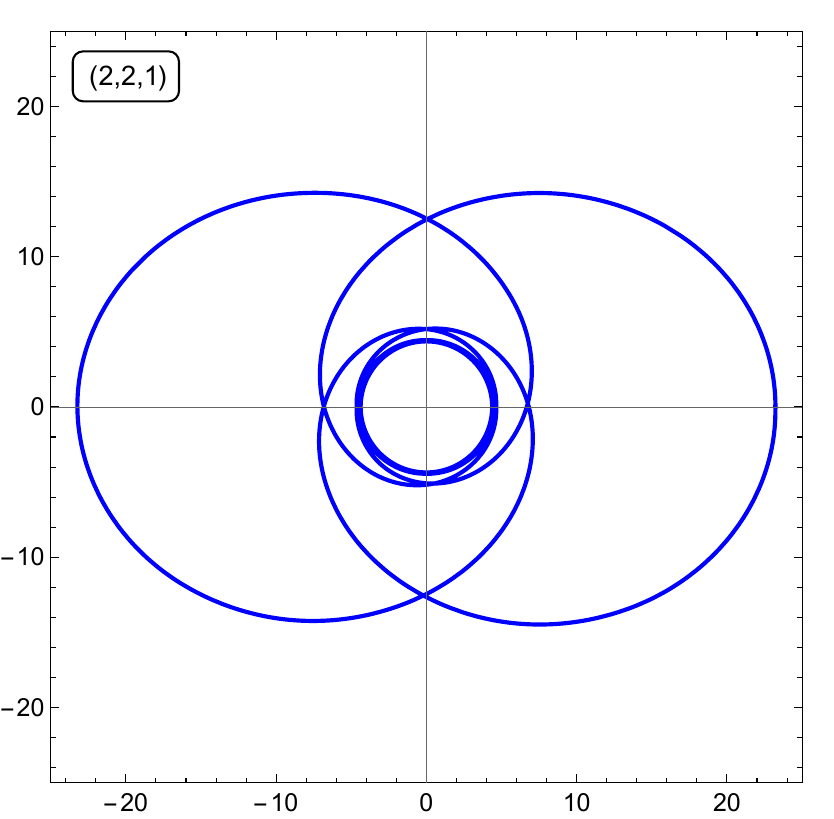}\label{subfig:B23}}
\subfigure[\hspace{0.1cm}$E=0.9679312$]{\includegraphics[width=0.3\textwidth]{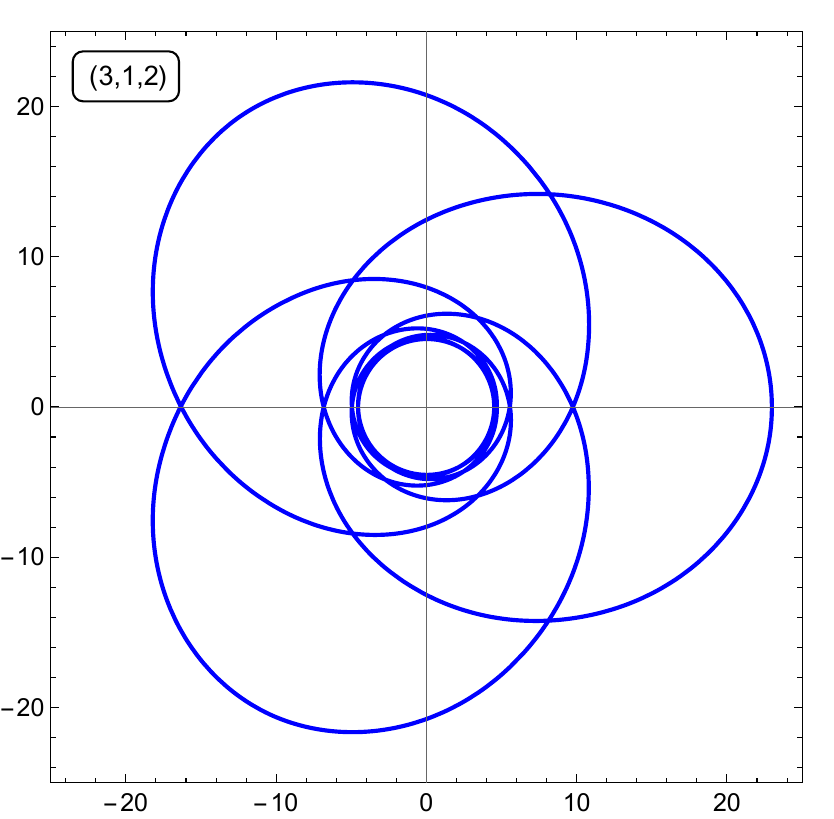}\label{subfig:B24}}
\subfigure[\hspace{0.1cm}$E=0.9681565$]{\includegraphics[width=0.3\textwidth]{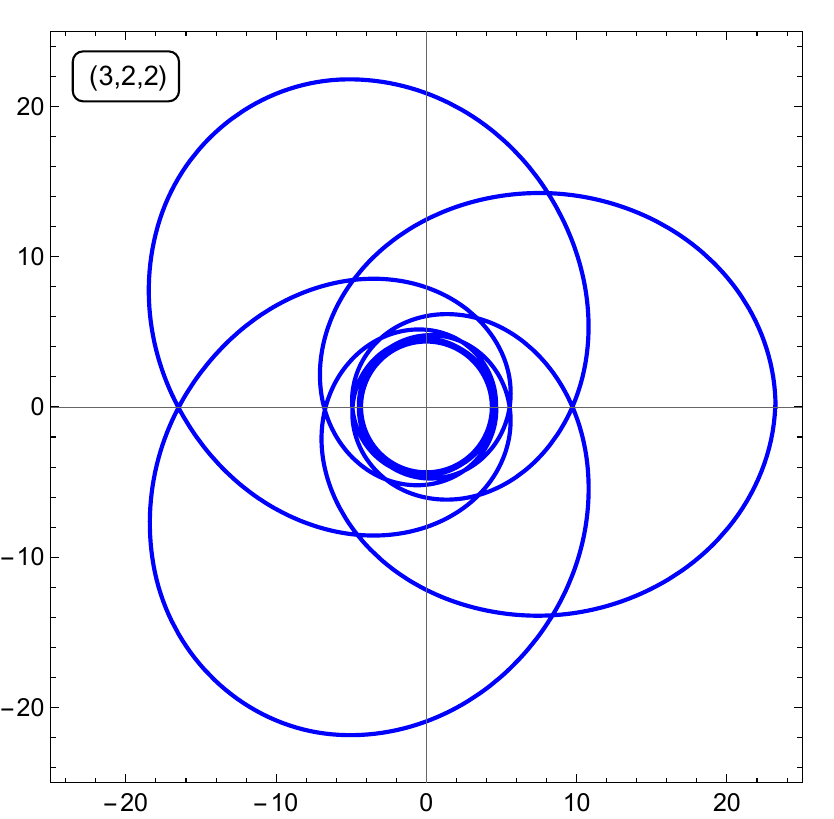}\label{subfig:B25}}
\subfigure[\hspace{0.1cm}$E=0.9679943$]{\includegraphics[width=0.3\textwidth]{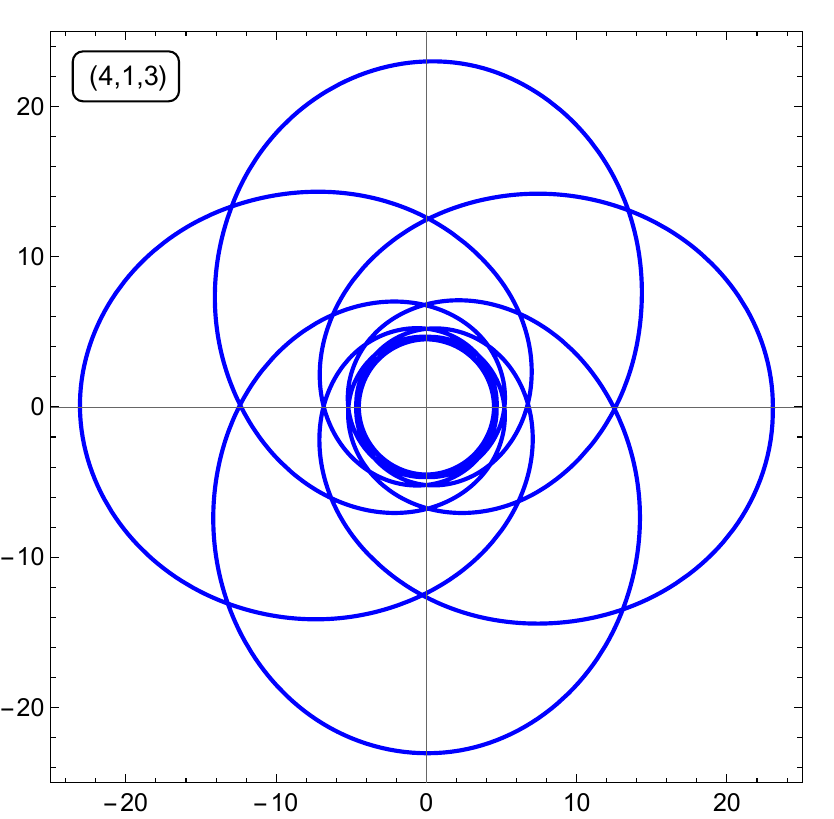}\label{subfig:B26}}
\caption{Plots of the periodic orbits characterised by different values of $(z,w,v)$ around the Bardeen regular black hole for $g=0.2M$ and $\epsilon=0.5$.}
\label{fig:B_orbits}
\end{figure}

\noindent We consider the Bardeen mass function to analyse the periodic orbits around such a spacetime. By incorporating Eqs.~\eqref{3.n1} and~\eqref{3.2} with metric function in Eq.~\eqref{2.2}, one may identify a relation between the rational number $q$ and the conserved energy $E$ (or angular momentum $L$) for some chosen values of metric parameters, numerically.
\begin{figure}[h]
\centering
\includegraphics[width=0.45\textwidth]{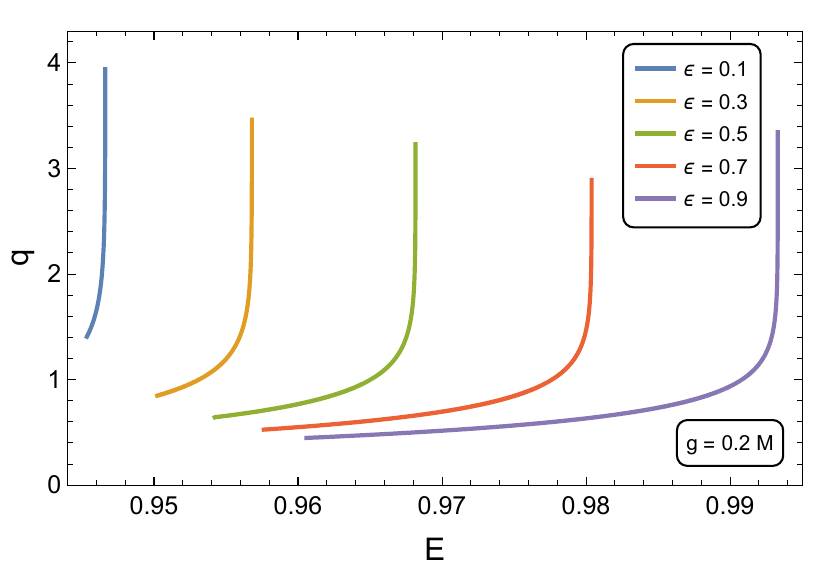}
\includegraphics[width=0.45\textwidth]{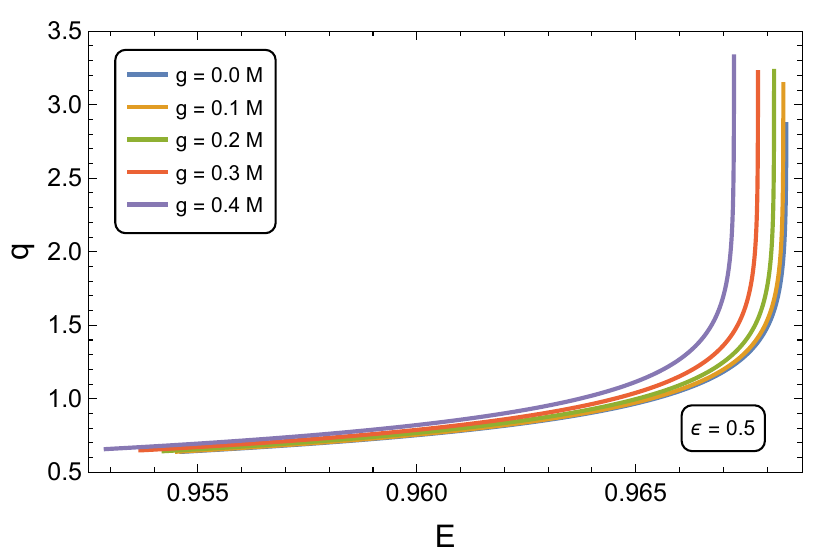}
\caption{Variation of $q$ with $E$ for different values of $\epsilon$ (left) and $g$ (right) for the Bardeen regular black hole.}
\label{fig:BqvsE}
\end{figure}
In Fig.~\ref{fig:BqvsE} (left), we present the variation of $q$ with $E$ for the chosen value of $g=0.2M$ corresponding to some selected values of the parameter $\epsilon$ (or angular momentum $L$).
As expected, $E$ is bounded by the upper limit of $1$. It is observed that $q$ increases slowly with $E$ at first, and a rapid rise occurs once the maximum energy is approached. Moreover, for small values of angular momentum $L$ (or, equivalently, $\epsilon$), the allowed range of $E$ that admits periodic orbits is narrower compared to the case of high $L$.
In Fig.~\ref{fig:BqvsE} (right), the variation of $q$ with $E$ is shown corresponding to different chosen values of $g$ with $\epsilon=0.5$. It is found that the maximum energy (where $q$ increases rapidly) changes significantly under the variation of $g$ with fixed $\epsilon$. For large values of $g$, the maximum energy is comparatively smaller than the $g=0$ case (Schwarzschild black hole). In contrast, the behaviour of $q$ with $L/M$ for different chosen values of $g$ is shown in Fig.~\ref{fig:BqvsL} (left) for a fixed $E$.
 Fig.~\ref{fig:BqvsL} (right) shows the relation of $q$ with $L/M$ for different values of $E$ with $g=0.2M$. We observe that $q$ decreases as the angular momentum $L$ increases. Moreover, as $g$ increases, periodic orbits with higher $q$ value occur for lower angular momentum.
\noindent For fixed values of energy and angular momentum, the variation of $q$ with the `regularisation parameter' $g$ is demonstrated in Fig.~\ref{fig:Bqvsg}. It is found that $q$ decreases monotonically as $g$ increases. 
We choose three different values of angular momentum, $L=3.6534056M$, $3.6553345M$, and $3.6575956M$ all with energy $E=0.96$. These correspond to $(1,2,0)$,$(3,1,2)$, and $(2,1,1)$ periodic orbits, respectively, in the $g=0$ case.
We then run our numerical code in {\it Mathematica 13.0} and identify several additional values $q$ for different values of $g$, each associated with different periodic orbits for the same $E$ and $L$.
\begin{table}[h]
\centering
\begin{tabular}{ |c|c|c|c|c|c|c|} 
\hline
$L/M$ &  $g/M_{(1,1,0)}$ & $g/M_{(1,2,0)}$ & $g/M_{(2,1,1)}$ & $g/M_{(3,1,1)}$ & $g/M_{(3,1,2)}$ & $g/M_{(4,0,3)}$ \\
\hline
\hline
$3.6534056$ &  $0.33895$ & $-0.00002^{\dagger}$ & $0.12619$ &$0.17793$  & $0.08559$ & $0.54875$ \\
\hline
$3.6553345 $ & $0.32832$ & $-$ & $0.09281$ & $0.15615$ & $-0.00001^{\dagger}$ & $0.54264$ \\
\hline
$3.6575956 $ & $0.31533$ & $-$ & $0.00003^{\dagger}$ & $0.12572$ & $-$ & $0.53536$ \\
\hline
\end{tabular}
\caption{Different value of $g$ representing different periodic orbits with the fixed values of $E$ and $L$. Here, energy is $E=0.96$ for all three cases . The symbols $\dagger$ and $-$ are discussed in the text.}
\label{table_Bqvsg}
\end{table}
In Table~\ref{table_Bqvsg}, we list some of such chosen values of $g$ associated with the $q$ for the same energy and angular momentum.
One may note that our code yields accurate results up to four decimal places. For the selected values of energy and angular momentum, the code should return $g=0$ for the $(1,2,0)$, $(3,1,2)$, and $(2,1,1)$ periodic orbits, respectively. However, our numerical result instead gives $g=-0.00002M$, $g=-0.00001M$, and $g=0.00003M$, respectively. These values are marked with a `$\dagger$' in Table~\ref{table_Bqvsg}. The symbol `$-$' indicates that no such values of $g$ corresponding to the given $q$ exist for the chosen values of $E$ and $L$.
\noindent In Fig.~\ref{fig:B_orbits}, we illustrate the trajectories of certain selected periodic orbits characterised by the integer triplet $(z,w,v)$ in the $r-\phi$ plane for $g=0.2M$ and $\epsilon=0.5$ around the Bardeen regular black hole. 
The vertical and horizontal axes represent the $r\,\sin{\phi}$ and $r\,\cos{\phi}$, respectively. For each periodic orbit, the integer triplet $(z,w,v)$ and the particle's energy $E$ are mentioned in the subfigures. One may note that the integer $z$ denotes the number of leaf-like patterns in the orbit. As $z$ increases, the number of leaves increases, and the trajectory becomes more complex. 
We demonstrate periodic orbits with relatively small values of $q$. Higher values of $q$ lead to more complicated whirl-dominated motion near the horizon. Such orbits are generally short-lived and therefore are less relevant in an astrophysical context. Moreover, orbits with high $q$ are more sensitive to external perturbations.
\noindent For a given fixed periodic orbit, as $g$ increases, the shape of the orbit shrinks more relative to the $g=0$ case. This shrinking results in a reduction of the orbital time period. We demonstrate this effect in the next section in the context of gravitational waves. Further, in Table~\ref{table_q_E_g}, a list of the energy $E$ for different periodic orbits with different selected values of $g$ is presented for the $\epsilon=0.5$. One may note that an increase in $g$ reduces the particle energy $E$ corresponding to a fixed orbit ($q$). Thus, we find the maximum value of $E$ for a periodic orbit when the system reduces to the Schwarzschild spacetime. 
A similar list for different values of angular momentum (or $\epsilon$) is shown in the Table~\ref{table_q_E_epsilon} with $g=0.2M$. 
\begin{table}[h]
\centering
\begin{tabular}{ |c|c|c|c|c|c|c|} 
\hline
$g/M$ &  $E_{(1,2,0)}$ & $E_{(2,1,1)}$ & $E_{(2,2,1)}$ & $E_{(3,1,2)}$ & $E_{(3,2,2)}$  & $E_{(4,1,3)}$ \\
\hline
\hline
$0.0$ &  $0.9683828$ & $0.9680265$ & $0.9684343$ & $0.9682249$ & $0.9684385$ & $0.9682850$\\
\hline
$0.1 $ & $0.9683124$ & $0.9679520$ & $0.9683648$ & $0.9681526$ & $0.9683690$ &$0.9682134$ \\
\hline
$0.2 $ & $0.9680973$ & $0.9677238$ &$0.9681520$ & $0.9679312$ & $0.9681565$ & $0.9679943$ \\
\hline
$0.3$ & $0.9677239$ & $0.9673264$ & $0.9677831$ & $0.9675465$ &$0.9677881$ & $0.9676136$ \\
\hline
$0.4$ & $0.9671666$ & $0.9667303$ & $0.9672333$ & $0.9669706$ & $0.9672390$ & $0.9670445$\\
\hline
\end{tabular}
\caption{The conserved energy $E$ for different periodic orbits around the Bardeen regular black hole, where $\epsilon=0.5$}
\label{table_q_E_g}
\end{table}

\begin{table}[h]
\centering
\begin{tabular}{ |c|c|c|c|c|c|c|} 
\hline
$\epsilon$ &  $E_{(1,2,0)}$ & $E_{(2,1,1)}$ & $E_{(2,2,1)}$ & $E_{(3,1,2)}$ & $E_{(3,2,2)}$  & $E_{(4,1,3)}$ \\
\hline
\hline
$0.1$ &  $0.9463860$ & $0.9456481$ & $0.9465617$ &$0.9460210$  & $0.9465838$ & $0.9461478$ \\
\hline
$0.3 $ & $0.9566917$ & $0.9561922$ & $0.9567801$ & $0.9564597$ & $0.9567886$ & $0.9565448$\\
\hline
$0.5 $ & $0.9680973$ & $0.9677238$ & $0.9681520$ & $0.9679312$ & $0.9681565$ & $0.9679943$ \\
\hline
$0.7$ & $0.9803455$ & $0.9800482$ & $0.9803835$ & $0.9802177$ & $0.9803863$ & $0.9802674$ \\
\hline
$0.9$ & $0.9932810$ & $0.9930339$ & $0.9933092$ & $0.9931775$ & $0.9933113$ & $0.9932186$ \\
\hline
\end{tabular}
\caption{The conserved energy $E$ for different periodic orbits in the Bardeen regular black hole with $g=0.2M$}
\label{table_q_E_epsilon}
\end{table}

\subsection{Hayward regular black hole}
\begin{figure}[h]
\centering
\includegraphics[width=0.45\textwidth]{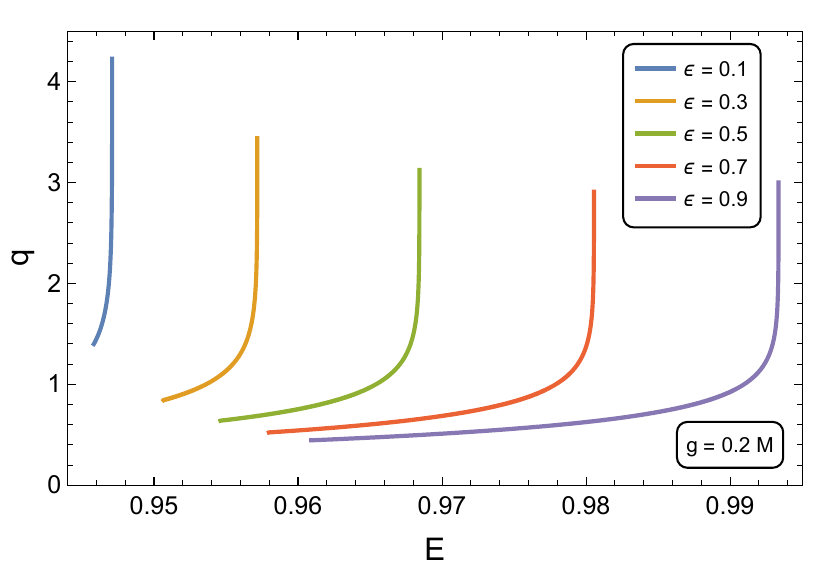}
\includegraphics[width=0.45\textwidth]{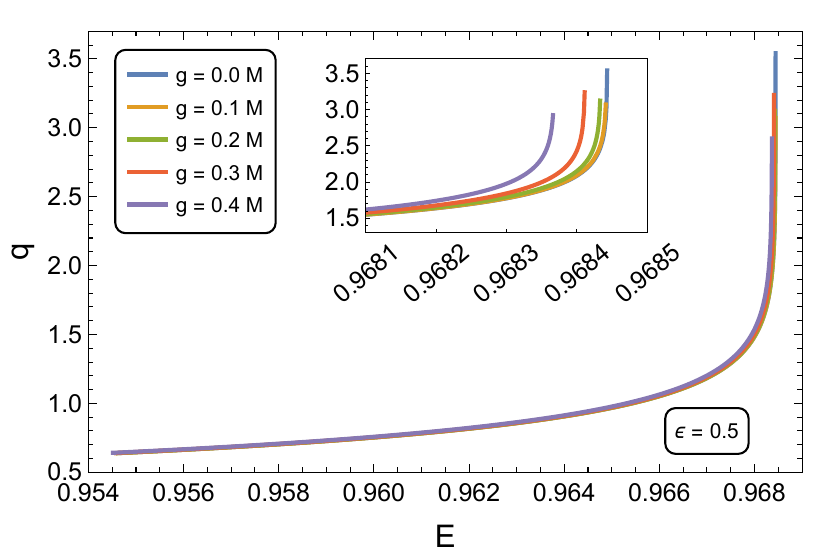}
\caption{Variation of $q$ with $E$ for different values of $\epsilon$ (left) and $g$ (right) for the Hayward regular black hole}
\label{fig:HqvsE1}
\end{figure}

\noindent A similar study of periodic orbits is performed by considering the Hayward mass function in Eq.~\eqref{2.3}. In this case, the dependency of $q$ with $E$ for different selected values of $\epsilon$ is shown in Fig.~\ref{fig:HqvsE1} (left) with $g=0.2M$. The Fig.~\ref{fig:HqvsE1} (right) shows the variation of $q$ with $E$ for different values of $g$ with $\epsilon=0.5$. The behaviour of $q$ is similar to the Bardeen case. We have the maximum energy when the spacetime is the Schwarzschild $(g=0)$. We observe that the influence of the parameter $g$ on $q$ is relatively less than in the Bardeen case.
\begin{figure}[h]
\centering
\includegraphics[width=0.45\textwidth]{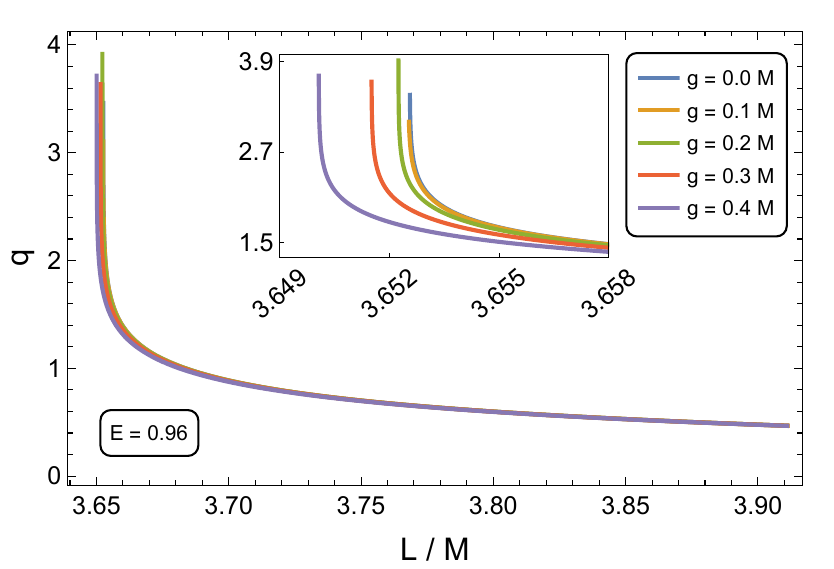}
\includegraphics[width=0.45\textwidth]{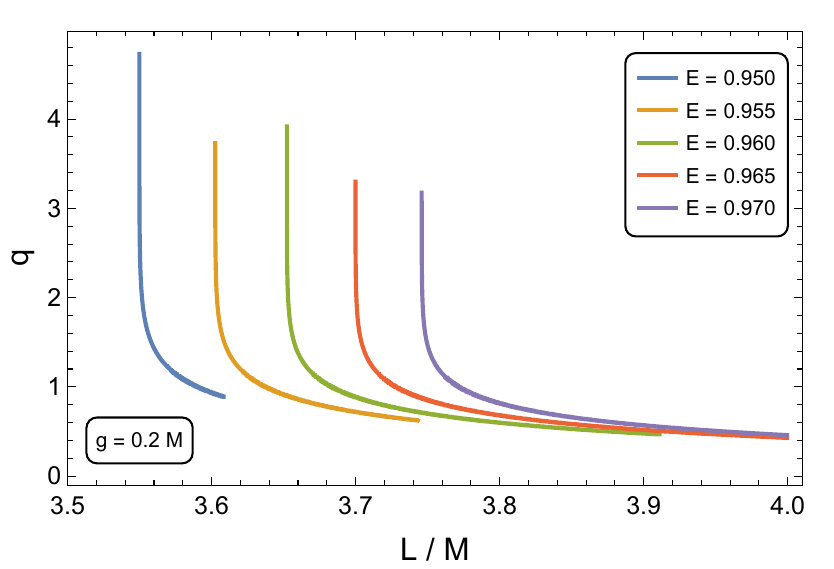}
\caption{Variation of $q$ with $L$ for different values of $g$ (left) and $E$ (right) for the Hayward regular black hole}
\label{fig:HqvsL}
\end{figure}
The variation of $q$ with the angular momentum $L$ is presented in Fig.~\ref{fig:HqvsL}. The Fig.~\ref{fig:HqvsL} (left) shows the variation for different values of $g$ with $E=0.96$, whereas Fig.~\ref{fig:HqvsL} (right) demonstrates it for different values of $E$ with $g=0.2M$. It is found that for lower values of $E$, we require less angular momentum to construct periodic orbits of high $q$ value.

\begin{figure}[h]
\centering
\includegraphics[width=0.5\textwidth]{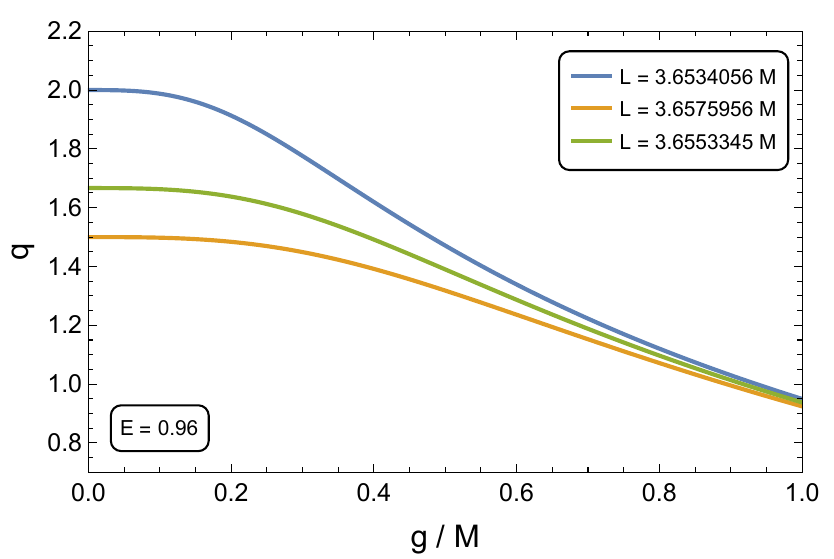}
\caption{Plot of $q$ with $g$ for a fixed value of energy $E$ and angular momentum $L$ of the Hayward regular metric.}
\label{fig:Hqvsg}
\end{figure}

\noindent The variation of $q$ with $g/M$ for fixed values of energy and angular momentum is shown in Fig.~\ref{fig:Hqvsg}. Similar to the Bardeen case, the $q$ value decreases monotonically with $g/M$. Here as well, we set the values of energy and angular momentum corresponding to the periodic orbits $(1,2,0)$, $(2,1,1)$ and $(3,1,2)$ of the $g=0$ spacetime. Then, we identify several other periodic orbits with different $g$ values for the same energy and angular momentum.

\section{Gravitational wave radiation from periodic orbits}\label{IV}
\begin{figure}[h]
\centering
\subfigure[\hspace{0.1cm} Periodic orbit]{\includegraphics[width=0.3\textwidth]{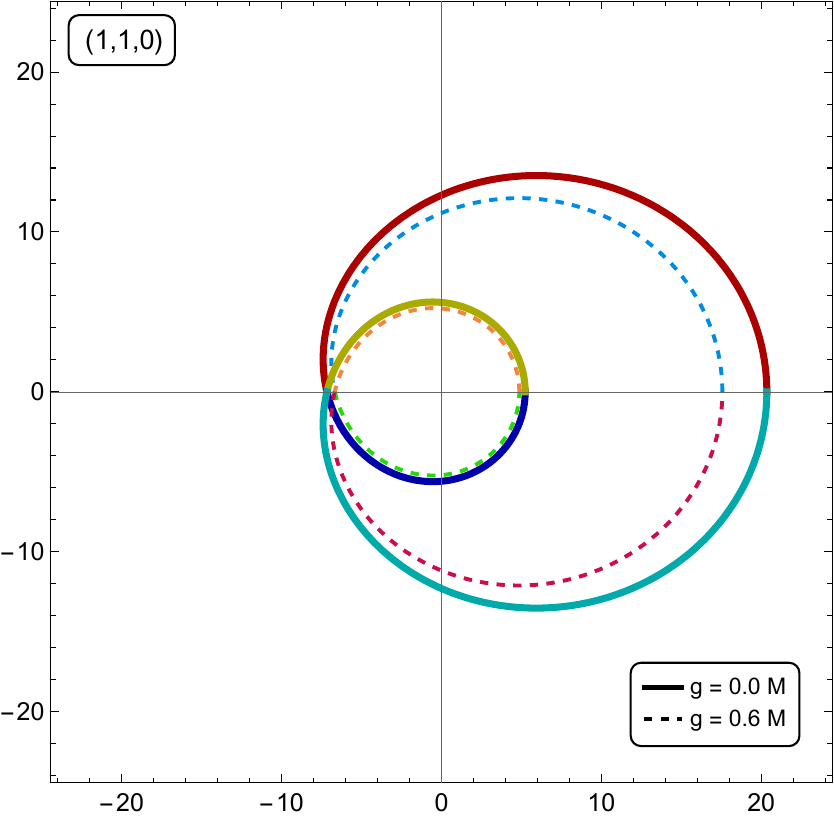}\label{subfig:B110orbit}}
\subfigure[\hspace{0.2cm} Gravitational waveforms]{\includegraphics[width=0.6\textwidth]{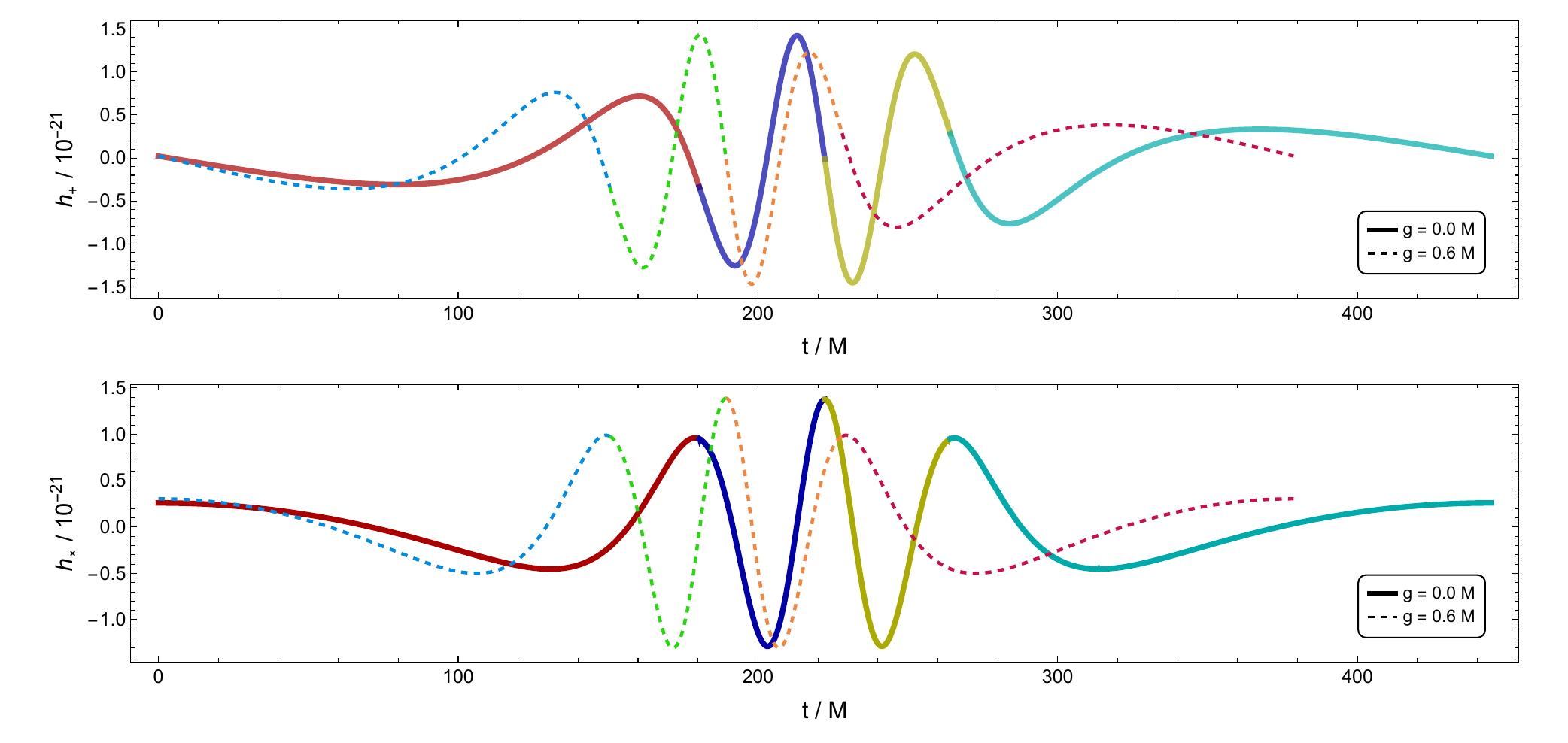}\label{subfig:B110GW}}
\caption{Gravitational waves for $(1,1,0)$ orbit around Bardeen black hole with different segments in different colours. $\epsilon=0.5$, $E_{g=0}=0.9654253$, $E_{g=0.6M}=0.96089$, $M=10^6\,M_{\odot}$, $m=10\,M_{\odot}$ and $D_L=200$ Mpc}
\label{fig:B_colored_orbit}
\end{figure}

\noindent In this section, we study gravitational waves emitted by a test object (star/small mass black hole) orbiting in a periodic orbit around a central supermassive regular black hole. These systems, known as Extreme Mass Ratio Inspirals (EMRIs), are among the most important sources for future space-based GW observations. The gravitational waves emitted from these systems may contain information about strong-field gravity and supermassive black holes.
In the literature, several studies have been conducted that utilise the adiabatic approximation to obtain gravitational waves from EMRI systems~\cite{Hughes:2001jr, Glampedakis:2002ya, Hughes:2005qb, Drasco:2005is, Gair:2005ih, Glampedakis:2005cf, Drasco:2005kz, Sundararajan:2007jg, Sundararajan:2008zm, Miller:2020bft, Isoyama:2021jjd}. Such an approximation is useful because the inspiral timescale is much longer than the orbital period in an EMRI system.
Under an adiabatic approximation, the energy and angular momentum loss due to the motion of the smaller mass object is negligible over a few periods in comparison to the total energy of the system. 
Therefore, the motion of the smaller object can be approximated as a geodesic in the background regular black hole metric.
Given the short orbital evolution, the radiation reaction or back-reaction of the emitted gravitational waves on the motion of the test object can be ignored (i.e., the motion of the object is considered slow).
\noindent The gravitational waveform emitted from such systems can be explored through the numerical Kludge method~\cite{Babak:2006uv}. This method mainly consists of two steps: (a) the motion of the test object is obtained by solving the geodesic equation in the regular spacetime in the Boyer-Lindquist coordinate, which then projected to the polar coordinates of a spherically symmetric flat spacetime, and (b) the associated gravitational waveform is built using the standard symmetric and trace-free mass quadrupolar equation of gravitational radiation~\cite{Thorne:1980ru, gravity-book}. Thus, we view the Boyer-Lindquist coordinates as hypothetical spherical polar coordinates and project the geodesic of the test object on to the Cartesian coordinates as~\cite{Babak:2006uv}
 \begin{equation}
     x= r\sin \theta \cos \phi ,\quad~ y= r\sin \theta \sin \phi, \quad~ z= r\cos \theta.
 \end{equation}
Hence, we construct the geodesic of the test object in a pseudo-flat spacetime and apply the standard gravitational wave equation in a flat spacetime. In the weak-field, linear assumption and in the Lorentz gauge, the field equation is expressed as,
\begin{equation}
     \Box \bar{h}^{\mu\nu}=-16\pi \tau^{\mu\nu}.
\end{equation}
where $\Box$ denotes the standard flat-spacetime wave operator, $\bar{h}^{\mu\nu}\equiv h^{\mu\nu}-(1/2)h\,\eta^{\mu\nu}$, $h^{\mu\nu}$ is a metric perturbation corresponding to the gravitational wave field, $\eta^{\mu\nu}$ is the flat spacetime metric and $\tau^{\mu\nu}$ is the stress-energy tensor of the source. Using the Green's function method, the known retarded solution of the above equation is,
\begin{equation}
     \bar{h}^{\mu\nu}= 4 \int d^3x' \frac{\tau^{\mu\nu}(\vec{x'},t-\vert \vec{x}-\vec{x'}\vert )}{\vert \vec{x}-\vec{x'} \vert}.
\end{equation}
where $\vec{x'}$ and $\vec{x}$ represent the positions of the source and observer, respectively. In the slow-motion approximation (source motion negligible), the above solution reduces to \cite{Thorne:1980ru}
\begin{equation}
    \bar{h}^{ij}= \frac{2}{r}\left[ \ddot{I}(t')\right]_{t'=t-r}\,,
\end{equation}
where
\begin{equation}
    I^{ij}=\int x'^ix'^j T^{tt}(t',x'^i)d^3x'\,,
\end{equation}
is the source mass quadrupole moment. The $T^{tt}$ component of the stress-energy tensor of the small object with trajectory $Z^i(t)$ in a spacetime is
\begin{equation}
T^{tt}(t',x'^i)=m\delta^3(x'^i-Z^i(t')),
\end{equation}
where $m$ is the mass of the test object. Rearranging the above equations, the gravitational waveforms are obtained from the second time derivative of the mass quadrupolar moment as
 \begin{equation}\label{4.7}
     h_{ij}= \frac{2}{D_L}\frac{d^2I_{ij}}{dt^2}=\frac{2m}{D_L}\left(a_i x_j + a_j x_i+ 2 v_iv_j\right),
 \end{equation}
where $a_i$ is the spatial acceleration, $v_i$ is the spatial velocity of the test object, and $D_L$ is the Luminosity distance from the EMRIs to the detector. This formalism follows the standard numerical Kludge waveform method, which provides approximate and physically consistent waveforms for EMRI systems. 
\noindent In order to analyse the gravitational wave signal observed by a detector, a detector-adapted coordinate system is constructed~\cite{gravity-book}. The origin of the new coordinate system coincides with the origin of the original source frame $(x,y,z)$, both of which are centred on the regular black hole.
We consider the orthonormal basis of spherical coordinates at the observer's position,
\begin{equation}
    \hat{e}_r= \frac{\partial}{\partial r}, \quad~ \hat{e}_{\Theta}= \frac{1}{r}\frac{\partial}{\partial \Theta}, \quad~ \hat{e}_{\Phi}= \frac{1}{r\sin \Theta}\frac{\partial }{\partial \Phi},
\end{equation}
where $ \Theta$ is the inclination angle of the orbital plane of the test object to the observer and $\Phi$ is the longitude of the pericenter measured in the orbital plane. The wave propagates along the unit vector $N^i=\lbrace \sin \Theta \cos \Phi, \sin \Theta \sin \Phi, \cos \Theta \rbrace$ directed towards the observer from the source.
The transverse-traceless (TT) part of the metric perturbation is, formally~\cite{gravity-book},
\begin{equation}
    h^{jk}_{TT}= (TT)^{jk}_{pq}h^{pq},
\end{equation}
where the $(TT)$ projection operator is $(TT)^{jk}_{pq}= P^j{}_p P^{k}{}_q - \frac{1}{2}P^{jk}P_{pq}$
with the transverse projector $P^j{}_{k}= \delta^j_{k}- N^jN^k$.
We consider the two unit vectors transverse to $N^i$ as $ u^i= \lbrace \cos \Theta \cos \Phi, \cos \Theta \sin \Phi, -\sin \Theta \rbrace$ and $v^i = \lbrace -\sin \Phi,\cos \Phi, 0 \rbrace$,
which satisfy $N^jN^k + u^ju^k + v^jv^k= \delta^{jk}$.
Hence, the transverse projector becomes $P^{jk}= u^ju^k + v^jv^k.$
Therefore, $h^{jk}_{TT}$ can be decomposed in the tensorial basis built on the vectors $u^i$ and $v^i$ as
\begin{equation}
    h^{jk}_{TT}= h_{+}\left(u^ju^k -v^jv^k\right)+ h_{\times} \left(u^jv^k + u^k v^j\right),
\end{equation}
where $h_{+}$ and $h_{\times}$ are two polarizations of $h^{jk}_{TT}$.
Contracting with the polarisation tensor, we have~\cite{gravity-book}
\begin{equation}
     h_{+}=  \frac{1}{2}\left(u_ju_k-v_jv_k \right)h^{jk},
     \qquad
    h_{\times} = \frac{1}{2} \left(u_jv_k +u_kv_j\right)h^{jk},
\end{equation}
where we take into account that the tensorial operators acting on $h^{jk}_{TT}$ are already transverse and traceless.
In the orthonormal spherical coordinates, the TT metric perturbation takes the following form
\begin{equation}
    h^{jk}_{TT}= \frac{1}{2} \begin{pmatrix}
        0& 0 & 0 \\
        0& h^{\Theta \Theta}-h^{\Phi \Phi} & 2 h^{\Theta \Phi} \\
        0 & 2h^{\Theta \Phi} & h^{\Phi \Phi}-h^{\Theta \Theta}
    \end{pmatrix}.
\end{equation}
Hence, in this frame, the polarisations are simply 
\begin{eqnarray}
h_{+}= \frac{1}{2}\left(h^{\Theta\Theta}-h^{\Phi\Phi}\right), \quad
h_{\times} = h^{\Theta\Phi},
\end{eqnarray}
where the components are
\begin{eqnarray}
    h^{\Theta \Phi}&=& \cos \Theta \left[- \frac{1}{2}h^{xx}\sin(2\Phi)+ h^{xy}\cos(2\Phi) + \frac{1}{2}h^{yy}\sin(2\Phi)\right] 
     + \sin \Theta \left[h^{xz}\sin \Phi -h^{yz}\cos \Phi\right] ,\\
    h^{\Theta\Theta}
    &= & \cos^2 \Theta \left[h^{xx}\cos^2 \Phi + h^{yy}\sin^2 \Phi + h^{xy}\sin (2\phi)\right] + h^{zz}\sin^2 \Theta - \sin(2\Theta) \left[\cos \Phi h^{xz} + \sin \Phi h^{yz}\right], \\
    h^{\Phi\Phi}
    &=& \sin^2\Phi h^{xx} + \cos^2 \Phi h^{yy} - \sin(2\Phi)h^{xy}.
\end{eqnarray}
\noindent To demonstrate the gravitational waveforms from different periodic orbits and to show the effect of the `regularisation parameter' $g$, we take an EMRI system where the mass of the test object is $m=10\,M_{\odot}$ and the mass of the regular black hole $M=10^6\,M_{\odot}$ ($M_{\odot}$ is the solar mass). We consider the inclination angle $\Theta=\pi/4$, the longitude of pericenter $\Phi=\pi/4$, and the luminosity distance $D_L=200$ Mpc. In Fig.~\ref{fig:B_colored_orbit}, we consider the $(1,1,0)$ periodic orbit in the Bardeen regular black hole with $\epsilon=0.5$, $E=0.9654253$ and demonstrate the gravitational wave from this orbit for two values of $g$. The different colours are used to highlight the correspondence between the gravitational wave (Fig.~\ref{subfig:B110GW}) and the periodic orbit (Fig.~\ref{subfig:B110orbit}).
The zoom-whirl motion of the test object in the periodic orbit $(1,1,0)$ is clearly reflected in the associated waveform over a complete cycle.
The amplitude of the wave attains maxima when the test particle approaches perihelion and decreases as the particle moves far from perihelion.
Thus, we may infer that the smoothly varying waveform corresponds to the zoom phase of the orbit and the highly oscillating portion reflects the whirl motion. When the particle is far from the black hole, gravity is weak, and the amplitude of the gravitational wave is slowly varying. However, as the particle approaches the black hole, near the outer horizon, the gravitational wave amplitude and frequency increase significantly. The oscillation becomes more pronounced. On the other hand, as we increase the value of $g$, the orbit shrinks more, and the gravitational wave falls behind in phase compared to the $g=0$ case. The time period of one cycle decreases for higher values of $g$. Therefore, we may extract crucial information about the particle trajectory, the signature of $g$ and other properties of a regular black hole from the associated gravitational wave. Further, an analytically tractable expression of the gravitational waveform from a periodic orbit in Schwarzschild spacetime is provided in Appendix~\ref{Appendix1}, where it is also compared with our numerical results.
\noindent We quantify the above-discussed phase shift of the gravitational wave for finite values of $g$ relative to the $g=0$ case, by calculating the phase of the gravitational wave. The change in phase may be considered as a possible observable quantity which 
may quantify a deviation signaling the presence of a regular black hole as opposed to the singular Schwarzschild geometry. Under the adiabatic approximation, the phase of the gravitational wave is,
\begin{equation}
    \frac{d\phi_{GW}}{dt}= 2\frac{d\phi}{dt},   
\end{equation}
where $\frac{d\phi}{dt}$ is the instantaneous orbital angular frequency of the binary orbit. Thus, using the previous definitions of $\dot{t}$ and $\dot{\phi}$, we have
\begin{equation}
    \phi_{GW}(t)= \int^t_0 \frac{2L}{E}\frac{f(r(t'))}{r^2(t')}dt'\,.
\end{equation}
Therefore, the phase difference in the waveforms for a finite value of $g$ relative to  the $g=0$ case may be defined as
\begin{eqnarray}\label{eqphase}
    \Delta \phi_{GW}(t)= \phi_{GW}(g,t)- \phi_{GW}(0,t) 
    = \int^t_0 \left[\frac{2L_g}{E_g}\frac{f(r,g,t')}{r^2(g,t')}- \frac{2L_0}{E_0}\frac{f(r,0,t')}{r^2(0,t')}\right]dt' ,
\end{eqnarray}
where $L_g$ and $E_g$ are the angular momentum and energy of the test object around the regular black hole, and $L_0$ and $E_0$ are values corresponding to Schwarzschild geometry, for the same orbit. Note that the time period of the orbital motion for two different values of $g$ is different.
%

\begin{figure}[h]
\centering
\includegraphics[width=0.9\textwidth]{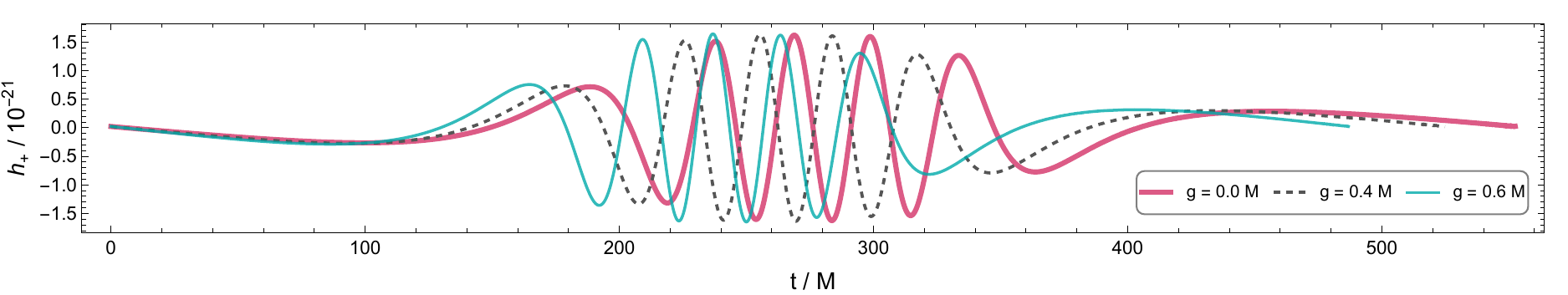}
\includegraphics[width=0.9\textwidth]{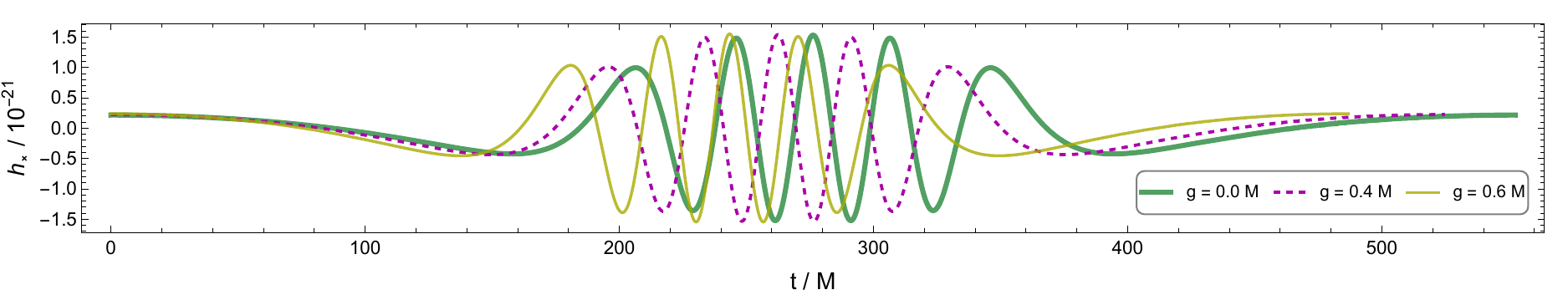}
\includegraphics[width=0.9\textwidth]{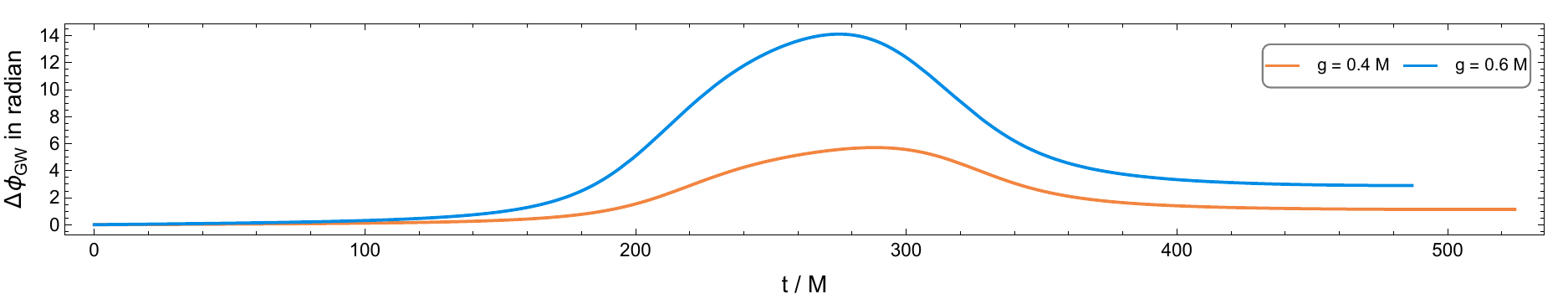}
\caption{The GW waveforms and phase shifts for $(1,2,0)$ orbit around Bardeen regular black hole. $\epsilon=0.5$, $E_{g=0}=0.9683828$, $E_{g=0.4M}=0.9671666$, $E_{g=0.6M}=0.9652875$, $M=10^6\,M_{\odot}$, $m=10\,M_{\odot}$ and $D_L=200$ Mpc}
\label{fig:B120_gw}
\end{figure}

%
\noindent In Fig.~\ref{fig:B120_gw}, we demonstrate the gravitational waveforms and phase shift for a $(1,2,0)$ orbit around the Bardeen regular black hole for three values of $g$. The increase in the number of sharp oscillations compared to the $(1,1,0)$ case reflects the increase in whirl number. We calculate the phase of the gravitational wave for each value of $g$ numerically and evaluate its difference relative to the $g=0$ case, at every time instant. Since the orbital time period decreases as $g$ increases, we evaluate the phase difference up to the time interval common to both cases.

\noindent From Fig.~\ref{fig:B120_gw}, we observe that a positive phase shift reaches a peak when the particle is closest to the black hole. In the zoom phase, where the motion occurs in the weak field region, the orbital frequencies are nearly the same, resulting in a comparatively smaller phase shift. However, during whirl motion in strong gravity, the change in $g$ significantly affects the orbital dynamics, leading to a rapid increase in phase shift. This decreases when the particle moves away from the black hole. As the phase shift defined in Eq.~\eqref{eqphase} is a cumulative time integral, a nonzero offset remains at the end of the evolution. 
\begin{figure}[h]
\centering
\includegraphics[width=0.9\textwidth]{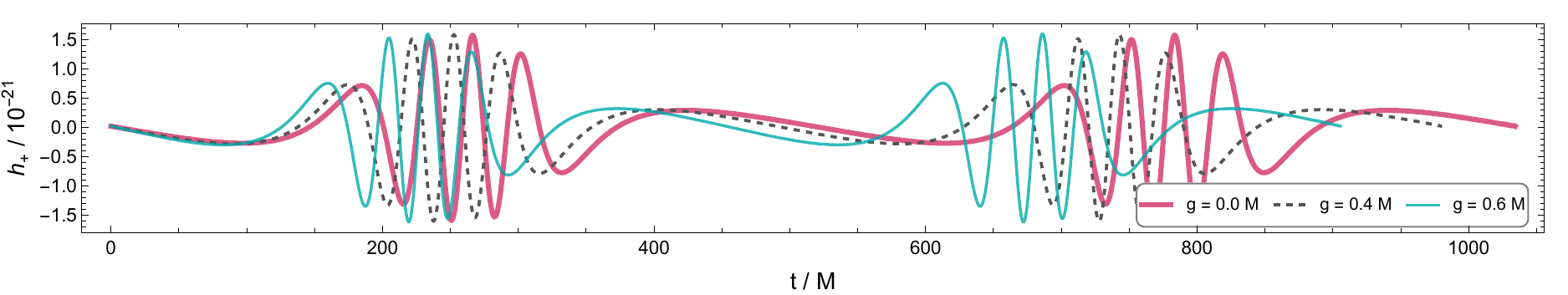}
\includegraphics[width=0.9\textwidth]{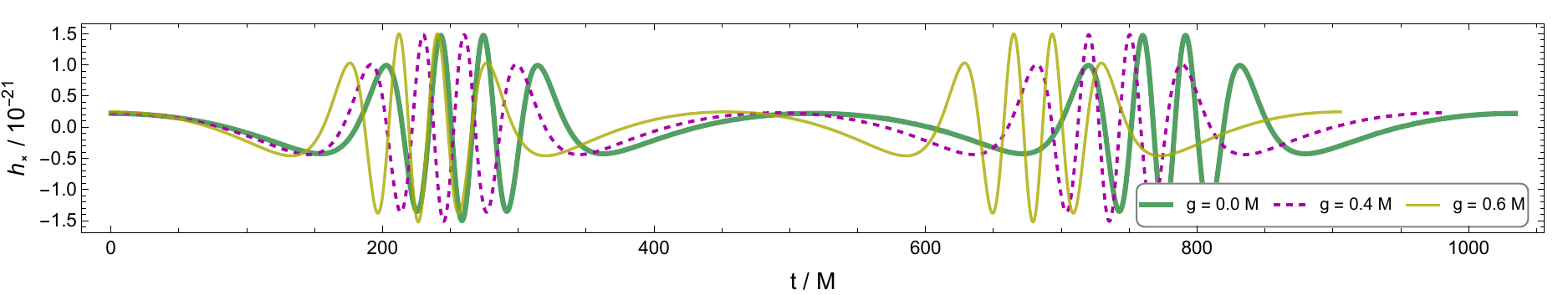}
\includegraphics[width=0.9\textwidth]{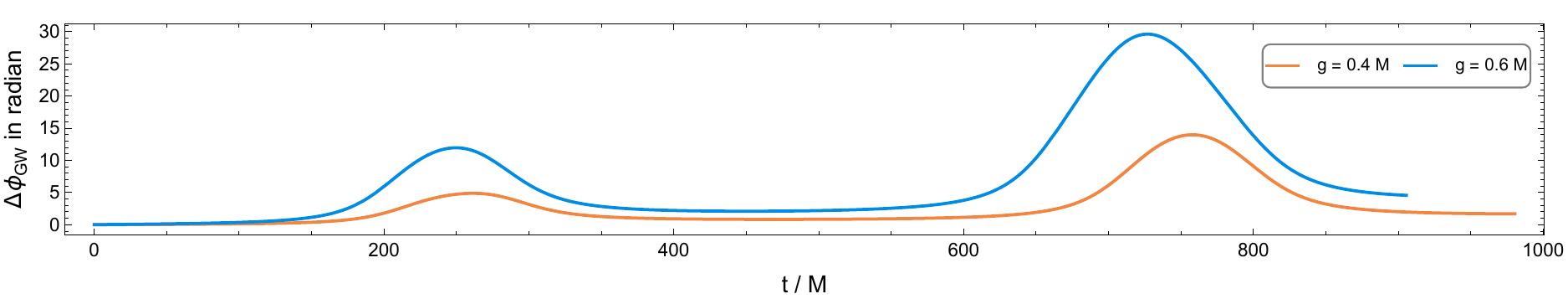}
\caption{The GW waveforms and phase shifts for $(2,1,1)$ orbit around Bardeen regular black hole. $\epsilon=0.5$, $E_{g=0}=0.9680264$, $E_{g=0.4M}=0.9667303$, $E_{g=0.6M}=0.9646878$, $M=10^6\,M_{\odot}$, $m=10\,M_{\odot}$ and $D_L=200$ Mpc}
\label{fig:B211gw}
\end{figure}
\begin{figure}[H]
\centering
\includegraphics[width=0.9\textwidth]{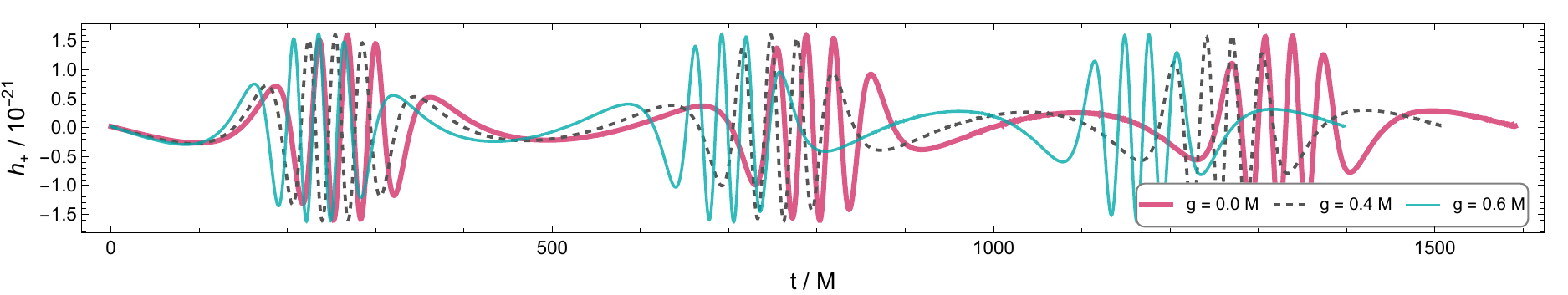}
\includegraphics[width=0.9\textwidth]{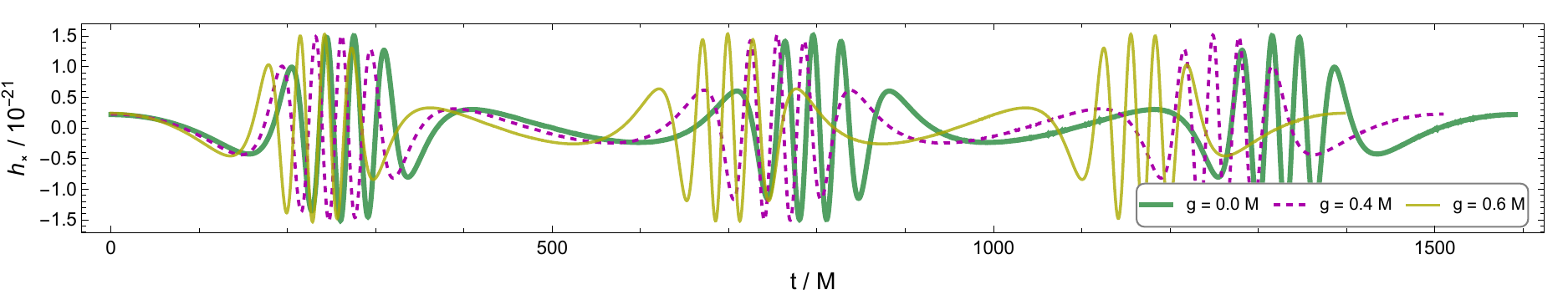}
\includegraphics[width=0.9\textwidth]{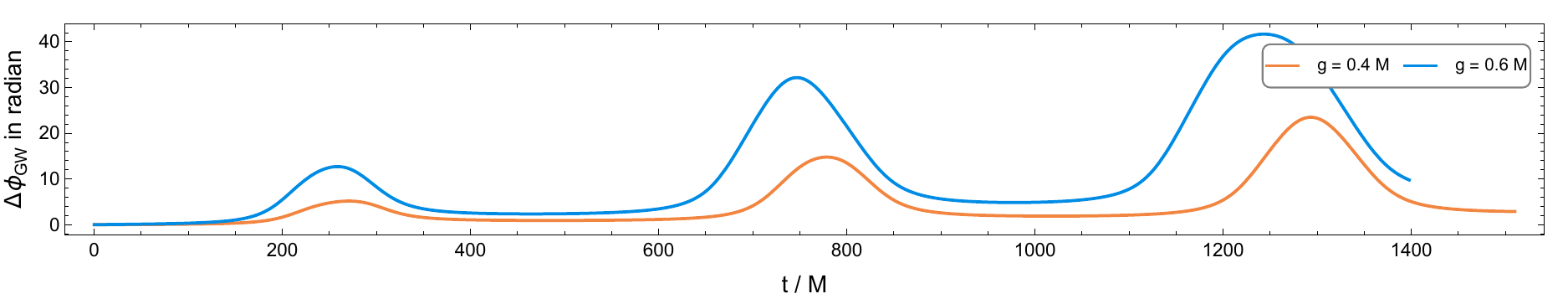}
\caption{The GW waveforms and phase shifts for $(3,1,2)$ orbit around Bardeen regular black hole. $\epsilon=0.5$, $E_{g=0}=0.9682249$, $E_{g=0.4M}=0.9669706$, $E_{g=0.6M}=0.9650121$, $M=10^6\,M_{\odot}$, $m=10\,M_{\odot}$ and $D_L=200$ Mpc}
\label{fig:B312gw}
\end{figure}

%
\noindent Similarly, the gravitational waveforms and phase shift can be calculated for other periodic orbits around the Bardeen black hole. In Figs.~\ref{fig:B211gw} and~\ref{fig:B312gw}, we present the gravitational wave and phase shift for $(2,1,1)$ and $(3,1,2)$ periodic orbits, respectively. In these cases, the sharp oscillations represent the whirl motion. The number of sharp oscillatory branches appearing within one orbital period matches the zoom number, as expected. Here, the phase shift again reaches its maxima during the whirl motion. Initially, when the particle is in the zoom phase, the phase shift is small. During the first whirl motion, the phase shift increases, and as the particle goes back to the zoom phase, the phase shift decreases with a finite offset. In the subsequent whirl motion, the phase shift increases further, exceeding its previous maxima. This reflects the cumulative nature of the time integral used to evaluate the phase shift. This behaviour continues for successive whirl phases.

\begin{figure}[h]
\centering
\includegraphics[width=0.9\textwidth]{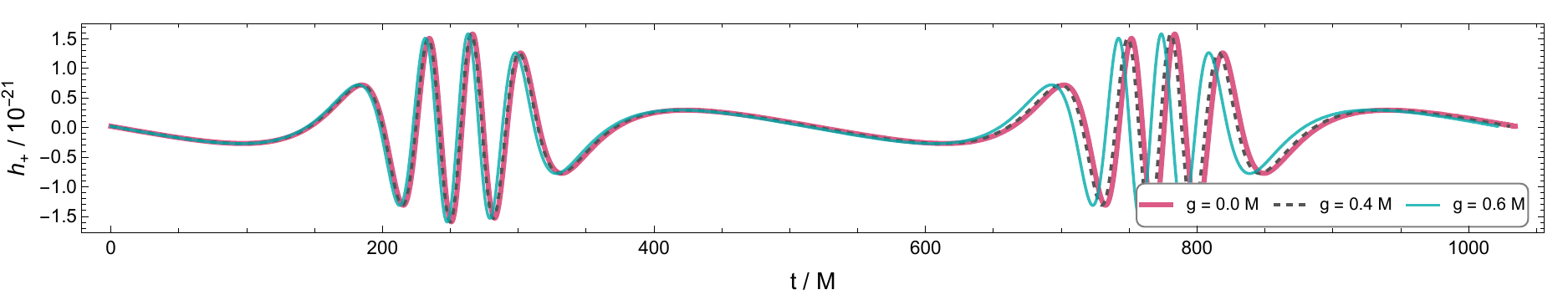}
\includegraphics[width=0.9\textwidth]{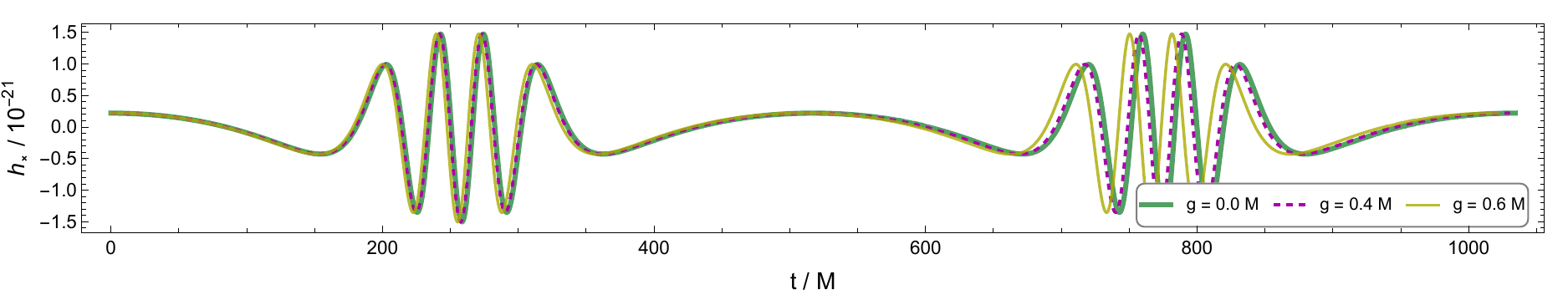}
\includegraphics[width=0.9\textwidth]{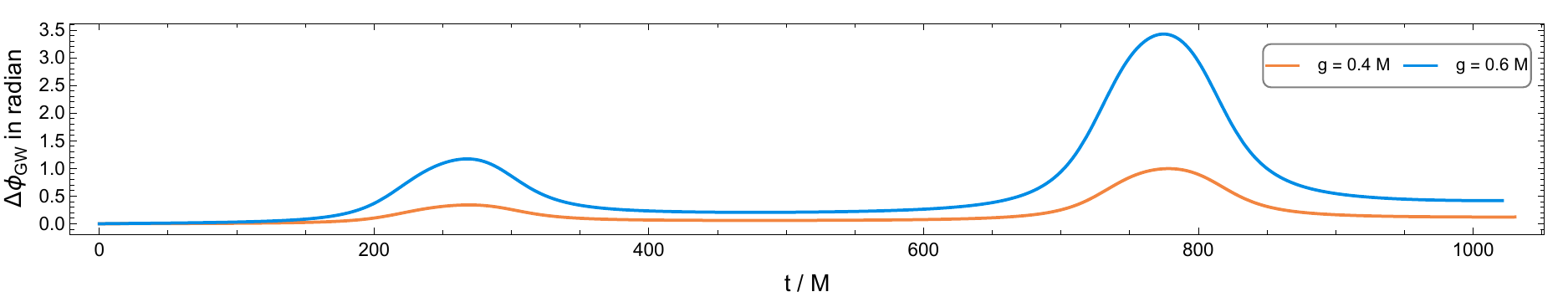}
\caption{The GW waveforms and phase shifts for $(2,1,1)$ orbit around Hayward regular black hole. $\epsilon=0.5$, $E_{g=0}=0.9683828$, $E_{g=0.4M}=0.9683055$, $E_{g=0.6M}=0.9681164$, $M=10^6\,M_{\odot}$, $m=10\,M_{\odot}$ and $D_L=200$ Mpc}
\label{fig:H211gw}
\end{figure}

\noindent In the same manner, gravitational waveforms and phase shifts from periodic orbits can be evaluated in the Hayward regular black hole. We obtain results similar to those of the Bardeen regular black hole. In Fig.~\ref{fig:H211gw}, the gravitational waves and phase shift for the $(2,1,1)$ orbit around the Hayward black hole are presented. In this case, the phase shift is relatively smaller, reflecting the weaker influence of parameter $g$ on the gravitational wave. Therefore, the essential characteristics of the zoom-whirl motion are captured in the gravitational waves emitted by the periodic orbits. The phase shift appearing from a finite value of $g$ is significant and may provide a means to distinguish a regular black hole from a singular one. 
%

\section{Power spectral density (PSD) and characteristic strain}\label{V}
\noindent In this section, we further analyse the affect of the regularisation parameter $g$ in the gravitational waveforms from periodic orbits, through their frequency spectra $|\Tilde{h}_{+,\times}(f)|$. The associated characteristic strain $h_c(f)$ defined as~\cite{Robson}
\begin{equation}\label{eq:characteristric_strain}
    h_c(f)=2f\left(|\Tilde{h}_{+}(f)|^2+|\Tilde{h}_{\times}(f)|^2\right)^{1/2}.
\end{equation}
To extract the frequency spectrum of the signal, we apply the discrete Fourier transform (DFT) to the time domain gravitational waveform.
This procedure decomposes the waveform into its constituent frequency components, allowing us to study how the power of the signal is distributed across different frequencies.
We are also able to examine the influence of the parameter $g$ on the frequency structure of the gravitational waveform from a particle in typical periodic orbits.
\noindent To this end, we consider a gravitational waveform of a particular periodic orbit with orbital time period $T$ and construct a time series of $N$ sample points with the time interval $\Delta t$. By applying the discrete Fourier transform to the time series, the Fourier-domain gravitational wave at each frequency value becomes,
\begin{equation}
    \Tilde{h}_{\,+,\times}^{(k)}=\Delta t\,\sum_{n=0}^{N-1}h_{\,+,\times}^{(n)}\, e^{-2\pi\mathrm{i}\,f_k\,n\Delta t},
\end{equation}
where $f_k=\frac{k}{N \Delta t}$ with $k=0,1,2,\ldots,N-1$, $\Tilde{h}^{(k)}=\Tilde{h}(f_k)$ and $h^{(n)}=h(n\Delta t)$. 
In our case, the gravitational signal due to a particle in a periodic orbit is a real-valued time series. Therefore, its Fourier transform satisfies the Hermitian symmetry $\Tilde{h}(-f)=\Tilde{h}(f)^*$, which makes the Fourier-domain profile symmetric around $f=0$. As a result, we only need the positive frequency modes of $f_k$ with $k=0,1,2,\ldots,\frac{N}{2}$ (given that $N$ is even). The remaining modes correspond to their negative frequency counterparts. The mode with $ k = \frac{N}{2}$ corresponds to the Nyquist frequency, $f_{k=\frac{N}{2}}$~\cite{Nyquist}. Thus, one has the one-sided power spectral density (PSD)
\begin{equation}
    P_{+,\times}^{(k)}=|\Tilde{h}^{(k)}_{\,+,\times}|^2, \hspace{1cm} k=0,1,2,\ldots\frac{N}{2}.
\end{equation}
Theoretically, one can study several power spectral density templates of the gravitational waves emitted from a periodic orbit by exploring the parameter spaces. In this work, we present two kinds of templates considering the gravitational waves emitted from periodic orbits with: (a) identical $(z,w,v)$, but corresponding to different $g$, $E$, $L$ and (b) different $(z,w,v)$, with each triplet of values associated with a different $g$, keeping $E$ and $L$ fixed.
We mention that we use the previously specified values for the particle mass, the mass of the central massive regular black hole, the luminosity distance, latitude and the inclination angle.

\subsection{PSD of gravitational wave from identical periodic orbits at different $g$}\label{VA}
\noindent Let us consider the gravitational waveform from the periodic orbit $(1,2,0)$ around the Bardeen regular black hole.
We set $\epsilon=0.5$ and $E=0.9683828$, $0.9671666$, $0.9652875$ to have a $(1,2,0)$ periodic orbit corresponding to $g=0$, $0.4M$ and $0.6M$ respectively.
\begin{figure}[h]
\centering
\subfigure[\hspace{0.1cm}Fourier spectra $|\Tilde{h}_{+}(f)|$ for different values of $g$.]{\includegraphics[width=0.45\textwidth]{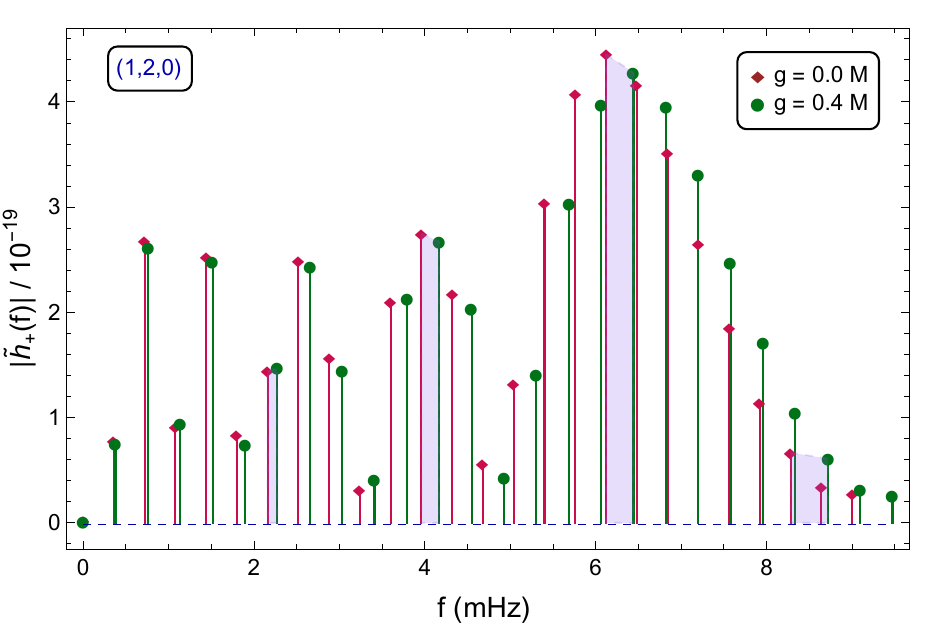}
\includegraphics[width=0.45\textwidth]{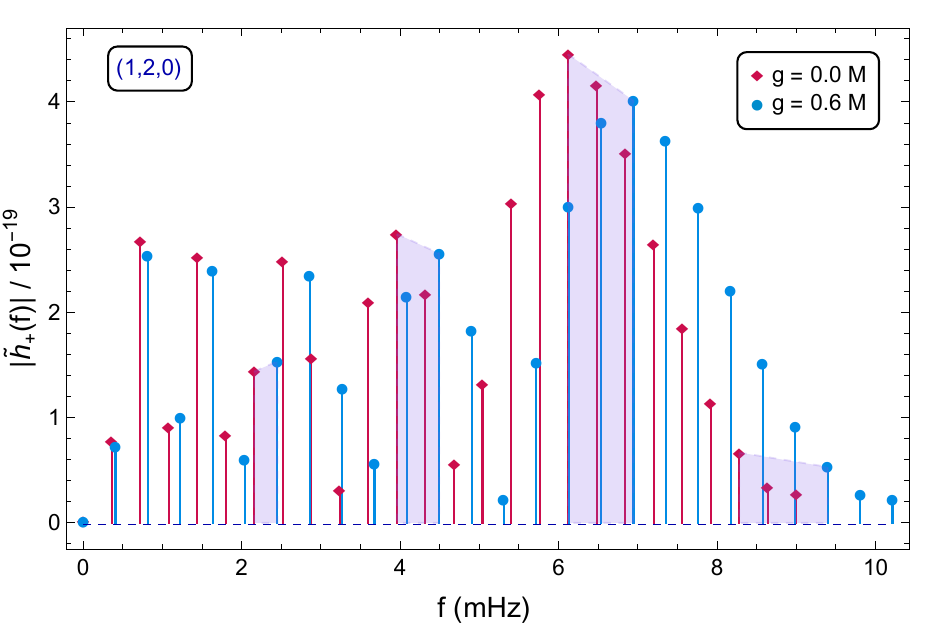}}
\subfigure[\hspace{0.1cm}Fourier spectra $|\Tilde{h}_{\times}(f)|$ for different values of $g$.]{\includegraphics[width=0.45\textwidth]{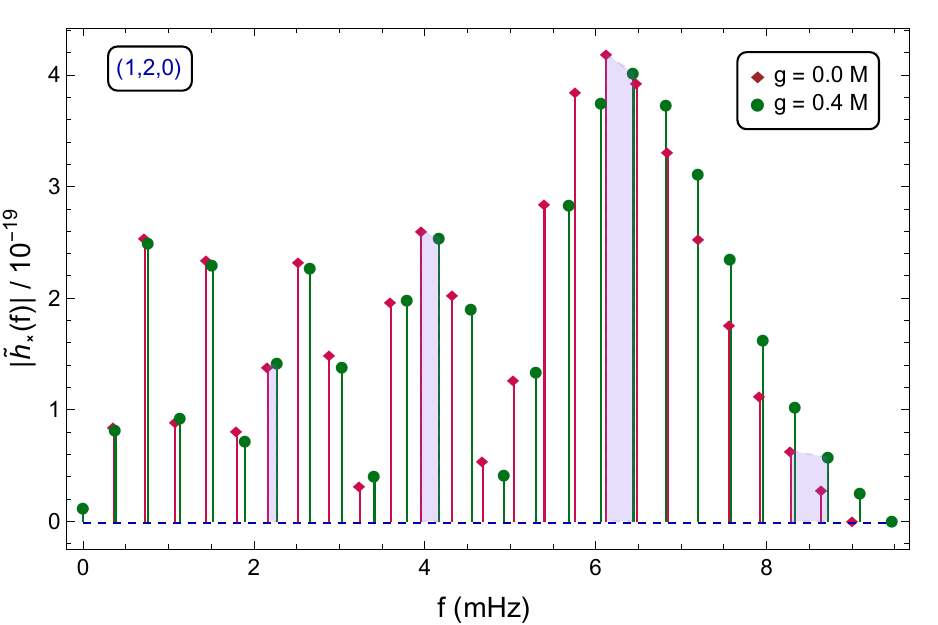}
\includegraphics[width=0.45\textwidth]{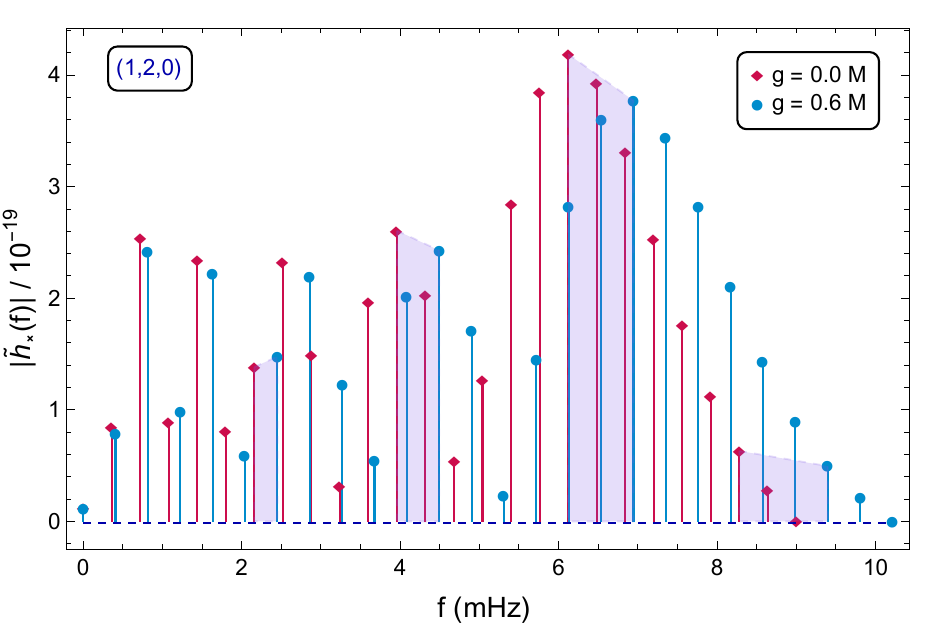}}
\caption{Fourier spectra $|\Tilde{h}_{+,\times}(f)|$ of the gravitational wave from a particle in $(1,2,0)$ periodic orbit around the Bardeen spacetime. The shaded regions highlight the frequency shift resulting from a finite $g$. $M=10^6\,M_{\odot}$, $m=10\,M_{\odot}$ and $D_L=200$ Mpc}
\label{fig:B_powerspectrum}
\end{figure}
The associated Fourier spectra for both $h_{+}$ and $h_{\times}$ are presented in Fig.~\ref{fig:B_powerspectrum}, which is obtained using  {Mathematica 13.0}. A cross-check of this numerical procedure is included in the Appendix~\ref{appendix2}.
The spectral frequencies fall primarily within the millihertz range. The power corresponding to the dominant frequencies is of the order $10^{-19}$. One may note that for a finite value of $g$, the Fourier spectrum is shifted right on the frequency axis relative to the $g=0$ case. The shift is shown in colored shades for a few chosen frequencies in the Fig.~\ref{fig:B_powerspectrum}. For a higher value of $g$, the shift is found to be high. Although a finite $g$ results in changes in the power of a particular mode, the change is not very significant.
Here, we take $N=50$ sample points in the full time period of the orbit $(1,2,0)$ for constructing the time series of the gravitational wave signals. The corresponding positive frequencies are listed in the Table~\ref{tab:frequencies1} for the two selected values of $g$ along with the Schwarzschild case.
\begin{table}[h]
\centering
\begin{tabular}{cc}
\begin{tabular}{|c|c|c|c|}
\hline
\multirow{2}{0.5cm}{$k$} & \multicolumn{3}{c|}{Absolute spectra frequency (mHz)} \\ \cline{2-4}
                        & $f_{g=0}$ & $f_{g=0.4M}$ & $f_{g=0.6M}$ \\ \hline\hline
0 & 0. & 0. & 0. \\ 
1 & 0.3600 & 0.3790 & 0.4086 \\
2 & 0.7200 & 0.7579 & 0.8171 \\
3 & 1.0801 & 1.1369 & 1.2257 \\
4 & 1.4401 & 1.5159 & 1.6343 \\
5 & 1.8001 & 1.8948 & 2.0429 \\
6 & 2.1601 & 2.2738 & 2.4514 \\
7 & 2.5202 & 2.6528 & 2.8600 \\
8 & 2.8802 & 3.0317 & 3.2686 \\
9 & 3.2402 & 3.4107 & 3.6772 \\
10 & 3.6002 & 3.7897 & 4.0857 \\
11 & 3.9603 & 4.1687 & 4.4943 \\
12 & 4.3203 & 4.5476 & 4.9029 \\
\hline
\end{tabular}
&
\begin{tabular}{|c|c|c|c|}
\hline
\multirow{2}{0.5cm}{$k$} & \multicolumn{3}{c|}{Absolute spectra frequency (mHz)} \\ \cline{2-4}
                        & $f_{g=0}$ & $f_{g=0.4M}$ & $f_{g=0.6M}$ \\ \hline\hline
13 & 4.6803 & 4.9266 & 5.3115 \\
14 & 5.0403 & 5.3056 & 5.7200 \\
15 & 5.4003 & 5.6845 & 6.1286 \\
16 & 5.7604 & 6.0635 & 6.5372 \\
17$^\dagger$ & 6.1204 & 6.4425 & 6.9458 \\
18 & 6.4804 & 6.8214 & 7.3543 \\
19 & 6.8405 & 7.2004 & 7.7629 \\
20 & 7.2005 & 7.5794 & 8.1715 \\
21 & 7.5605 & 7.9583 & 8.5801 \\
22 & 7.9205 & 8.3373 & 8.9886 \\
23 & 8.2806 & 8.7163 & 9.3972 \\
24 & 8.6406 & 9.0952 & 9.8058 \\
25 & 9.0006 & 9.4742 & 10.2144 \\
\hline
\end{tabular}
\end{tabular}
\caption{Spectral frequencies of the gravitational waveform for the particle in the $(1,2,0)$ periodic orbit around the Bardeen regular black hole for different values of $g$. Row marked with $\dagger$ corresponds to the most dominant mode.}
\label{tab:frequencies1}
\end{table}
The above-discussed spectral shift is clearly noticeable here. The mode with frequency $f_{k=17}$ corresponds to the mode with the highest power. The Nyquist frequency corresponds to $f_{k=25}$. 

\begin{figure}[h]
\centering
\includegraphics[width=0.5\textwidth]{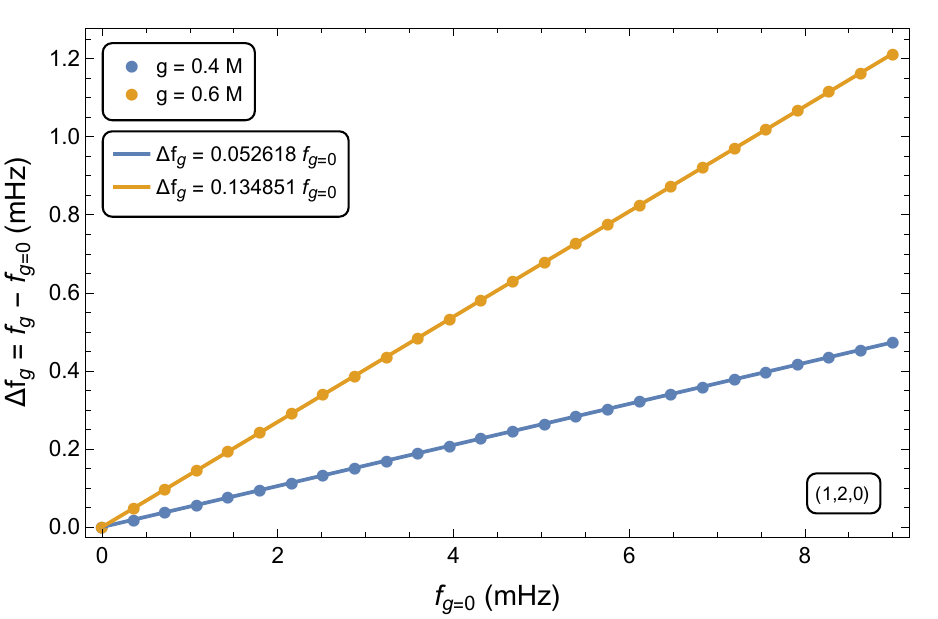}
\caption{Variation of the spectral frequency shift $\Delta f_g=f_g-f_{g=0}$ with the reference spectral frequency $f_{g=0}$ of the gravitational wave from the $(1,2,0)$ orbit around the Bardeen regular black hole for two values of $g$. Two sets of scattered points represent the corresponding $\Delta f_g$, and solid lines show their respective linear fits.}
\label{fig:frquency_shift}
\end{figure}

\noindent To quantify the shift in spectral frequency, we define the spectral frequency shift  $\Delta f_g=f_g-f_{g=0}$, which measures how the frequency changes for a finite value of $g$ relative to the Schwarzschild case. In Fig.~\ref{fig:frquency_shift}, the variation of spectral frequency shift $\Delta f_g$ with the reference spectral frequency $f_{g=0}$ is shown.
We observe that the shift $\Delta f_g$ grows linearly with $f_{g=0}$. This indicates that a finite value of $g$ produces a linearly increasing `{\sf blueshift}' in the frequency spectrum, which may be considered as a distinguishing feature of a regular black hole relative to its singular solution.
The primary reason behind this shift is explained as follows. When $g$ is nonzero, the time period of an identical periodic orbit $(z,w,v)$ is shorter compared to the $g=0$ case.
In Fig.~\ref{fig:B_colored_orbit}, we also discuss how the orbit shrinks for finite $g$. This contraction of the orbital size grows as $g$ increases. A shorter orbital time period leads to faster oscillatory motion. Therefore, when we evaluate the Fourier spectrum using the same number of bins $(N=50)$, the corresponding frequencies with each mode become higher for larger $g$.
We also examined the Fourier spectrum corresponding to the gravitational waveform from other periodic orbits, like $(2,1,1)$, $(3,1,2)$, etc., around the Bardeen spacetime (not included here). The spectral frequency also lies within the millihertz range. Similar to the $(1,2,0)$ orbit, we observe a linearly growing `{\sf blueshift}' in the frequency spectrum in these cases as well.

\subsection{PSD of gravitational wave from periodic orbits with constant $E$ and $L$ }
\noindent In this subsection, we study another template of power spectral density
\begin{figure}[h!]
\centering
\subfigure[\hspace{0.1cm}Fourier spectra $|\Tilde{h}_{+,\times}(f)|$ for $(1,2,0)$ orbit with $g=0.000M$, $E=0.96$ and $L=3.6534056M$.]{\includegraphics[width=0.45\textwidth]{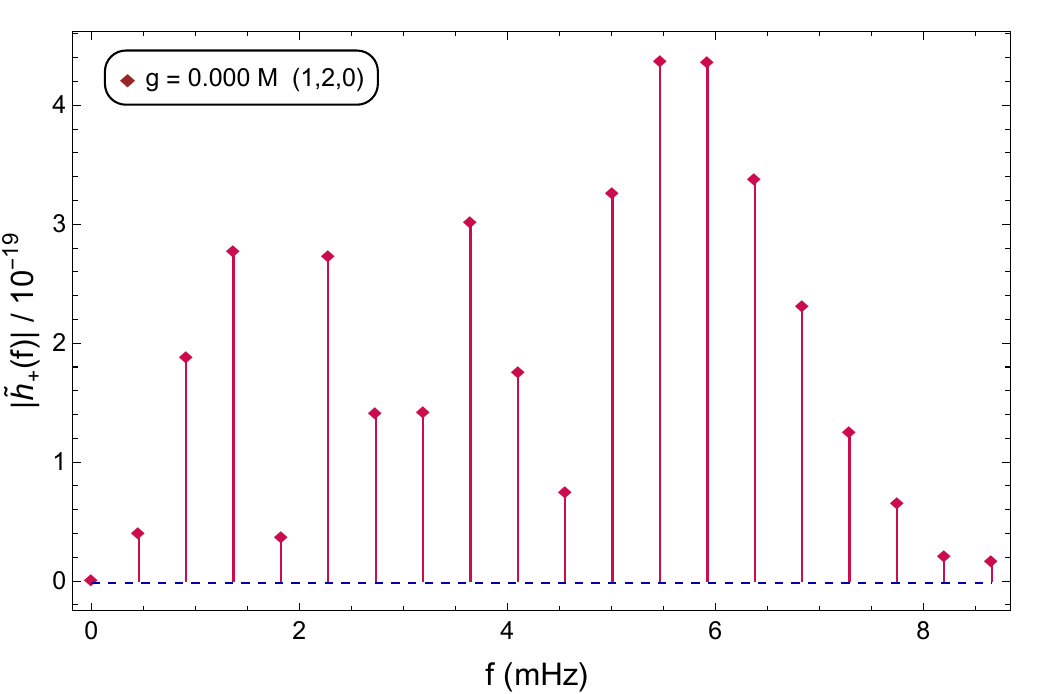}
\includegraphics[width=0.45\textwidth]{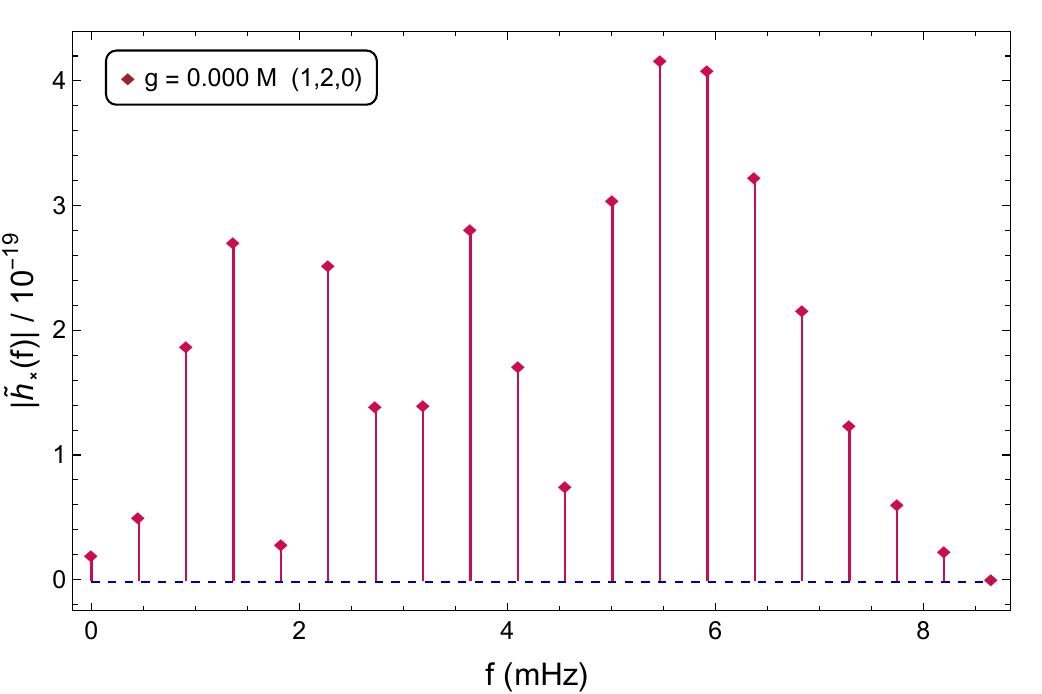}}
\subfigure[\hspace{0.1cm}Fourier spectra $|\Tilde{h}_{+,\times}(f)|$ for $(2,1,1)$ orbit with $g=0.126M$, $E=0.96$ and $L=3.6534056M$.]{\includegraphics[width=0.45\textwidth]{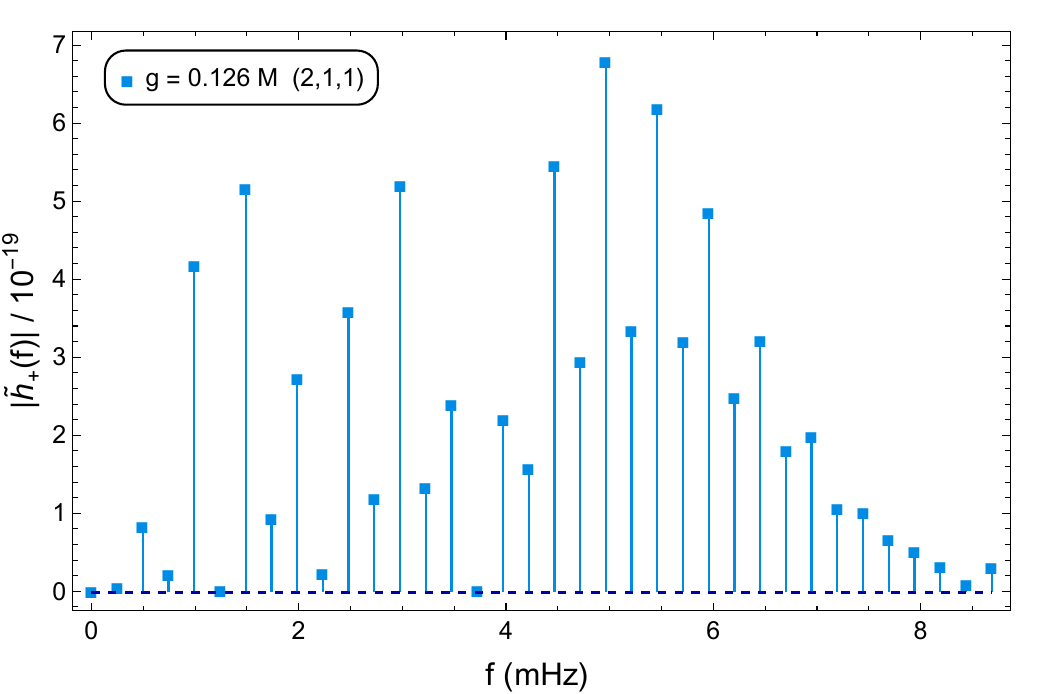}
\includegraphics[width=0.45\textwidth]{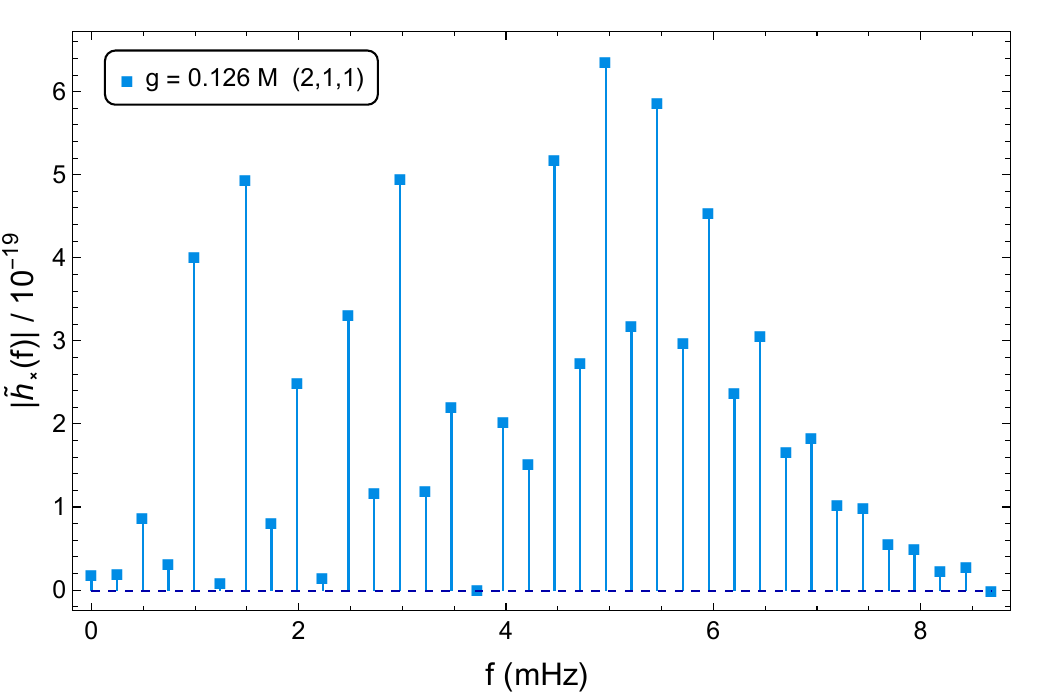}}
\subfigure[\hspace{0.1cm}Fourier spectra $|\Tilde{h}_{+,\times}(f)|$ for $(3,1,2)$ orbit with $g=0.086M$, $E=0.96$ and $L=3.6534056M$.]{\includegraphics[width=0.45\textwidth]{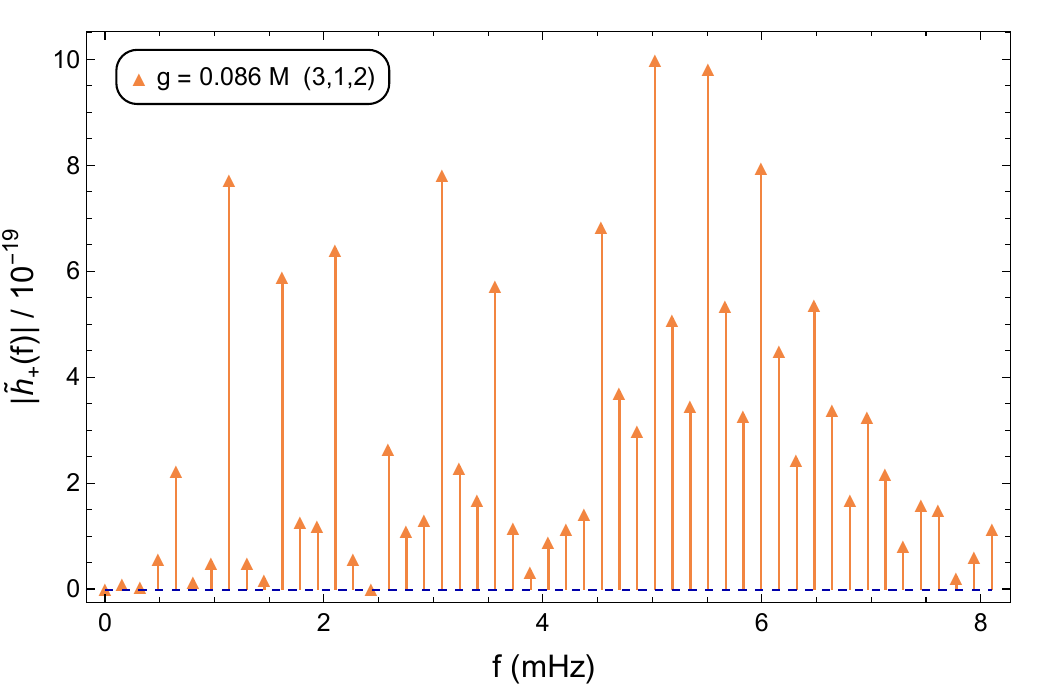}
\includegraphics[width=0.45\textwidth]{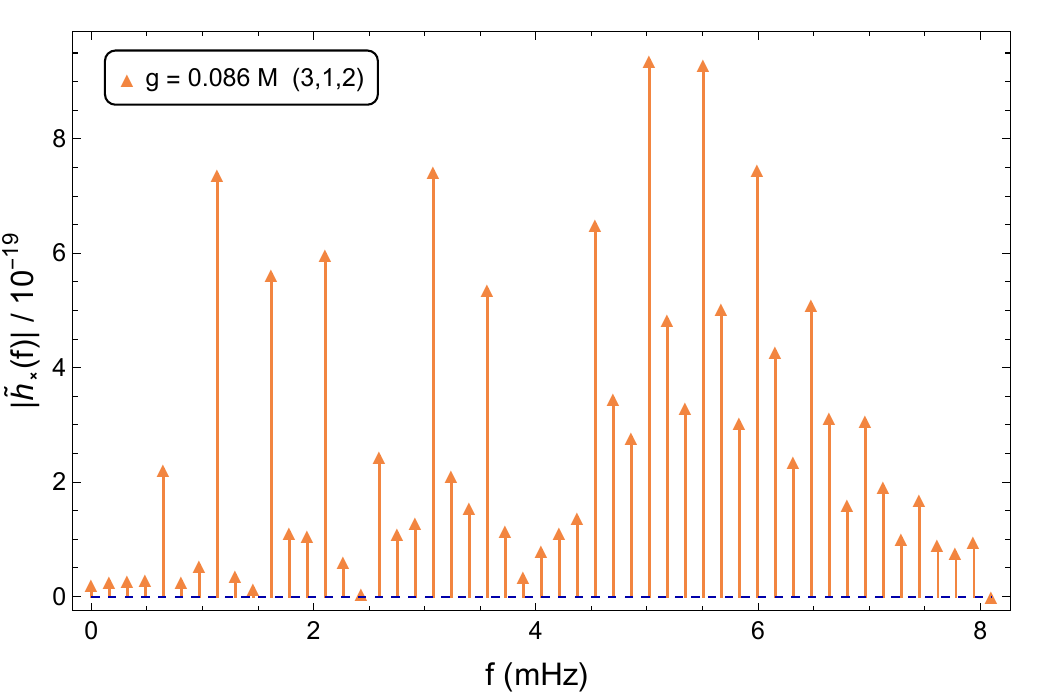}}
\caption{Plots of the Fourier spectra $|\Tilde{h}_{+,\times}(f)|$ of the gravitational wave from different periodic orbits corresponding to different $g$ values with fixed $E$ and $L$. $M=10^6\,M_{\odot}$, $m=10\,M_{\odot}$ and $D_L=200$ Mpc}
\label{fig:B_powerspectrum_E_L_constant}
\end{figure}
considering the gravitational wave from different periodic orbits associated with different $g$ values while keeping the energy and angular momentum of the orbiting particle fixed for the Bardeen regular black hole. We choose periodic orbits $(1,2,0)$, $(2,1,1)$ and $(3,1,2)$ corresponding to $g=0$, $0.126M$ and $0.86M$, respectively, all with the same energy $E=0.96$ and angular momentum $L=3.6534056M$ (as mentioned in Table~\ref{table_Bqvsg}). The associated Fourier spectra are shown in Fig.~\ref{fig:B_powerspectrum_E_L_constant}. The dominant spectral frequencies are found in the millihertz region. We observe that as $z$ increases, the power associated with the most dominant mode increases.
To construct the Fourier spectra, we take $N=38$, $70$, and $100$ sample points for the periodic orbit $(1,2,0)$, $(2,1,1)$ and $(3,1,2)$, respectively.
Although the highest resolved frequency in Fig.~\ref{fig:B_powerspectrum_E_L_constant} is about $8$ mHz for all orbits, the longer time period orbits require a larger number of sample points to resolve frequencies as compared to those for shorter time period orbits.
Thus, we explore only two cases to determine whether a given gravitational wave signal from a periodic orbit originates from a spacetime with finite $g$ or from the $g=0$ Schwarzschild geometry.

\subsection{Comparison with LISA characteristic sensitivity}
\noindent As discussed above, the characteristic frequencies of gravitational waves from periodic orbits fall within the millihertz range, making them suitable for observation in future space-based gravitational wave detectors, such as LISA (Laser Interferometer Space Antenna)~\cite{Gair:2004, LISA:2017}.
We assess the detectability of the gravitational waves from periodic orbits by comparing the dimensionless characteristic strain $h_c(f)$ defined in Eq.~\eqref{eq:characteristric_strain} with the LISA characteristic sensitivity. The reason for the characteristic strain of gravitational waves as the appropriate quantity for comparison against the LISA sensitivity is discussed in Ref.~\cite{Robson}. The analytical expression of the LISA sensitivity curve is also given in Ref.~\cite{Robson}. In this work, we consider the parameter values of the sensitivity curve corresponding to the four years of galactic-confusion-noise foreground removal, as mentioned in~\cite{Robson}. To capture the low-frequency content of the signal, we set the observation window to be five times the orbital time period of the specific periodic orbit. Higher frequency modes are resolved by setting a sufficiently small bin size.
\begin{figure}[h]
\centering
\includegraphics[width=0.45\textwidth]{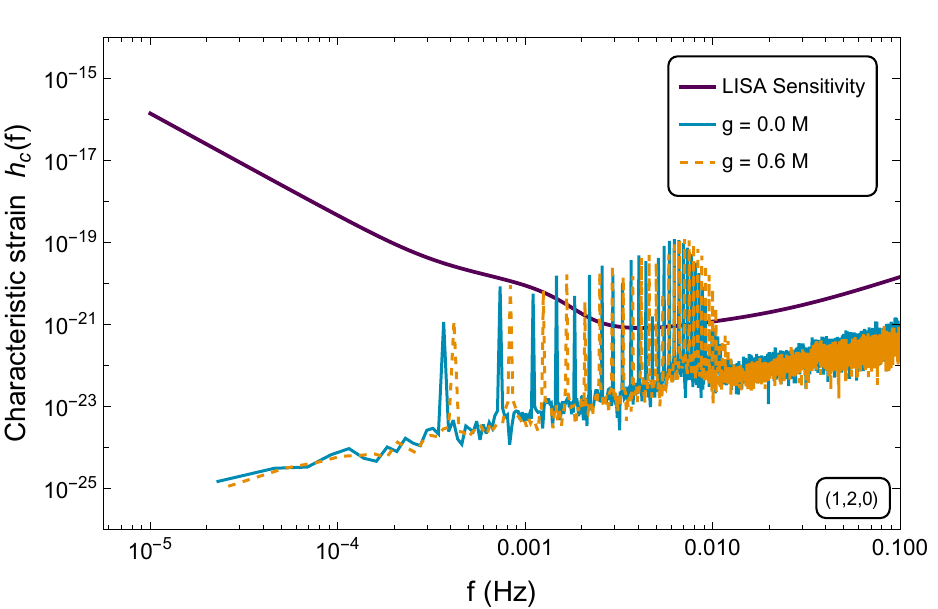}
\includegraphics[width=0.45\textwidth]{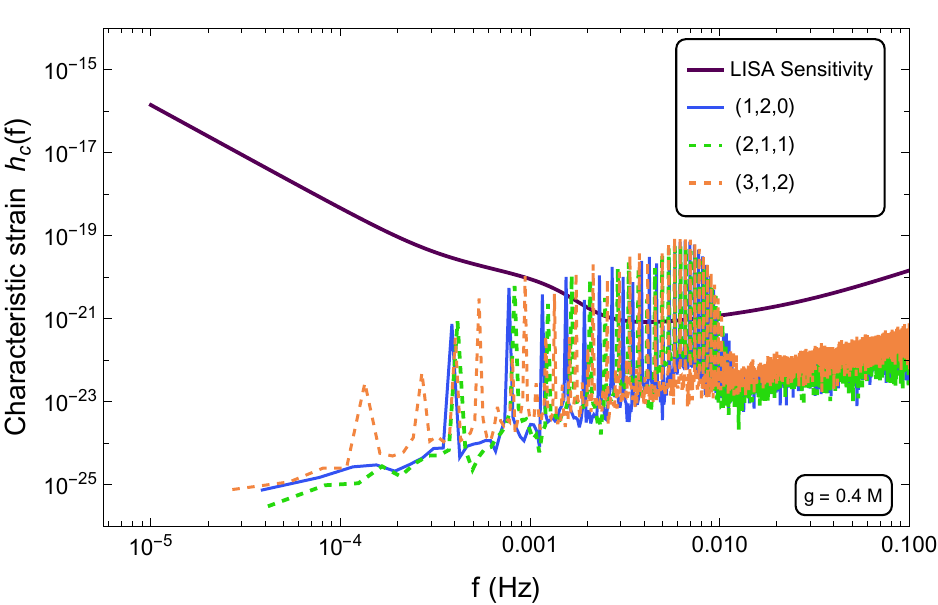}
\caption{Comparison of the characteristic strain $h_c(f)$ of the gravitational waveforms of the orbit $(1,2,0)$ for different values of $g$ (left), and for different orbits at $g=0.4M$ (right), with the LISA sensitivity for the Bardeen regular black hole. $M=10^6\,M_{\odot}$, $m=10\,M_{\odot}$ and $D_L=200$ Mpc}
\label{fig:lisa_sensitivity}
\end{figure}
In Fig.~\ref{fig:lisa_sensitivity} (left), we demonstrate a comparison between the characteristic strain $h_c(f)$ of the $(1,2,0)$ orbit and LISA sensitivity for different values of $g$. 
It is observed that some segment of $h_c(f)$ exceeds the LISA sensitivity, primarily in the millihertz range. For a fixed value of $g$, the characteristic strain of the different orbits is also compared in Fig.~\ref{fig:lisa_sensitivity} (right), which also falls within the LISA detectability region in the millihertz band. Here, $E$ and $L$ vary in each spectrum.
Thus, gravitational wave signals from different periodic orbits around the Bardeen regular black hole may be detectable in LISA.

\noindent In the same way, the power spectral density of the gravitational wave corresponding to different periodic orbits around the Hayward regular black hole can also be computed. We find that, similar to the Bardeen spacetime, the characteristic frequencies in this case also lie within the millihertz range (not shown here).
\begin{figure}[h]
\centering
\includegraphics[width=0.45\textwidth]{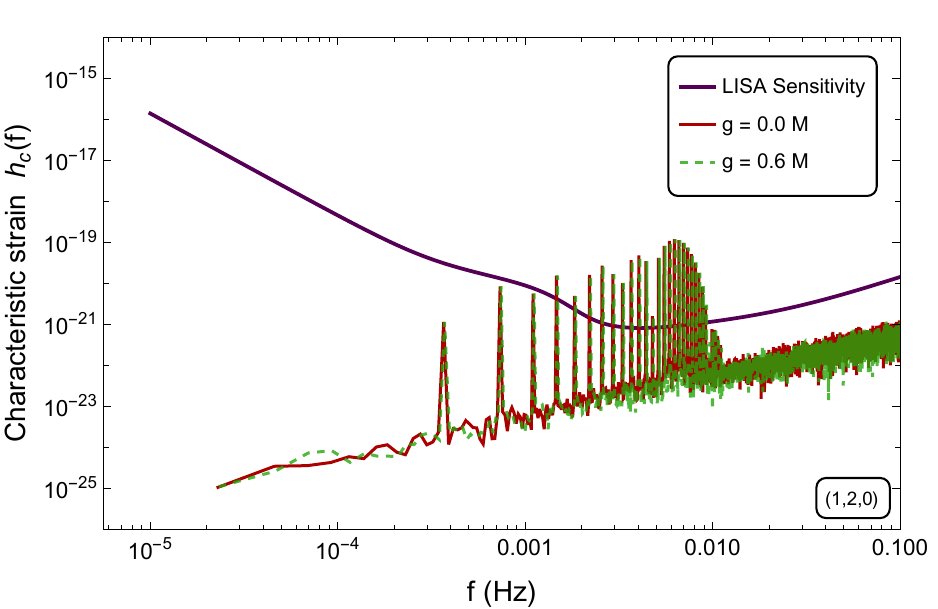}
\includegraphics[width=0.45\textwidth]{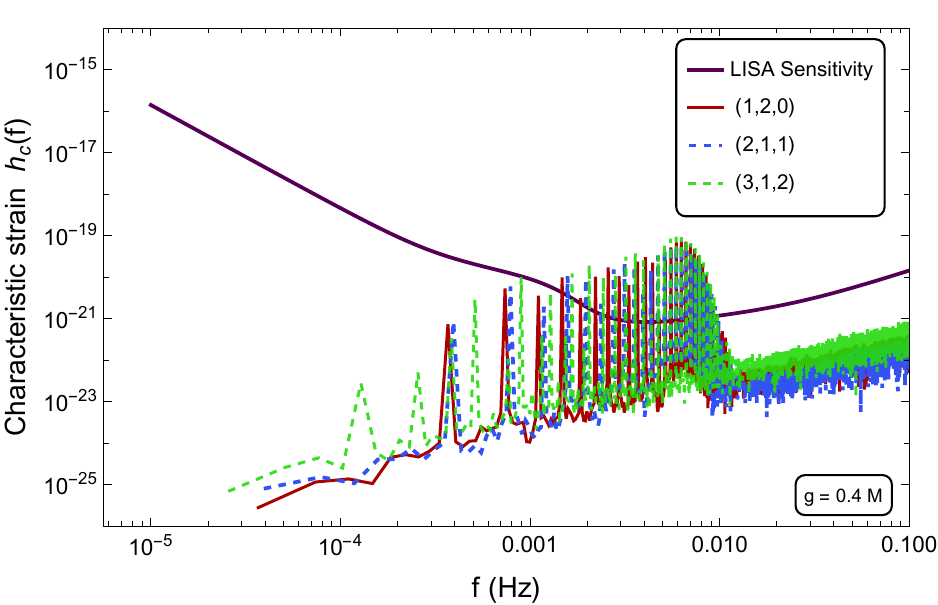}
\caption{Graph of the $h_c(f)$ of the gravitational waveforms from the orbit $(1,2,0)$ for two values of $g$ (left) and from different orbits at $g=0.4M$ (right) compared with the LISA sensitivity for the Hayward regular black hole. $M=10^6\,M_{\odot}$, $m=10\,M_{\odot}$ and $D_L=200$ Mpc}
\label{fig:lisa_sensitivity1}
\end{figure}
Fig.~\ref{fig:lisa_sensitivity1} compares the characteristic strain $h_c(f)$ of gravitational waves from a periodic orbit against the LISA sensitivity curve for different values of $g$ as well as for different periodic orbits. In this case, the influence of the parameter $g$ is relatively less prominent than in the Bardeen spacetime.
One can observe that, similar to the Bardeen case, the gravitational wave from periodic orbits around the Hayward regular black hole may also be trapped in LISA.

\section{Conclusion}\label{VI}
\noindent Let us now summarise our work with a few concluding remarks. We also discuss the limitations of our current work and outline several potential future directions.
\noindent First, we recall geodesic motion of test particles in two well-known regular black holes. By analysing the effective potential, we obtain the allowed values of energy and angular momentum of the orbiting particle that permit bounded motion. We demonstrate different periodic orbits that are characterised by the rational number $q$ in these two types of regular black hole spacetimes.
In particular, we study how $q$ behaves with the particle energy and angular momentum, for different values of the parameter $g$.
\noindent Second, we obtain the gravitational waveforms emitted from the periodic orbits around the regular black hole using the adiabatic approximation. To evaluate the waveforms, we use the numerical Kludge method. Our analysis shows that the presence of a finite value of $g$ induces a phase shift in the resulting gravitational waveform with respect to the singular $g=0$ case. An analytical expression of the gravitational wave from periodic orbits in the Schwarzschild spacetime is discussed in the Appendix~\ref{Appendix1}.
\noindent Finally, we obtain the power spectral density of the gravitational signals from periodic orbits using the discrete Fourier transform. We discuss two classes of waveform templates: one with a fixed periodic orbit configuration and another with fixed energy and angular momentum of the orbiting particle.
Given a fixed periodic orbit, we find that a finite $g$ value introduces a linearly growing {\sf blueshift} in the frequency spectrum relative to the Schwarzschild spacetime. We also demonstrate that the characteristic strain of gravitational wave signals from periodic orbits exceeds the LISA threshold in the millihertz range. In Appendix~\ref{appendix2}, we verify the numerical results via the reconstruction of the waveform using the inverse Fourier transform.
\noindent There are some limitations of our work. We adopt the adiabatic approximation to construct the waveforms, which neglects the back reaction due to the gravitational wave emission. This approximation provides reliable results over a few orbital time periods. However, for long-term evolution, it is not a suitable option. 
Moreover, we neglect multipole contributions of order higher than quadratic. For greater accuracy, the inclusion of higher order multipole terms is necessary.
\noindent In future work, we plan to address the above limitations in order to get more accurate results.
Our work can be extended further by including spin in regular black holes, as astrophysical black holes are expected to be rotating.
Another natural direction to explore is the inclusion of various
types of environmental effects which will surely influence the characteristics of gravitational wave emissions from periodic orbits in regular black holes. 

\section*{Acknowledgements} 
\noindent AK expresses gratitude to Minal Chhabra for several insightful and valuable discussions on the power spectrum, which significantly benefited this work. He also thanks IIT Kharagpur, India for the fellowship support. RA thanks 
Department of Physics, IIT Kharagpur, India where a part of this work was initiated during his BS dissertation, done under the supervision of one of the authors (SK) here. SJ acknowledges the warm hospitality and support provided during the visits at the Department of Physics, IIT Kharagpur when this research work was carried out.

\appendix
\section{Comparison between numerical and analytical waveforms}\label{Appendix1}
\begin{figure}[h]
\centering
\includegraphics[width=0.9\textwidth]{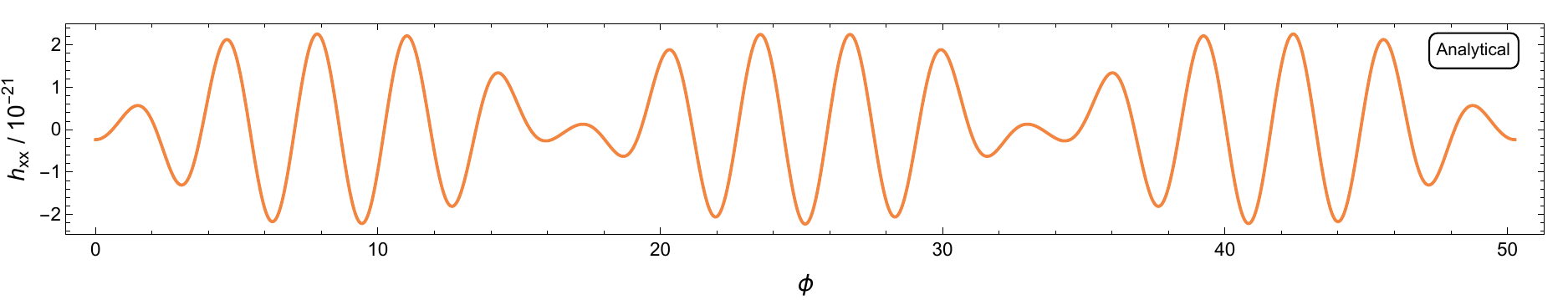}
\caption{Analytically computed $h_{xx}$ component of the waveform as a function of $\phi$ for the $(3,1,2)$ periodic orbit. $M=10^6\,M_{\odot}$, $m=10\,M_{\odot}$ and $D_L=200$ Mpc}
\label{fig:AppendixA_plot1}
\end{figure}

\begin{figure}[h]
\centering
\includegraphics[width=0.9\textwidth]{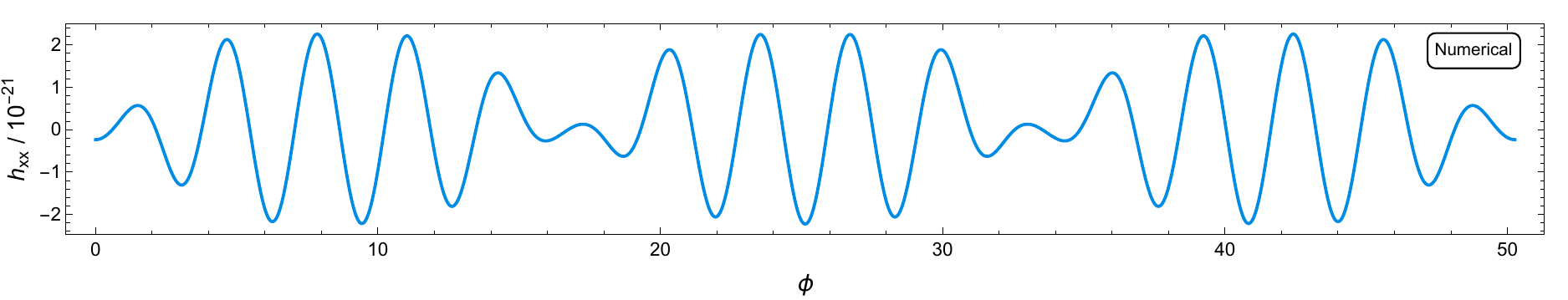}
\caption{Numerically computed $h_{xx}$ component of the waveform as a function of $\phi$ for the $(3,1,2)$ periodic orbit. $M=10^6\,M_{\odot}$, $m=10\,M_{\odot}$ and $D_L=200$ Mpc}
\label{fig:AppendixA_plot2}
\end{figure}

\noindent In this appendix, we compare the gravitational wave emitted from a periodic orbit around the Schwarzschild black hole $(g=0)$, computed using our numerical approach, with the waveform obtained analytically up to a tractable order. This will provide a consistency check of our numerical method. Following Ref.~\cite{Lim:2024}, the trajectory of the bound orbits that do not fall inside the Schwarzschild radius, can be expressed explicitly in terms of orbital parameters as,
\begin{equation}\label{A1}
    r(\phi)=\frac{\lambda}{1-e+2e\,\sn^2\left(\sqrt{1+\frac{2M(e-3)}{\lambda}}\,\frac{\phi}{2},\,\sqrt{\frac{4Me}{\lambda+2M(e-3)}}\right)}\,,
\end{equation}
where $e$ and $\lambda$ are the {\em eccentricity} of the orbit and the {\em latus rectum}, respectively. The sn$(\theta,k)$ is the Jacobi elliptic sine function~\cite{Gradshteyn}. The parameter pair $(e,\lambda)$ is related to the energy and angular momentum of the orbiting particle through the following relations~\cite{Lim:2024}
\begin{equation}\label{A2}
    E=\sqrt{\frac{\lambda^2-4M\lambda+4M^2-4M^2e^2}{\lambda(\lambda-3M-Me^2)}}\,, \qquad L=\lambda\sqrt{\frac{M}{\lambda-3M-Me^2}}\,.
\end{equation}
For periodic orbits in the Schwarzschild spacetime, Eqs.~\eqref{3.n1} and~\eqref{3.2} are derived fully in analytical form in Ref.~\cite{Lim:2024} and are given by
\begin{equation}\label{A3}
    q+1=\frac{\Delta\phi_{r}}{2\pi}=\frac{2}{\pi\sqrt{1+\frac{2M}{\lambda}(e-3)}}\,K\left(\sqrt{\frac{4Me}{\lambda+2M(e-3)}}\,\right)\,,
\end{equation}
where $K(k)$ is the complete elliptic integral of the first kind with elliptic modulus $k$. Thus, a periodic orbit (characterised by $q$) occurs if $K\left(\sqrt{\frac{4Me}{\lambda+2M(e-3)}}\,\right)$ is a rational multiple of $\left(\pi\sqrt{1+\frac{2M}{\lambda}(e-3)}\right)^{-1}$. As $K(k)$ is a monotonic function of $k$, we have to find the $\lambda$ for a given periodic orbit of $q$ with eccentricity $e$ by solving Eq.~\eqref{A3}.
\noindent The gravitational wave from periodic is obtained using the quadrupolar formula in Eq.~\eqref{4.7}, which is
\begin{equation}
     h_{ij}= \frac{2m}{D_L}\left(a_i x_j + a_j x_i+ 2 v_iv_j\right)\,.
 \end{equation}
We evaluate the $h_{xx}$ component analytically as a function of $\phi$ by expressing the acceleration and velocity of the particle (test object) as functions of orbital phase in the following way
\begin{equation}
    x(\phi)=r(\phi)\,\cos{\phi},\qquad v_{x}=\frac{dx}{d\phi}\,\frac{d\phi}{dt},\qquad a_x=\frac{d^2\phi}{dt^2}\,\frac{dx}{d\phi}+\left(\frac{d\phi}{dt}\right)^2\,\frac{d^2x}{d\phi^2}\,,
\end{equation}
where $\frac{d\phi}{dt}$ is calculated using geodesic equation. Therefore, $h_{xx}$ is obtained as
\begin{align}
    h_{xx}(\phi)&= \frac{2m}{D_L}\frac{M \left(M \left(e \left(4 \sn^2(\theta,k)-2\right)+2\right)-\lambda \right)}
{\lambda ^2 (2 (e+1) M-\lambda ) (2 (e-1) M+\lambda )}
\Big\lbrace A(\phi)\,\cos{(2\phi)}+B(\phi)\sin{(2\phi)}+C(\phi)\Big\rbrace
\end{align}
where
\begin{align*}
    A(\phi)&=
4 e M^2 \left(6 e \sn^4(\theta,k)-6 (e-1) \sn^2(\theta,k)+e-3\right)
\left(-2 e \sn^2(\theta,k)+e-1\right)^2  \nonumber\\
&\quad
+4 \lambda M
\left(1+e \left\{-1+\sn^2(\theta,k) \left(2+e \{\sn^2(\theta,k)-1\}
\left\{5+e (2 \sn^2(\theta,k)-1)\right\}\right)\right\}\right)
\nonumber\\
&\quad
-\lambda ^2 \left(2+e \left\{8 e \sn^4(\theta,k)-8 e \sn^2(\theta,k)+6 \sn^2(\theta,k)+e-3\right\}\right)\,,
\end{align*}
\begin{align*}
    B(\phi)&=4e \cn(\theta,k) \dn(\theta,k) \sn(\theta,k) \nonumber\\
    &\quad \sqrt{\lambda  (2 (e-3) M+\lambda )}\left(2 e \sn^2(\theta,k)-e+1\right)\left(M \left(-2 e \sn^2(\theta,k)+e-1\right)+\lambda \right)\,,
\end{align*}
\begin{align*}
    C(\phi)&=4e M^2 \left(6 e \sn^4(\theta,k)-6 (e-1) \sn^2(\theta,k)+e-3\right) \left(-2 e \sn^2(\theta,k)+e-1\right)^2-4 \lambda  Me \nonumber\\
    &\quad\left\{-2+4 \sn^2(\theta,k)+e \left\{3-e+6 e \sn^6(\theta,k)+(7-9 e) \sn^4(\theta,k)+(5 e-7) \sn^2(\theta,k)\right\}
    \right\}\nonumber\\
    &\quad+\lambda ^2e \left(e+2 \sn^2(\theta,k)-1\right).
\end{align*}
Here, $\cn(\theta,k)$ and $\dn(\theta,k)$ are Jacobi elliptic cosine function and Jacobi elliptic delta function, respectively with $\theta=\sqrt{1+\frac{2M(e-3)}{\lambda}}\,\frac{\phi}{2}$ and $k=\sqrt{\frac{4Me}{\lambda+2M(e-3)}}$. Similarly, $h_{yy}$ and $h_{xy}$ can be calculated in terms of the Jacobi elliptic functions, which we have not included here.
\noindent We consider the $(3,1,2)$ periodic orbit in Schwarzschild spacetime with eccentricity $e=0.8$. By solving Eq.~\eqref{A3}, we find $\lambda=7.82779M$. The corresponding energy and angular momentum of the test particle, calculated using Eq.~\eqref{A2}, are $E=0.978756$ and $L=3.82513M$, respectively. In Fig.~\ref{fig:AppendixA_plot1}, we present the $h_{xx}$ component corresponding to the $(3,1,2)$ orbit as a function of $\phi$, which shows nearly perfect agreement with the numerical result as shown in Fig.~\ref{fig:AppendixA_plot2}. This confirms the consistency of our numerical method.
To evaluate $h_{xx}$ component as a function of $t$, we first need to express $t$ as a function $\phi$ and then invert this relation. Even for the Schwarzschild case, this process is analytically hard. Therefore, we compare the numerical result with the analytically tractable expression.

\section{Waveform reconstruction via inverse Fourier transform}\label{appendix2}
\begin{figure}[h]
\centering
\includegraphics[width=0.9\textwidth]{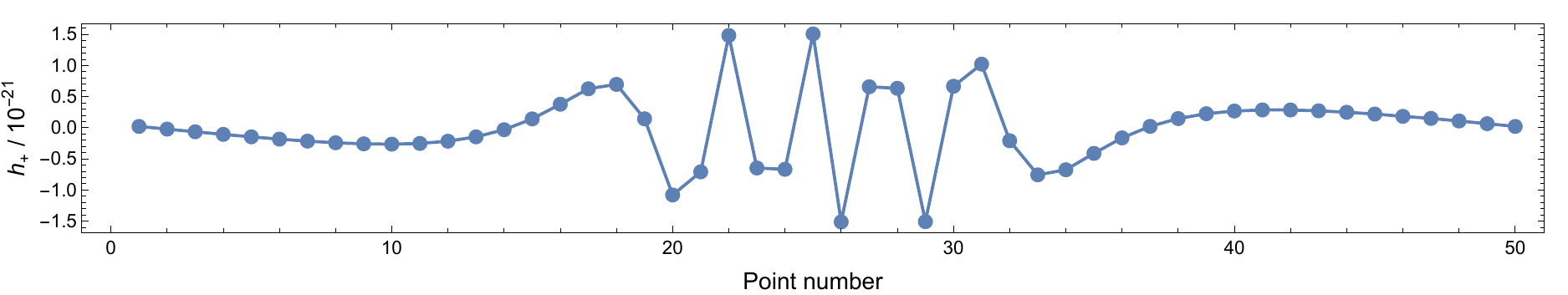}
\caption{The original theoretical gravitational wave form $(1,2,0)$ orbit corresponding to $g=0$ for $\epsilon=0.5$ and $E=0.9683828$ in the Bardeen spacetime with $50$ sample points. $M=10^6\,M_{\odot}$, $m=10\,M_{\odot}$ and $D_L=200$ Mpc}
\label{fig:Appendix_plot1}
\end{figure}

\begin{figure}[h]
\centering
\includegraphics[width=0.9\textwidth]{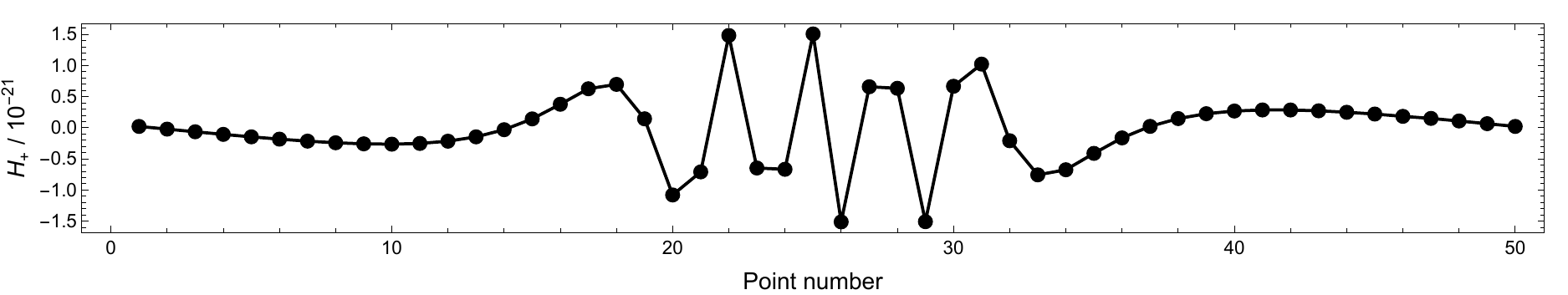}
\caption{The reconstructed signal $H_+$ form $(1,2,0)$ orbit corresponding to $g=0$ case.}
\label{fig:Appendix_plot2}
\end{figure}

\begin{figure}[h]
\centering
\includegraphics[width=0.9\textwidth]{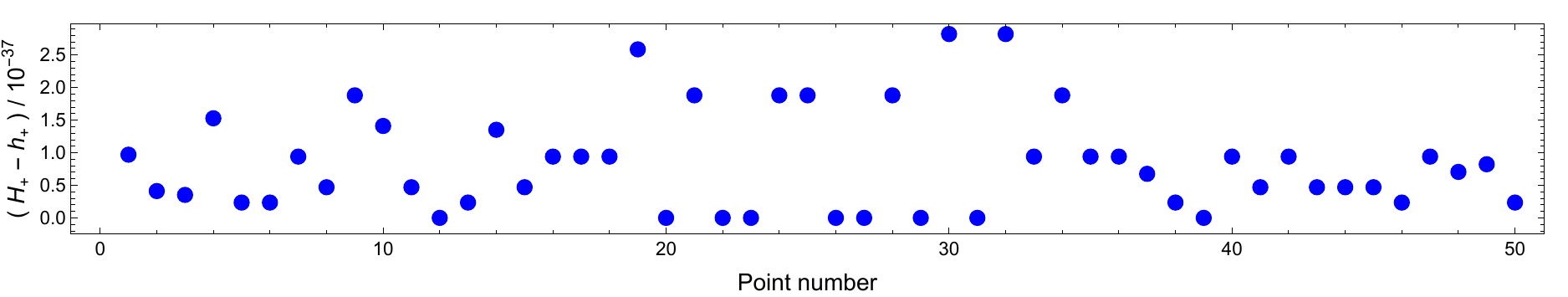}
\caption{Estimation of point-wise error of the original signal with the reconstructed one.}
\label{fig:Appendix_plot3}
\end{figure}

\noindent In the section \ref{VA}, we construct the Fourier spectrum of the gravitational wave from the $(1,2,0)$ orbit in Bardeen spacetime using \emph{Fourier} in {\it Mathematica 13.0}.
The sample point of the signal is $N=50$. Here, we reconstruct the time domain signal from the Fourier spectra by incorporating \emph{InverseFourier} in {\it Mathematica 13.0} and estimate the point-wise error in this process. For this analysis, we choose the signal $h_+$ corresponding to the $g=0$ case. 
The original theoretical signal is shown in Fig.~\ref{fig:Appendix_plot1}.
The reconstructed signal, denoted as $H_+$, from the Fourier spectrum is presented in Fig.~\ref{fig:Appendix_plot2}, where we only retain the real part of the output.
The imaginary component is of order $10^{-37}$ and is therefore neglected.
The point-wise error, defined as $H_+-h_+$, is shown in Fig.~\ref{fig:Appendix_plot3}. It is found to be of the order of $10^{-37}$, which is sufficiently small. Therefore, one may infer that the reconstruction is `accurate' and the results are consistent. 

\begin{thebibliography}{Rubinsteinetal}

\bibitem{Davis: 1971} M. Davis, R. Ruffini, W. H. Press, and
R. Price, \href{https://doi.org/10.1103/PhysRevLett.27.1466}{Phys. Rev. Lett. {\bf D 27}, 1466 (1971).}

\bibitem{Danzmann:1997} K. Danzmann, \href{https://doi.org/10.1088/0264-9381/14/6/002}{Class. Quant. Grav. {\bf 14}, 1399 (1997).}

\bibitem{Schutz:1999} B. F. Schutz, \href{https://doi.org/10.1088/0264-9381/16/12A/307}{Class. Quant. Grav. {\bf 16}, A131 (1999).}

\bibitem{Gair:2004} J. R. Gair, L. Barack, T. Creighton, C. Cutler, S. L. Larson, E. S. Phinney, and M. Vallisneri, \href{https://doi.org/10.1088/0264-9381/21/20/003}{Class. Quant. Grav. {\bf 21}, S1595 (2004).}

\bibitem{LISA:2017} P. Amaro-Seoane et al. (LISA), (2017), \href{https://doi.org/10.48550/arXiv.1702.00786}{arXiv:1702.00786 [astro-ph.IM].}

\bibitem{Maselli:2022} A. Maselli, N. Franchini, L. Gualtieri, T. P. Sotiriou, S. Barsanti, and P. Pani, \href{https://doi.org/10.1038/s41550-021-01589-5}{Nature Astron. {\bf 6}, 464 (2022).}

\bibitem{Luo:2016} J. Luo et al. (TianQin), \href{https://doi.org/10.1088/0264-9381/33/3/035010}{Class. Quant. Grav. {\bf 33}, 035010 (2016).}

\bibitem{Gong:2021} Y. Gong, J. Luo, and B. Wang, \href{https://doi.org/10.1038/s41550-021-01480-3}{Nature Astron. {\bf 5}, 881 (2021).}

\bibitem{Hu:2017} W.-R. Hu and Y.-L. Wu, \href{https://doi.org/10.1093/nsr/nwx116}{Natl. Sci. Rev. {\bf 4}, 685 (2017).}

\bibitem{Bardeen:1968} J. M. Bardeen, {\em in Proceedings of International Conference GR5, Tbilisi, USSR, 1968}, p. 174.

\bibitem{Hayward:2006} S. A. Hayward, \href{https://link.aps.org/doi/10.1103/PhysRevLett.96.031103}{Phys. Rev. Lett. \textbf{96}, 031103 (2006).}

\bibitem{Fan:2016} Z. Y. Fan and X. Wang, \href{https://doi.org/10.1103/PhysRevD.94.124027}{Phys. Rev. D \textbf{94}, 124027 (2016).}

\bibitem{Dymnikova:1992} I. Dymnikova, \href{https://ui.adsabs.harvard.edu/link_gateway/1992GReGr..24..235D/doi:10.1007/BF00760226}{Gen. Relativ. Gravit. \textbf{24}, 235 (1992).}


\bibitem{Roman:1983} T.A. Roman and P.G. Bergmann, \href{https://link.aps.org/doi/10.1103/PhysRevD.28.1265}{Phys. Rev. D \textbf{28} 1265 (1983)}.

\bibitem{Ayon:1998} E. Ayon-Beato and A. Garcia, \href{https://doi.org/10.1103/PhysRevLett.80.5056}{Phys. Rev. Lett. \textbf{80}, 5056 (1998).}

\bibitem{Ayon:1999}  E. Ayon-Beato and A. Garcia, \href{https://doi.org/10.1016/S0370-2693%2899%2901038-2}{Phys. Lett. B \textbf{464}, 25 (1999).}

\bibitem{Ayon:1999grg} E. Ayon-Beato and A. Garcia, \href{https://doi.org/10.1023/A:1026640911319}{Gen. Relativ. Gravit. \textbf{31}, 629 (1999).}

\bibitem{Ayon:2000} E. Ayon-Beato and A. Garcia, \href{https://doi.org/10.1016/S0370-2693%2800%2901125-4}{Phys. Lett. B \textbf{493}, 149 (2000).}

\bibitem{Bronnikov:2001} K. A. Bronnikov, \href{https://link.aps.org/doi/10.1103/PhysRevD.63.044005}
{Phys. Rev. D \textbf{63}, 044005 (2001).}

\bibitem{Ayon:2005} E. Ayon-Beato and A. Garcia, \href{https://doi.org/10.1007/s10714-005-0050-y}{Gen. Relativ. Gravit. \textbf{37}, 635 (2005).}

\bibitem{Balart:2014} L. Balart and E. C. Vagenas, \href{https://link.aps.org/doi/10.1103/PhysRevD.90.124045}{Phys. Rev. D \textbf{90}, 124045 (2014).}


\bibitem{Bronnikov:2018}K. A. Bronnikov, \href{https://doi.org/10.1142/S0218271818410055}{Int. J. Mod. Phys. D \textbf{27}, 1841005 (2018).}




\bibitem{Poshteh:2021} M. B. Jahani Poshteh and N. Riazi, \href{https://doi.org/10.1142/S0218271821500796}{Int. J. Mod. Phys. D 30,
2150079 (2021).}

\bibitem{Bronnikov:2024} K. A. Bronnikov, \href{https://doi.org/10.1103/PhysRevD.110.024021}{Phys. Rev. D \textbf{110}, 024021 (2024).}

\bibitem{Kar:2024} A. Kar and S. Kar, \href{ https://doi.org/10.1007/s10714-024-03238-4}{Gen. Relativ. Gravit. \textbf{56}, 52 (2024).}

\bibitem{Kar:2024kk} A. Kar, \href{https://doi.org/10.1140/epjc/s10052-024-13603-x}{Eur. Phys. J. C {\bf 84}, 1246 (2024)}



\bibitem{Borissova:2025} J. Borissova, S. Liberati and M. Visser, \href{https://doi.org/10.1103/PhysRevD.111.104054}{Phys. Rev. D \textbf{111}, 104054 (2025).}

\bibitem{Bueno:2025} P. Bueno, P. A. Cano, R. A. Hennigar, \'A. J. Murcia, A. Vicente-Cano, \href{https://doi.org/10.1103/qrbb-mdvm}{Phys. Rev. D {\bf 112}, 064039 (2025).}

\bibitem{Eichhorn:2025} A. Eichhorn and P. G. S. Fernandes, \href{https://doi.org/10.48550/arXiv.2508.00686}{arXiv:2508.00686 [gr-qc].}

\bibitem{Muniz:2025} C.R. Muniz, J. A. Rebouças, L. T. d. Oliveira, F. T. B. Sampaio and F. B. Lustosa, \href{https://arxiv.org/abs/2511.11419}{arxiv:2511.11419 [gr-qc].}


\bibitem{Konoplya:2025} R. A. Konoplya and A. Zhidenko, \href{https://arxiv.org/abs/2511.03066}{arxiv:2511.03066 [gr-qc].}

\bibitem{Santos:2025} L.C.N. Santos, \href{https://arxiv.org/abs/2510.21037}{arxiv:2510.21037 [gr-qc].}

\bibitem{Aydiner:2025} E. Aydiner, E. Sucu and İ. Sakallı, \href{https://doi.org/10.1016/j.dark.2025.102164}{Phys. Dark  Univ. {\bf 50}, 102164 (2025).}

\bibitem{Santos:2025prd} L.C.N. Santos, \href{https://doi.org/10.1103/PhysRevD.111.064032}{Phys. Rev. D {\bf 111}, 064032 (2025)}


\bibitem{Kar:2025} A. Kar and S. Kar, \href{https://doi.org/10.1140/epjc/s10052-025-14483-5}{ Eur. Phys. J. C \textbf{85}, 773 (2025).}

\bibitem{Myung:2009} Y. S. Myung, Y. W. Kim and Y. J. Park, \href{https://doi.org/10.1007/s10714-008-0690-9}{ Gen Relativ Gravit {\bf 41}, 1051–1067 (2009).}

\bibitem{Wu:2025} M. H. Wu, H. Guo and X. M. Kuang, \href{https://doi.org/10.48550/arXiv.2509.26270}{arXiv:2509.26270 [gr-qc].}

\bibitem{Bhattacharjee:2025} C. Bhattacharjee, S. Sau and A. Mukherjee, \href{https://doi.org/10.1140/epjc/s10052-025-14725-6}{Eur. Phys. J. C {\bf 85}, 1071 (2025).}

\bibitem{Kar:2025xx} A. Kar, A. Dey and S. Kar, \href{https://doi.org/10.48550/arXiv.2510.11364}{arXiv:2510.11364 [gr-qc].}

\bibitem{Bolokhov:2025} S. V. Bolokhov, \href{https://doi.org/10.48550/arXiv.2511.12859}{arXiv:2511.12859 [gr-qc].}

\bibitem{Wang:2025} R. Wang, Q. L. Shi, W. Xiong and P. C. Li, \href{https://doi.org/10.48550/arXiv.2512.05767}{arXiv:2512.05767 [gr-qc].}

\bibitem{Saka:2025} E. U. Saka, \href{https://doi.org/10.48550/arXiv.2512.08904}{arXiv:2512.08904 [gr-qc].}

\bibitem{Arbelaez:2026} J. P. Arbelaez, \href{https://doi.org/10.48550/arXiv.2601.22340}{arXiv:2601.22340 [gr-qc].}

\bibitem{Acharyya:2026} R. Acharyya and S. Kar, \href{https://doi.org/10.48550/arXiv.2602.15523}{arXiv.2602.15523 [gr-qc].}

\bibitem{Levin:2009sk}
J. Levin, \href{https://doi.org/10.1088/0264-9381/26/23/235010}{Class. Quant. Grav. \textbf{26}, 235010 (2009).}

\bibitem{Levin:2008ci}
J. Levin and B. Grossman, \href{https://doi.org/10.1103/PhysRevD.79.043016}{Phys. Rev. D \textbf{79}, 043016 (2009).}

\bibitem{Rana:2019bsn}
P. Rana and A. Mangalam, \href{https://doi.org/10.1088/1361-6382/ab004c}{Class. Quant. Grav. \textbf{36}, 045009 (2019).}

\bibitem{Bambhaniya:2020zno}
P. Bambhaniya, D. N. Solanki, D. Dey, A. B. Joshi, P. S. Joshi and V. Patel, \href{https://doi.org/10.1140/epjc/s10052-021-08997-x}{
Eur. Phys. J. C \textbf{81}, 205 (2021).}

\bibitem{Misra:2010pu}
V. Misra and J. Levin, \href{https://doi.org/10.1103/PhysRevD.82.083001}{
Phys. Rev. D \textbf{82}, 083001 (2010).}

\bibitem{Liu:2018vea}
C. Liu, C. Ding and J. Jing, \href{https://doi.org/10.1088/0253-6102/71/12/1461}{
Commun. Theor. Phys. \textbf{71}, 1461 (2019).}



\bibitem{Healy:2009zm}
J. Healy, J. Levin and D. Shoemaker, \href{https://doi.org/10.1103/PhysRevLett.103.131101}{
Phys. Rev. Lett. \textbf{103}, 131101 (2009).}

\bibitem{Babar:2017gsg}
G. Z. Babar, A. Z. Babar and Y. K. Lim, \href{https://doi.org/10.1103/PhysRevD.96.084052}{
Phys. Rev. D \textbf{96}, 084052 (2017).}

\bibitem{Pugliese:2013xfa}
D. Pugliese, H. Quevedo and R. Ruffini, \href{https://doi.org/10.1140/epjc/s10052-017-4769-x}{
Eur. Phys. J. C \textbf{77}, 206 (2017).}

\bibitem{Wei:2019zdf}
S. W. Wei, J. Yang and Y. X. Liu, \href{https://doi.org/10.1103/PhysRevD.99.104016}{
Phys. Rev. D \textbf{99}, 104016 (2019).}

\bibitem{Zhou:2020zys}
T. Y. Zhou and Y. Xie, \href{ https://doi.org/10.1140/epjc/s10052-020-08661-w}{
Eur. Phys. J. C \textbf{80}, 1070 (2020).}

\bibitem{Gao:2020wjz}
B. Gao and X. M. Deng, \href{https://doi.org/10.1016/j.aop.2020.168194}{
Annals Phys. \textbf{418}, 168194 (2020).}

\bibitem{Deng:2020hxw}
X. M. Deng, \href{https://doi.org/10.1140/epjc/s10052-020-8067-7}{
Eur. Phys. J. C \textbf{80}, 489 (2020).}

\bibitem{Azreg-Ainou:2020bfl}
M. Azreg-A\"\i{}nou, Z. Chen, B. Deng, M. Jamil, T. Zhu, Q. Wu and Y. K. Lim, \href{https://doi.org/10.1103/PhysRevD.102.044028}{
Phys. Rev. D \textbf{102}, 044028 (2020).}

\bibitem{Gao:2021arw}
B. Gao and X. M. Deng, \href{https://doi.org/10.1142/S0217732321502370}{
Mod. Phys. Lett. A \textbf{36}, 2150237 (2021).}

\bibitem{Lin:2021noq}
H. Y. Lin and X. M. Deng, \href{https://doi.org/10.1016/j.dark.2020.100745}{
Phys. Dark Univ. \textbf{31}, 100745 (2021).}

\bibitem{Lin:2022llz}
H. Y. Lin and X. M. Deng, \href{https://doi.org/10.3390/universe8050278}{
Universe \textbf{8}, 278 (2022).}

\bibitem{Wang:2022tfo}
R. Wang, F. Gao and H. Chen, \href{https://doi.org/10.1016/j.aop.2022.169167}{
Annals Phys. \textbf{447}, 169167 (2022).}

\bibitem{Zhang:2022psr}
J. Zhang and Y. Xie, \href{https://doi.org/10.1007/s10509-022-04046-5}{
Astrophys. Space Sci. \textbf{367}, 17 (2022)}

\bibitem{Lin:2022wda}
H. Y. Lin and X. M. Deng, \href{https://doi.org/10.1140/epjp/s13360-022-02391-6}{Eur. Phys. J. Plus \textbf{137}, 176 (2022).}

\bibitem{Zhang:2022zox}
J. Zhang and Y. Xie, \href{https://doi.org/10.1140/epjc/s10052-022-10846-4}{
Eur. Phys. J. C \textbf{82}, 854 (2022).}

\bibitem{Yao:2023ziq}
J. T. Yao and X. Li, \href{https://doi.org/10.1103/PhysRevD.108.084067}{
Phys. Rev. D \textbf{108}, 084067 (2023).}

\bibitem{Lin:2023rmo}
H. Y. Lin and X. M. Deng, \href{https://doi.org/10.1140/epjc/s10052-023-11487-x}{
Eur. Phys. J. C \textbf{83}, 311 (2023).}

\bibitem{Habibina:2022ztd}
A. S. Habibina and H. S. Ramadhan, \href{https://doi.org/10.1016/j.aop.2022.169169}{
Annals Phys. \textbf{448}, 169169 (2023).}

\bibitem{Chan:2025ocy}
Z. C. S. Chan and Y. K. Lim, \href{https://doi.org/10.1007/s10714-025-03368-3}{
Gen. Rel. Grav. \textbf{57}, 35 (2025).}

\bibitem{Al-Badawi:2025yum}
A. Al-Badawi, F. Ahmed, T. Xamidov, S. Shaymatov and {\.I}. Sakall{\i}, \href{https://doi.org/10.48550/arXiv.2503.18027}{
arXiv:2503.18027 [gr-qc].}

\bibitem{Wang:2025wob}
C. H. Wang, Y. P. Zhang, T. Zhu and S. W. Wei, \href{https://doi.org/10.48550/arXiv.2508.20558}{arXiv:2508.20558 [gr-qc].}

\bibitem{Sharipov:2025yfw}
J. Sharipov, T. Xamidov, Q. Wu, S. Shaymatov and T. Zhu, \href{https://doi.org/10.48550/arXiv.2511.10043}{
arXiv:2511.10043 [gr-qc].}

\bibitem{Wei:2025qlh}
Z. L. Wei, J. Zhang, Y. Xie and P. L. Yin, \href{https://doi.org/10.1140/epjc/s10052-025-14437-x}{
Eur. Phys. J. C \textbf{85}, 698 (2025).}




\bibitem{Tu:2023xab}
Z. Y. Tu, T. Zhu and A. Wang, \href{https://doi.org/10.1103/PhysRevD.108.024035}{Phys. Rev. D \textbf{108}, 024035 (2023).}

\bibitem{Li:2024tld}
Y. Z. Li, X. M. Kuang and Y. Sang, \href{https://doi.org/10.1140/epjc/s10052-024-12895-3}{ Eur. Phys. J. C 84, \textbf{529} (2024).}

\bibitem{QiQi:2024dwc}
Q. Qi, X. M. Kuang, Y. Z. Li and Y. Sang, \href{https://doi.org/10.1140/epjc/s10052-024-12989-y}{ Eur. Phys. J. C \textbf{84}, 645 (2024).}

\bibitem{Yang:2024lmj}
S. Yang, Y. P. Zhang, T. Zhu, L. Zhao and Y. X. Liu, \href{https://doi.org/10.1088/1475-7516/2025/01/091}{JCAP \textbf{01}, 091 (2025).}

\bibitem{Shabbir:2025kqh}
O. Shabbir, M. Jamil and M. Azreg-A\"\i{}nou, \href{https://doi.org/10.1016/j.dark.2025.101816}{Phys. Dark Univ. \textbf{47}, 101816 (2025).}

\bibitem{Junior:2024tmi}
E. L. B. Junior, J. T. S. S. Junior, F. S. N. Lobo, M. E. Rodrigues, D. Rubiera-Garcia, L. F. D. da Silva and H. A. Vieira, \href{https://doi.org/10.1140/epjc/s10052-025-14299-3}{ Eur. Phys. J. C \textbf{85}, 557 (2025).}

\bibitem{Zhao:2024exh}
L. Zhao, M. Tang and Z. Xu, \href{https://doi.org/10.1140/epjc/s10052-025-13767-0}{ Eur. Phys. J. C \textbf{85}, 36 (2025).}


\bibitem{Meng:2024cnq}
L. Meng, Z. Xu and M. Tang, \href{https://doi.org/10.1140/epjc/s10052-025-14032-0}{ Eur. Phys. J. C \textbf{85}, 306 (2025). }

\bibitem{Haroon:2025rzx}
S. Haroon and T. Zhu, \href{https://doi.org/10.1103/ckdt-wtsl}{Phys. Rev. D \textbf{112}, 044046 (2025).}

\bibitem{Lu:2025cxx}
S. Lu and T. Zhu, \href{https://doi.org/10.1016/j.dark.2025.102141}{Phys. Dark Univ. \textbf{50}, 102141 (2025).}

\bibitem{Chen:2025aqh}
J. Chen and J. Yang, \href{https://doi.org/10.1140/epjc/s10052-025-14457-7}{ Eur. Phys. J. C \textbf{85}, 726 (2025).}

\bibitem{Choudhury:2025qsh}
S. Choudhury, M. K. Hossain, G. Bauyrzhan and K. Yerzhanov, \href{https://doi.org/10.48550/arXiv.2507.00904}{arXiv:2507.00904 [gr-qc].}

\bibitem{Alloqulov:2025ucf}
M. Alloqulov, T. Xamidov, S. Shaymatov and B. Ahmedov, \href{https://doi.org/10.1140/epjc/s10052-025-14529-8}{Eur. Phys. J. C \textbf{85}, 798 (2025).}

\bibitem{Wang:2025hla}
C. H. Wang, X. C. Meng, Y. P. Zhang, T. Zhu and S. W. Wei, \href{https://doi.org/10.1088/1475-7516/2025/07/021}{JCAP \textbf{07}, 021 (2025).}

\bibitem{Alloqulov:2025bxh}
M. Alloqulov, S. Shaymatov, B. Ahmedov and T. Zhu, \href{https://doi.org/10.48550/arXiv.2508.05245}{arXiv:2508.05245 [gr-qc].}

\bibitem{Li:2025sfe}
Y. Z. Li and X. M. Kuang, \href{https://doi.org/10.48550/arXiv.2509.07333}{arXiv:2509.07333 [gr-qc].}

\bibitem{Gong:2025mne}
H. Gong, S. Long, X. J. Wang, Z. Xia, J. P. Wu and Q. Pan, \href{https://doi.org/10.48550/arXiv.2509.23318}{arXiv:2509.23318 [gr-qc].}

\bibitem{Zare:2025aek}
S. Zare, T. Zhu, L. M. Nieto, S. Lu and H. Hassanabadi, \href{https://doi.org/10.48550/arXiv.2510.05166}{arXiv:2510.05166 [gr-qc].}

\bibitem{Zahra:2025tdo}
T. Zahra, O. Shabbir, B. Majeed, M. Jamil, J. Rayimbaev and A. Shermatov, \href{https://doi.org/10.1140/epjc/s10052-025-15000-4}{Eur. Phys. J. C \textbf{85}, 1340 (2025).}

\bibitem{Li:2025eln}
G. H. Li, C. K. Qiao and J. Tao, \href{https://doi.org/10.48550/arXiv.2510.24989}{arXiv:2510.24989 [gr-qc].}

\bibitem{Deng:2025wzz}
W. Deng, S. Long, Q. Tan and J. Jing, \href{https://doi.org/10.48550/arXiv.2510.24468}{arXiv:2510.24468 [gr-qc].}

\bibitem{Ahmed:2025azu}
F. Ahmed, Q. Wu, S. G. Ghosh and T. Zhu, \href{https://doi.org/10.48550/arXiv.2511.08456}{arXiv:2511.08456 [gr-qc].}

\bibitem{Zhang:2025wni}
C. Zhang and T. Zhu, \href{https://doi.org/10.48550/arXiv.2511.14080}{arXiv:2511.14080 [gr-qc].}

\bibitem{Alloqulov:2025dqi}
M. Alloqulov, S. Shaymatov, B. Ahmedov and T. Zhu, \href{https://doi.org/10.48550/arXiv.2511.15237}{arXiv:2511.15237 [gr-qc].}

\bibitem{Lu:2025xxx} S. Lu, H. J. Lin, T. Zhu, Y. X. Liu and X. Zhang, \href{https://arxiv.org/abs/2512.11911}{arXiv:2512.11911 [gr-qc].}

\bibitem{Ahmed:2025xxx} F. Ahmed, Q. Wu, S. G. Ghosh and T. Zhu, \href{https://arxiv.org/abs/2512.24036}{arxiv:2512.24036 [gr-qc].}

\bibitem{Chen:2025xxx} R T. Chen, G. Fu, D. Zhang and J. P. Wu, \href{https://arxiv.org/abs/2601.00185}{arxiv:2601.00185 [gr-qc].}

\bibitem{Hua:2025xxx} Z. Hua, Z. T. He, J. Jiao, J. Q. Lai and Yu Tian, \href{https://arxiv.org/abs/2601.00550}{arxiv:2601.00550 [gr-qc].}

\bibitem{Gaete:2026} M. B. Gaete, J. Lin, Y. Liu and X. Zhang, \href{https://doi.org/10.48550/arXiv.2602.15609}{arXiv.2602.15609 [gr-qc].}




\bibitem{Chandrasekhar:1985kt}
S. Chandrasekhar, The mathematical theory of black holes, Oxford University Press, 1985.

\bibitem{Levin:2008} J. Levin and G. Perez-Giz, \href{https://journals.aps.org/prd/abstract/10.1103/PhysRevD.77.103005}{Phys. Rev. D \textbf{77}, 103005 (2008).}

\bibitem{Hughes:2001jr}
S. A. Hughes,
\href{https://journals.aps.org/prd/abstract/10.1103/PhysRevD.64.064004}{Phys. Rev. D \textbf{64}, 064004 (2001).}

\bibitem{Glampedakis:2002ya}
K. Glampedakis and D. Kennefick, 
\href{https://journals.aps.org/prd/abstract/10.1103/PhysRevD.66.044002}{Phys. Rev. D \textbf{66}, 044002 (2002).}
	
\bibitem{Hughes:2005qb}
S. A. Hughes, S. Drasco, E. E. Flanagan, and J. Franklin,
\href{https://journals.aps.org/prl/abstract/10.1103/PhysRevLett.94.221101}{Phys. Rev. Lett. \textbf{94}, 221101 (2005).}
	
\bibitem{Drasco:2005is}
S. Drasco, E. E. Flanagan, and S. A. Hughes,\href{https://iopscience.iop.org/article/10.1088/0264-9381/22/15/011}{Class. Quant. Grav. \textbf{22}, S801-846 (2005).}

\bibitem{Gair:2005ih}
J. R. Gair and K. Glampedakis,
\href{https://journals.aps.org/prd/abstract/10.1103/PhysRevD.73.064037}{Phys. Rev. D \textbf{73}, 064037 (2006).}

\bibitem{Glampedakis:2005cf}
K. Glampedakis and S. Babak,
\href{https://iopscience.iop.org/article/10.1088/0264-9381/23/12/013}{Class. Quant. Grav. \textbf{23}, 4167-4188 (2006).}

\bibitem{Drasco:2005kz}
S. Drasco and S. A. Hughes,
\href{https://journals.aps.org/prd/abstract/10.1103/PhysRevD.73.024027}{Phys. Rev. D \textbf{73}, 2, 024027 (2006).}

\bibitem{Sundararajan:2007jg}
P. A. Sundararajan, G. Khanna, and S. A. Hughes,
\href{https://journals.aps.org/prd/abstract/10.1103/PhysRevD.76.104005}{Phys. Rev. D \textbf{76}, 104005 (2007).}

\bibitem{Sundararajan:2008zm}
P. A. Sundararajan, G. Khanna, S. A. Hughes, and S. Drasco, 
\href{https://journals.aps.org/prd/abstract/10.1103/PhysRevD.78.024022}{Phys. Rev. D \textbf{78}, 024022 (2008).}

\bibitem{Miller:2020bft}
J. Miller and A. Pound, 
\href{https://journals.aps.org/prd/abstract/10.1103/PhysRevD.103.064048}{Phys. Rev. D \textbf{103}, 6, 064048 (2021).}

\bibitem{Isoyama:2021jjd}
S. Isoyama, R. Fujita, A. J. K. Chua, H. Nakano, A. Pound, and N. Sago,
\href{https://journals.aps.org/prl/abstract/10.1103/PhysRevLett.128.231101}{Phys. Rev. Lett. \textbf{128}, 23, 231101 (2022).}

\bibitem{Babak:2006uv}
S. Babak, H. Fang, J. R. Gair, K. Glampedakis, and S. A. Hughes, 
\href{https://journals.aps.org/prd/abstract/10.1103/PhysRevD.75.024005}{Phys. Rev. D \textbf{75}, 024005 (2007).}

\bibitem{Thorne:1980ru}
K. S. Thorne,
\href{https://doi.org/10.1103/RevModPhys.52.299}{Rev. Mod. Phys. \textbf{52}, 299-339 (1980).}

\bibitem{gravity-book}
E. Poisson and C. M. Will, Gravity: Newtonian, Post-Newtonian, Relativistic (Cambridge University Press,
Cambridge, England, 2014).

\bibitem{Robson} T. Robson, N. J. Cornish, and C. Liu, \href{https://doi.org/10.1088/1361-6382/ab1101}{Class. Quant. Grav. \textbf{36}, 10,
105011 (2019).}

\bibitem{Nyquist}A. V. Oppenheim and R. W. Schafer, {\it Discrete-Time Signal Processing}, Prentice-Hall, Englewood Cliffs, NJ, 1989. 

\bibitem{Lim:2024} 
Y. K. Lim and Z. C. Yeo, \href{https://doi.org/10.1103/PhysRevD.109.024037}{Phys. Rev. D \textbf{109}, 024037 (2024).}

\bibitem{Gradshteyn}I. Gradshteyn and I. Ryzhik, Table of Integrals, Series, and
Products (Elsevier Science, Amsterdam, 2002).


\end{thebibliography}
\end{document}